\documentclass[twocolumn,twocolappendix]{aastex631}
\usepackage{amsmath}
\usepackage{hyperref}
\received{15-Aug-2022}
\revised{06-Sep-2022}
\accepted{07-Sep-2022}
\submitjournal{Astronomical Journal}
\shorttitle{High Velocity Stars in SDSS/APOGEE DR17}
\shortauthors{Quispe-Huaynasi et al.}
\graphicspath{{./}{figures/}}
\begin{document}

\title{High Velocity Stars in SDSS/APOGEE DR17}

\author[0000-0001-8741-8642]{F. Quispe-Huaynasi}
\affiliation{Observatorio Nacional, Rio de Janeiro, RJ 20921-400, Brazil}

\author[0000-0001-7059-5116]{F. Roig}
\affiliation{Observatorio Nacional, Rio de Janeiro, RJ 20921-400, Brazil}

\author[0000-0002-9448-6261]{D. J. McDonald}
\affiliation{Department of Astronomy, University of Virginia, Charlottesville, VA 22904, USA}

\author{V. Loaiza-Tacuri}
\affiliation{Observatorio Nacional, Rio de Janeiro, RJ 20921-400, Brazil}

\author[0000-0003-2025-3147]{S. R. Majewski}
\affiliation{Department of Astronomy, University of Virginia, Charlottesville, VA 22904, USA}

\author{F. C. Wanderley}
\affiliation{Observatorio Nacional, Rio de Janeiro, RJ 20921-400, Brazil}

\author{K. Cunha}
\affiliation{Steward Observatory, University of Arizona, Tucson, AZ 85721, USA}

\author{C. B. Pereira}
\affiliation{Observatorio Nacional, Rio de Janeiro, RJ 20921-400, Brazil}

\author{S. Hasselquist}
\affiliation{Space Telescope Science Institute, Baltimore, MD 21218, USA}

\author{S. Daflon}
\affiliation{Observatorio Nacional, Rio de Janeiro, RJ 20921-400, Brazil}




\begin{abstract}
We report 23 stars having Galactocentric velocities larger than $450~\mathrm{km\,s}^{-1}$ in the final data release of the APOGEE survey.
This sample was generated using space velocities derived by complementing the high quality radial velocities from the APOGEE project in Sloan Digital Sky Survey's Data Release 17 (DR17) with distances and proper motions from Gaia early Data Release 3 (eDR3). 
We analyze the observed kinematics and derived dynamics of these stars, considering different potential models for the Galaxy. We find that three stars could be unbound depending on the adopted potential, but in general all of the stars show typical kinematics of halo stars.
The APOGEE DR17 spectroscopic results and Gaia eDR3 photometry are used to assess the stellar parameters and chemical properties of the stars.
All of the stars belong to the red giant branch, and, in general, they follow the abundance pattern of typical halo stars. There are a few exceptions that would deserve further analysis through high-resolution spectroscopy. In particular, we identify a high velocity Carbon-Enhanced Metal-Poor (CEMP) star, with Galactocentric velocity of 482 km\,s$^{-1}$.
We do not confirm any hypervelocity star in the sample, but this result is very sensitive to the adopted distances, and less sensitive to the Galactic potential.
\end{abstract}


\section{Introduction} \label{sec:intro}
High velocity stars in the Milky Way can be classified as bound or unbound stars to the Galaxy's gravitational potential. In the literature, such stars are also classified as runaway stars, hyper-runaway stars, hypervelocity stars, and high velocity halo stars. The runaway stars are bound to the Galaxy and, as defined by \citet{Blaauw1961}, are OB type stars with peculiar velocities faster than $40~\mathrm{km\,s}^{-1}$ and with an origin in OB associations located in the Galactic disk. There are two main mechanism for the production of runaway stars:
supernova explosions in binary systems \citep{Blaauw1961}, and dynamical interaction between massive stars in young clusters \citep{Poveda1967}.

Hyper-runaway or unbound runaway stars are stars that exceed the escape velocity of the Galaxy \citep{Przybilla2008}. Thermonuclear explosions of a white dwarf star orbiting another white dwarf (i.e., the ``double-degenerate scenario'') is one of the most likely channel to generate hyper-runaway stars \citep{Shen2018}. The first confirmed hyper-runaway star is HD 271791 with Galactic rest frame velocity of $630~\mathrm{km\,s}^{-1}$, and an apparent origin in the outer disk \citep{Heber2008}. 

Unlike runaway/hyper-runaway stars with origin in the Galactic disk, hypervelocity stars (HVS), as coined by \citet{Hills1988}, in principle originate in the center of the Galaxy, as a result of the interaction of binary stars with the supermassive black hole (Sgr A*) located in the center of the Milky Way (the ``Hills mechanism''). Actually, these stars were theoretically predicted as
evidence for the presence of a black hole in the center of the Galaxy \citep{Hills1988}. 
The first hypervelocity star (HVS1) was observed by \citet{Brown2005} while searching for Blue Horizontal Branch (BHB) star candidates in the first
data release of the Sloan Digital Sky Survey (SDSS), with the goal of 
spectroscopic follow-up observations with the $6.5~\mathrm{m}$ Multiple Mirror Telescope (MMT). 
HVS1 is a B-type star with a Galactic rest frame velocity of $673~\mathrm{km/s}$ located in the halo at a current Galactocentric distance of $\sim 107~\mathrm{kpc}$ \citep{Brown2014}.

Another interesting predicted population of high velocity stars are those
with an origin in globular clusters \citep{CapuzzoDolcetta2015}, and high velocity stars with an origin in the Large Magellanic Cloud \citep{Boubert2016}. 
Finally, high velocity halo stars represent the
extreme tail of the velocity distribution of this Galactic stellar population.
In general, these could be an in-situ population of stars (i.e., stars formed within Milky Way) \citep{DiMatteo2019, Belokurov2020}, or stars accreted 
during the numerous minor mergers that happened during the Galactic history formation \citep{Pereira2012, Belokurov2018, Helmi2018, Myeong2019, Koppelman2019}. \citet{Abadi2009} showed, through numerical simulation, that some halo stars acquire extreme kinematics
as a result of near-radial infalls of parent dwarf galaxies that have subsequently been disrupted by tidal forces.

As we can see, the high velocity stars are linked to extreme astrophysical phenomena in our Galaxy. Due to its kinematical characteristics and large range of Galactocentric radii, this population of stars has been
proposed as a sensitive dynamical tracer 
of the structure, shape, and dynamics of the Galaxy and its underlying, dark-matter-dominated gravitational potential
\citep{Gnedin2005, Yu2007, Unwin2008, Piffl2014, Hattori2018A, Hattori2018}. Nonetheless, until recently, carrying out such analyses was challenged by
limited access to precision measurements of all of the astrometric parameters (positions, proper motions, parallaxes) as well as radial velocities. 
As a consequence, the origin of the confirmed and candidate 
hypervelocity stars was in debate, and some stars were found to be misidentified as 
hypervelocity stars \citep{Boubert2018}. 

This situation has been changing in recent years with the release of data from the Gaia astrometric mission \citep{GaiaCol2016} and ground-based spectroscopic surveys, such as the Apache Point Observatory Galactic Evolution Experiment \citep[APOGEE,][]{Majewski2017}, 
GALactic Archaeology with HERMES \citep[GALAH,][]{Martell2017}, and 
Large Sky Area Multi-Object Fiber Spectroscopic Telescope \citep[LAMOST,][]{Cui2012}. 
On one hand, Gaia provides high precision positions, trigonometric parallaxes, and proper motions, which together provide high quality estimates (depending on a star's distance) of five out of six stellar phase space coordinates. 
On the other hand, the spectroscopic surveys, in addition to providing information on stellar atmospheric parameters and radial velocities, yield
chemical abundance measurements that can be used to constrain the origin of the high velocity stars via the concept of \textit{chemical tagging} \citep{Hawkins2018}.

After the second Gaia data release \citep[Gaia DR2;][]{GaiaCol2018}, several new searches for high velocity stars were published \citep{Marchetti2019, Hattori2018A, Li2021}, and some previous studies were revisited in light of the new astrometric data \citep{Irrgang2018, Boubert2018, Brown2018, Kreuzer2020}. Because, as these previous studies have shown, the synergy between Gaia and large-scale spectroscopic surveys allows for both discovery and a more complete and accurate kinematical and chemical characterization of high velocity stars, we propose here to perform a similar investigation of the fastest stars in the APOGEE Data Release 17 \citep[APOGEE DR17;][]{Abdurrouf2022}.  To do so, we exploit the Gaia early Data Release 3 \citep[Gaia eDR3;][]{GaiaCol2021} as an essential part to verifying whether these stars belong to one of the high-velocity star populations previously mentioned.  The data and quality control cuts used to select sources with reliable astrometric and spectroscopic measurements are described in Sect. \ref{sec:data}. In Sect. \ref{sec:sample} we present the process for determining positions and velocities in Galactocentric coordinates, and the selection of the high velocity star candidates. In Sect. \ref{sec:kinematic}, we focus on assessment of the orbital properties of this sample. The observed stellar properties are presented in Sect. \ref{sec:stellar_param}. As a means to assess potential origins for these high velocity stars, we look at their detailed chemical abundances in Sect. \ref{sec:chemical}. The main results are discussed in Sect. \ref{sec:discussion}. Finally, our conclusions are summarized in Sect. \ref{sec:conclusion}.

\section{Data}
\label{sec:data}
The chemical and astrometric data used in this work to search for and to characterize
high velocity stars come from the APOGEE DR17 and the Gaia eDR3, respectively. Since Gaia does not provide radial velocities for all stars, we use the radial velocities from APOGEE.

\subsection{Gaia eDR3}
Astrometric and photometric data used in this work come from observations of the Gaia satellite \citep{GaiaCol2016}. This is an astrometric mission that aims to measure positions and velocities of stars with precision on the order of tens of microarcseconds. Beside astrometric information, Gaia provides photometric data in the $G$, $G_{BP}$ and $G_{RP}$ bands, and radial velocities for the brightest stars ($G_{RVS} \lesssim 16.2$), measured with a radial velocity spectrograph with a resolution of $\sim 11\,000$ \citep{Soubiran2018}. More in-depth information about Gaia can be found in \citet{Brown2021}. In this work we use the Gaia eDR3 \citep{GaiaCol2021}, which provides positions, parallaxes and proper motions for $\sim 1.4$ billion stars, of which $\sim 7$ million have radial velocities already determined in Gaia DR2. 

\subsection{APOGEE DR17}
All chemical information and radial velocities used to analyze the high velocity stars come from the near-infrared ($1.51-1.70~\mu\mathrm{m}$), multi-object, stellar spectroscopic survey APOGEE \citep{Majewski2017}, which is one of the main observational programs of the of the SDSS-III \citep{Eisenstein2011} and SDSS-IV \citep{Blanton2017} surveys. The APOGEE program corresponding to SDSS-IV is designated as APOGEE-2 and the APOGEE program corresponding to SDSS-III is designated simply as APOGEE. Unlike APOGEE,
which conducted observations only from the Northern Hemisphere using the Sloan 2.5-m Telescope \citep{Gunn2006}, the installation of a second APOGEE spectrograph \citep{Wilson2019} on the duPont 2.5-m Telescope \citep{Bowen1973} enabled APOGEE-2 to conduct
observations from both the Northern (APOGEE-2N) and the Southern (APOGEE-2S) Hemispheres. 

Target selection for the two APOGEE surveys is described in \citet{Zasowski2013,Zasowski2017},  \citet{Beaton2021}, and \citet{Santana2021}.  
Data products from spectra taken in both APOGEE surveys are included in SDSS DR17 \citep{Abdurrouf2022}, after being automatically recalculated \citep{Nidever2015,Holtzman2015} using the latest APOGEE data reduction pipeline,
which uses an updated algorithm in DR17, {\texttt{Doppler}}\footnote{\url{https://github.com/dnidever/doppler}}, for radial velocity determination; this algorithm has improved the derivation of radial velocities for fainter sources.  
The APOGEE Stellar Parameters and Chemical Abundances Pipeline \citep[\texttt{ASPCAP};][]{GarciaPerez2016}, which is rooted in the {\texttt{FERRE}}\footnote{\url{https://github.com/callendeprieto/ferre}} code of \citet{AllendePrieto2006}, provides atmospheric parameters ($T_\mathrm{eff}$, $\log g$, [Fe/H]) and chemical abundances measurements for up to 20 chemical species for 733\,901 sources.
\texttt{ASPCAP} estimates the stellar atmospheric parameters by comparing observed spectra against a 
library of MARCS stellar atmospheres \citep{Meszaros2012,Jonsson2020},
generated using an $H$-band line list from \citet{Smith2021}, that updates the earlier APOGEE line list presented in \citet{Shetrone2015}
to include the Ce and Nd line identifications from \citet{Cunha2017} and \citet{Hasselquist2016}, respectively. 

For the purpose of selecting stars with reliable radial velocity measurements, we perform a series of quality cuts in the APOGEE DR17 data.
First, for any duplicated sources in the catalog, we select the entry with the higher signal-to-noise ratio (SNR) spectrum, $\mathtt{SNR} > 10$, and we only select stars with a
SNR-weighted velocity uncertainty $\mathtt{VERR} < 1~\mathrm{km\,s}^{-1}$. 
To reduce the impact of binary stars in our sample, 
we consider sources having a number of visits $\mathtt{NVISITS} > 1$, and only consider stars with $\mathtt{VSCATTER} < 1~\mathrm{km\,s}^{-1}$. Also 
the label \texttt{ASPCAPFLAG STAR\_BAD} (bit 23) was used to remove all the stars that are marked as STAR BAD. Finally, the \texttt{STARFLAG} bitmasks (bits 0, 3, 9, 4, 12, 13, 19, 22) are used to remove stars with issues associated with their radial velocity determination. These cuts reduce our sample to 370\,301 sources. 

\section{High velocity stars selection}
\label{sec:sample}

\begin{table*}[ht]
\caption{Equatorial coordinates, paralaxes from Gaia eDR3, and distances from \citet{BailerJones2021} for the  HiVel stars. The last three rows list the stars that exceed the escape velocity in the MWPotential2014 model (unbound candidates; Sect. \ref{subsec:unb_cand}).}
\label{tab:hvs_astrometric_parameter1}
\footnotesize
\centering
\begin{tabular}{lcccc}
\hline\hline
APOGEE ID & $\alpha$ & $\delta$ & $\varpi$ & $d_\mathrm{{BJ}}$ \\
& ($\mathrm{deg \pm mas}$) &
($\mathrm{deg \pm mas}$) &
($\mathrm{mas}$) &   
($\mathrm{kpc}$) \\
\hline
2M18333156-3439135 & 278.381 $\pm$ 0.013 & -34.654 $\pm$ 0.012 & 0.077 $\pm$ 0.015 & $7.134_{-0.576}^{+1.024}$ \\
2M17183052+2300281 & 259.627 $\pm$ 0.006 &  23.008 $\pm$ 0.009 & 0.110 $\pm$ 0.011 & $7.992_{-0.728}^{+0.995}$ \\
2M00465509-0022516 &  11.730 $\pm$ 0.021 &  -0.381 $\pm$ 0.013 & 0.384 $\pm$ 0.024 & $2.389_{-0.089}^{+0.099}$ \\
2M17223795-2451372 & 260.658 $\pm$ 0.016 & -24.860 $\pm$ 0.010 & 0.138 $\pm$ 0.018 & $6.452_{-0.646}^{+0.962}$ \\
2M18562350-2948361 & 284.098 $\pm$ 0.016 & -29.810 $\pm$ 0.015 & 0.124 $\pm$ 0.021 & $6.043_{-0.532}^{+0.641}$ \\
2M17472865+6118530 & 266.869 $\pm$ 0.011 &  61.315 $\pm$ 0.011 & 0.172 $\pm$ 0.011 & $5.348_{-0.221}^{+0.241}$ \\
2M18070909-3716087 & 271.788 $\pm$ 0.015 & -37.269 $\pm$ 0.014 & 0.127 $\pm$ 0.017 & $6.550_{-0.532}^{+0.510}$ \\
2M14473273-0018111 & 221.886 $\pm$ 0.029 &  -0.303 $\pm$ 0.028 & 0.062 $\pm$ 0.037 & $8.766_{-1.431}^{+1.162}$ \\
2M17145903-2457509 & 258.746 $\pm$ 0.018 & -24.964 $\pm$ 0.011 & 0.079 $\pm$ 0.020 & $6.963_{-0.728}^{+0.823}$ \\
2M16323360-1200297 & 248.140 $\pm$ 0.024 & -12.008 $\pm$ 0.013 & 0.090 $\pm$ 0.027 & $6.968_{-1.038}^{+0.963}$ \\
2M17122912-2411516 & 258.121 $\pm$ 0.054 & -24.198 $\pm$ 0.033 & 0.035 $\pm$ 0.061 & $8.469_{-1.109}^{+0.964}$ \\
2M15191912+0202334 & 229.830 $\pm$ 0.014 &   2.043 $\pm$ 0.012 & 0.075 $\pm$ 0.016 & $10.113_{-1.815}^{+2.653}$ \\
2M17054467-2540270 & 256.436 $\pm$ 0.023 & -25.674 $\pm$ 0.013 & 0.078 $\pm$ 0.026 & $7.504_{-0.836}^{+1.236}$ \\
2M16344515-1900280 & 248.688 $\pm$ 0.028 & -19.008 $\pm$ 0.015 & 0.107 $\pm$ 0.035 & $6.428_{-1.069}^{+1.097}$ \\
2M22242563-0438021 & 336.107 $\pm$ 0.017 &  -4.634 $\pm$ 0.013 & 0.063 $\pm$ 0.018 & $10.760_{-0.879}^{+0.903}$ \\
2M18051096-3001402 & 271.296 $\pm$ 0.020 & -30.028 $\pm$ 0.017 & 0.039 $\pm$ 0.022 & $7.939_{-0.809}^{+0.787}$ \\
2M18364421-3418367 & 279.184 $\pm$ 0.021 & -34.310 $\pm$ 0.020 & 0.065 $\pm$ 0.028 & $6.096_{-0.453}^{+0.594}$ \\
2M17065425-2606471 & 256.726 $\pm$ 0.022 & -26.113 $\pm$ 0.013 & 0.101 $\pm$ 0.024 & $6.828_{-0.793}^{+1.240}$ \\
2M17191361-2407018 & 259.807 $\pm$ 0.078 & -24.117 $\pm$ 0.050 & 0.280 $\pm$ 0.094 & $6.923_{-2.083}^{+1.677}$ \\
2M17412026-3431349 & 265.334 $\pm$ 0.026 & -34.526 $\pm$ 0.026 & 0.115 $\pm$ 0.041 & $5.626_{-1.321}^{+2.178}$ \\
\hline
2M14503361+4921331 & 222.640 $\pm$ 0.011 &  49.359 $\pm$ 0.014 & 0.103 $\pm$ 0.014 & $7.457_{-0.689}^{+0.675}$ \\
2M15180013+0209292 & 229.501 $\pm$ 0.014 &   2.158 $\pm$ 0.013 & 0.047 $\pm$ 0.015 & $11.714_{-1.449}^{+2.777}$ \\
2M19284379-0005176 & 292.182 $\pm$ 0.015 &  -0.088 $\pm$ 0.014 & 0.078 $\pm$ 0.019 & $8.055_{-0.943}^{+1.170}$ \\
\hline
\end{tabular}
\end{table*}

\begin{table*}[ht]
\caption{Proper motions from Gaia eDR3, radial velocities from APOGEE DR17, radial velocity from Gaia eDR3 (when available), and estimated Galactocentric velocity for the HiVel stars. The last three rows correspond to the unbound candidates.}
\label{tab:hvs_astrometric_parameter2}
\footnotesize
\centering
\begin{tabular}{lccccc}
\hline\hline
APOGEE ID &
$\mu_{\alpha}\cos\delta$ &
$\mu_{\delta}$ &   
$v_{\mathrm{rad}_\mathrm{APOGEE}}$ &
$v_{\mathrm{rad}_\mathrm{Gaia}}$ &  
$v_\mathrm{GC}$ \\
& ($\mathrm{mas\,yr^{-1}}$) &
($\mathrm{mas\,yr^{-1}}$) &   
($\mathrm{km\,s^{-1}}$) &
($\mathrm{km\,s^{-1}}$) &  
($\mathrm{km\,s^{-1}}$) \\
\hline
2M18333156-3439135 & -11.640 $\pm$ 0.017 &  -4.019 $\pm$ 0.014 & -334.832 $\pm$ 0.779 &       \nodata        & $450.574_{-17.739}^{+37.677}$  \\
2M17183052+2300281 &  -4.145 $\pm$ 0.008 & -14.259 $\pm$ 0.010 &   18.794 $\pm$ 0.205 &   20.466 $\pm$ 0.645 & $451.366_{-44.668}^{+59.797}$  \\
2M00465509-0022516 &  -5.623 $\pm$ 0.031 &  17.908 $\pm$ 0.031 &   70.881 $\pm$ 0.083 &   70.188 $\pm$ 1.285 & $463.276_{-7.151}^{+8.803}$    \\
2M17223795-2451372 &  -1.433 $\pm$ 0.021 &  -5.171 $\pm$ 0.014 &  435.777 $\pm$ 0.096 &  435.660 $\pm$ 1.098 & $463.408_{-3.003}^{+2.867}$    \\
2M18562350-2948361 &  -0.471 $\pm$ 0.021 &   0.279 $\pm$ 0.018 &  356.485 $\pm$ 0.157 &       \nodata        & $464.402_{-0.358}^{+0.380}$    \\
2M17472865+6118530 &  10.871 $\pm$ 0.017 &  -9.406 $\pm$ 0.015 & -212.967 $\pm$ 0.110 & -213.278 $\pm$ 0.839 & $469.810_{-14.870}^{+16.624}$  \\
2M18070909-3716087 &  11.487 $\pm$ 0.018 &  -4.047 $\pm$ 0.014 &   91.719 $\pm$ 0.204 &   90.634 $\pm$ 1.756 & $482.896_{-23.709}^{+23.944}$  \\
2M14473273-0018111 &   2.733 $\pm$ 0.039 &  -1.712 $\pm$ 0.039 &  342.836 $\pm$ 0.592 &       \nodata        & $453.063_{-7.773}^{+7.665}$    \\
2M17145903-2457509 &  -3.426 $\pm$ 0.022 &  -6.802 $\pm$ 0.015 &  440.678 $\pm$ 0.622 &  441.401 $\pm$ 1.373 & $454.279_{-0.833}^{+1.370}$    \\
2M16323360-1200297 &  -8.011 $\pm$ 0.033 &  -2.397 $\pm$ 0.023 & -463.083 $\pm$ 0.307 &       \nodata        & $459.954_{-4.847}^{+7.701}$    \\
2M17122912-2411516 &   3.258 $\pm$ 0.070 &  -2.919 $\pm$ 0.049 & -378.209 $\pm$ 0.718 &       \nodata        & $460.280_{-6.593}^{+6.625}$    \\
2M15191912+0202334 &   4.066 $\pm$ 0.019 &  -9.831 $\pm$ 0.017 &   54.762 $\pm$ 0.122 &   55.099 $\pm$ 1.105 & $470.407_{-73.310}^{+108.852}$ \\
2M17054467-2540270 & -15.615 $\pm$ 0.030 &   0.717 $\pm$ 0.019 &   55.665 $\pm$ 0.260 &       \nodata        & $478.116_{-54.718}^{+88.233}$  \\
2M16344515-1900280 &   5.948 $\pm$ 0.038 & -11.929 $\pm$ 0.024 &  295.646 $\pm$ 0.239 &       \nodata        & $481.088_{-38.416}^{+42.192}$  \\
2M22242563-0438021 &   1.565 $\pm$ 0.021 & -13.190 $\pm$ 0.015 & -180.219 $\pm$ 0.076 &       \nodata        & $481.810_{-56.742}^{+57.497}$  \\
2M18051096-3001402 &   1.347 $\pm$ 0.027 &  -9.845 $\pm$ 0.019 &  416.464 $\pm$ 0.131 &       \nodata        & $487.275_{-12.253}^{+13.960}$  \\
2M18364421-3418367 &  -2.512 $\pm$ 0.027 &  -2.976 $\pm$ 0.023 &  458.399 $\pm$ 0.171 &       \nodata        & $492.254_{-2.691}^{+2.015}$    \\
2M17065425-2606471 &  -3.017 $\pm$ 0.029 & -14.325 $\pm$ 0.019 & -437.308 $\pm$ 0.024 &       \nodata        & $507.591_{-23.769}^{+50.363}$  \\
2M17191361-2407018 &   5.908 $\pm$ 0.102 &  -3.076 $\pm$ 0.072 &  357.739 $\pm$ 0.144 &       \nodata        & $508.340_{-27.027}^{+28.941}$  \\
2M17412026-3431349 &   3.407 $\pm$ 0.034 &  -1.289 $\pm$ 0.024 &  456.415 $\pm$ 0.020 &       \nodata        & $526.611_{-5.270}^{+10.700}$   \\
\hline
2M14503361+4921331 & -15.377 $\pm$ 0.013 & -13.192 $\pm$ 0.018 & -120.268 $\pm$ 0.008 & -123.121 $\pm$ 0.403 & $512.687_{-63.112}^{+59.956}$  \\
2M15180013+0209292 &   4.194 $\pm$ 0.019 &  -9.785 $\pm$ 0.017 &   54.910 $\pm$ 0.644 &   54.354 $\pm$ 0.804 & $546.256_{-61.675}^{+118.447}$ \\
2M19284379-0005176 &  -1.037 $\pm$ 0.020 & -21.443 $\pm$ 0.017 &  -30.883 $\pm$ 0.234 &       \nodata        & $649.449_{-90.470}^{+110.097}$ \\
\hline
\end{tabular}
\end{table*}

Because our goal is to characterize stars with high velocity within the Milky Way, we consider stars with space velocity in Galactocentric coordinates (GC) greater than $450~\mathrm{km\,s}^{-1}$, hereafter \textbf{HiVel stars}, close to the value considered in \citet{Li2021}. To determine the positions and velocities in GC, it is necessary to have the full astrometric parameters (positions, proper motions, distances and radial velocities). In our case, radial velocities are obtained from APOGEE DR17, and positions and proper motions from Gaia eDR3. At variance with the works of \citet{Hattori2018A}, \citet{Marchetti2019}, and \citet{Li2021}, which use distances computed as the inverse of the parallax for sources with positive parallax ($\varpi > 0$) and low fractional parallax error ($f = \sigma_{\varpi}/\varpi \leq 0.1, 0.2$), here we use the photogeometric distances estimated by \citet{BailerJones2021}. These distances were determined using a Bayesian probabilistic approach, in which the parallax is used to determine the geometric distances for nearby sources, and the magnitude and color given by the Gaia catalogue is included to determine the photogeometric distances for distant sources. 
Alternative distances for APOGEE stars have been provided through two SDSS Value Added Catalogs: the StarHorse catalog, that
uses photometric information from other catalogs to estimate the distance applying also a Bayesian approach \citep{Anders2019, Anders2022}, and
the AstroNN catalog that gives distances determined using a deep neural network \citep{Leung2019}. Since the distance has a direct effect on the GC velocity, we compare the distances obtained by the three catalogs and discuss their possible effect on our sample in Appendix \ref{app:distances}.

Since we do not consider the distance cutoff imposed by $f \leq 0.2$, our sample of stars will not only include nearby stars but also more distant stars. However, the GC velocities will depend on the priors assumed in the determination of the distances.
Then, with the full astrometric parameters, the transformation from the International Celestial Reference System (ICRS) to GC coordinates is performed. The uncertainty propagation during the transformation is computed with 1\,000 Monte Carlo (MC) realizations using multivariate normal distributions $\mathcal{N}(\boldsymbol{\theta}, \boldsymbol{\Sigma})$, where $\boldsymbol{\theta} = \left(\alpha, \delta, \varpi, \mu_{\alpha}^{*}, \mu_{\delta}, v_{r}\right)$ are the observable parameters, and $\boldsymbol{\Sigma}$ is a $6 \times 6$ covariance matrix composed of the uncertainties and the correlation coefficients between the astrometric parameters provided by Gaia\footnote{Correlation coefficients between radial velocity and the other parameters are assumed to be zero, because they were measured with different instruments.}. Then, each realization is transformed to GC coordinates, and we use the median, the 16th and 84th percentiles over the distributions in GC coordinates to get the position and velocity ($v_\mathrm{GC}$) for each star. This process was performed using the software \texttt{Astropy} \citep{astropy:2013, astropy:2018} and the \texttt{Pyia} package \citep{adrian-price-whelan_2018}. The parameters used in the transformation were the parameters set by default in version 4.0 of the \texttt{Astropy} Galactocentric frame, where the distance from the Sun to the Galactic center is $8.122~\mathrm{kpc}$ \citep{GravityCol2018}, the distance from the Sun to the Galactic midplane is $20.8~\mathrm{pc}$ \citep{Bennett2019}, and the cartesian velocity of the Sun in the Galactocentric frame is $(12.9,\, 245.6,\, 7.78)~\mathrm{km~s}^{-1}$ \citep{Drimmel2018}. 

After this process, we get a sample of 70 stars with velocities larger than $450~\mathrm{km~s}^{-1}$. However, this sample of stars was obtained without considering any cut on the Gaia data. In order to select a reliable sample of stars in Gaia, we only consider stars with positive parallax ($\varpi > 0$). We also use the parameter $\mathtt{ruwe} < 1.4$ to select sources with good astrometric solution. Besides, using the catalog gedr3spur.main \citep{Rybizki2021}, hosted at the German Astrophysical Virtual Observatory (GAVO), we select stars with $\mathtt{fidelity\_v2} > 0.5$, classified as good sources, and $\mathtt{norm\_dG} < -3$ to consider stars with good color measurement. Finally, as adopted by \citet{Marchetti2019}, we select stars with relative error in GC velocity $\sigma_v/v_\mathrm{GC} < 30 \%$. After all these considerations, we are left with a sample of 26 stars.

We further analyze the APOGEE spectra of these 26 stars, and we discard 3 stars that show problems in their spectra, not allowing to obtain a reliable radial velocity. This left us with 23 stars, that constitute our final HiVel sample. Within this sample, nine stars have radial velocities determined both in Gaia eDR3 and APOGEE DR17, and we verified that they are in good agreement. Astrometric parameters and velocities for the HiVel stars are displayed in Tables \ref{tab:hvs_astrometric_parameter1} and \ref{tab:hvs_astrometric_parameter2}.

\subsection{Unbound candidates}
\label{subsec:unb_cand}

\begin{figure}
    \centering
    \includegraphics[width=\columnwidth]{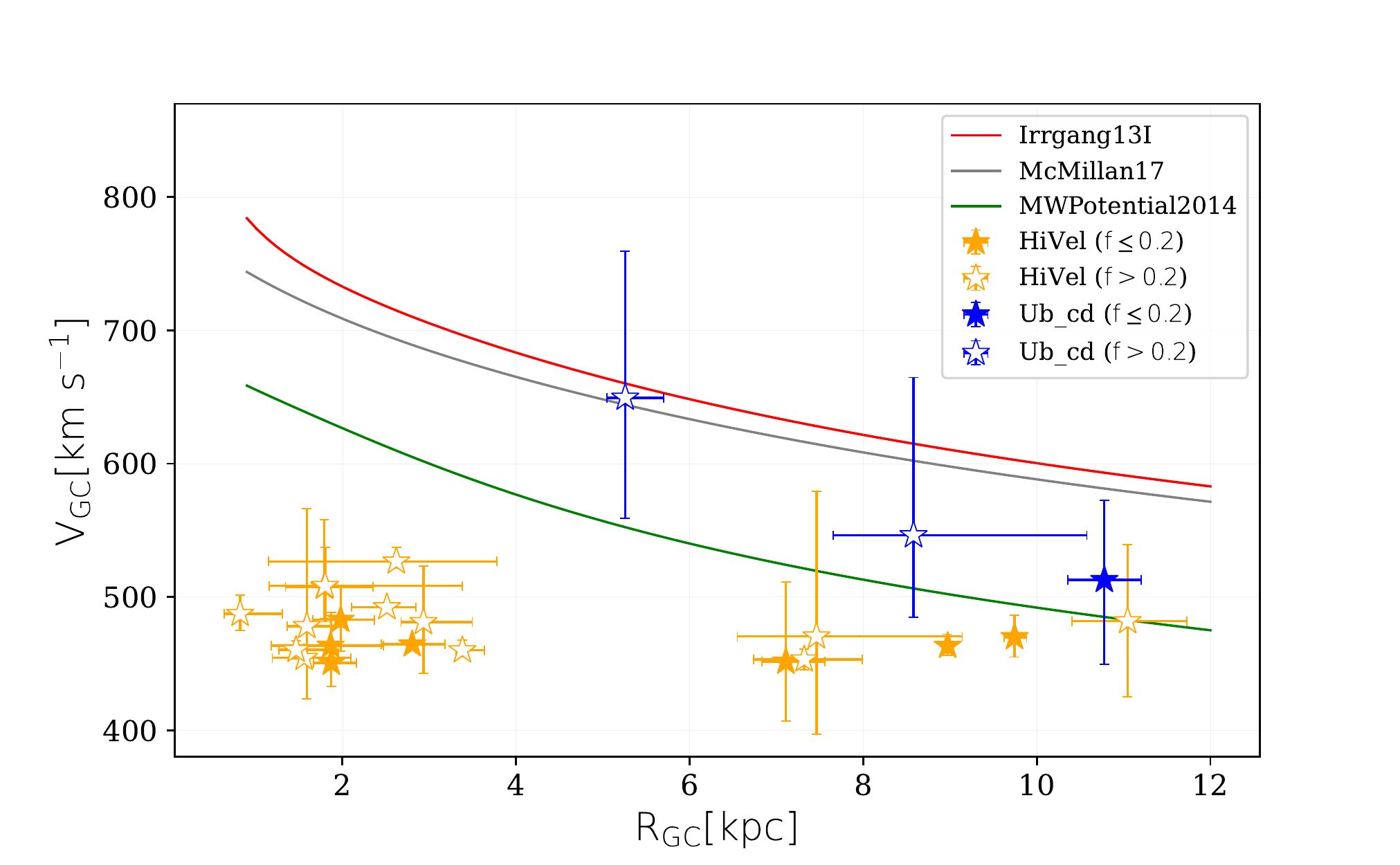}
    \caption{Galactocentric velocity as a function of the Galactocentric distance. Filled markers have $f \leq 0.2$ while hollow markers have $f > 0.2$. Orange markers represent stars with velocities $>450~\mathrm{km\,s}^{-1}$. Blue markers correspond to stars that exceed the escape velocity of the MWPotential2014 potential (unbound candidates). Red, gray and green lines are the Galactic escape velocity curves from the Irrgang13I, McMillan17, and MWPotential2014 potentials, respectively.}
    \label{fig:identifying_unbound_candidates}
\end{figure}

To identify if the HiVel stars are unbound candidates, we calculate for each star the probability of being unbound as the ratio of 1\,000 MC realizations resulting in a GC velocity $v_\mathrm{GC}$ higher than the escape speed from the Galaxy at the current position of the star
\[
p_{ub} = n(v_\mathrm{GC} > v_\mathrm{esc})/1000,
\]
as defined in \cite{Marchetti2021}. This is computed for the gravitational potential models of Irrgang Model I \citep{Irrgang2013}, McMillan \citep{McMillan2017}, and MWPotential2014 \citep{Bovy2015}, which are fully implemented in the \texttt{galpy} library \citep{Bovy2015}. The red, gray and green lines in Fig. \ref{fig:identifying_unbound_candidates} represent the escape velocity curves calculated using the above mentioned potentials. Filled markers correspond to stars with $f \leq 0.2$, which are the stars with more precise parallax. Open markers correspond to stars with $f > 0.2$. Orange markers are stars with GC velocities larger than $450~\mathrm{km~s}^{-1}$, and blue markers correspond to stars with velocities that exceed the escape velocity curve of the MWPotential2014 model, but do not exceed the escape velocity of the other potentials. We refer to such three stars as the \textit{unbound candidates}. 
Note, however, that the excess with respect to the MWPotential2014 escape velocity is within the $1\sigma$ uncertainties of the candidates' velocities.

\section{Kinematics and orbit integrations}
\label{sec:kinematic}
\begin{figure*}
    \centering
    \includegraphics[width=0.97\columnwidth]{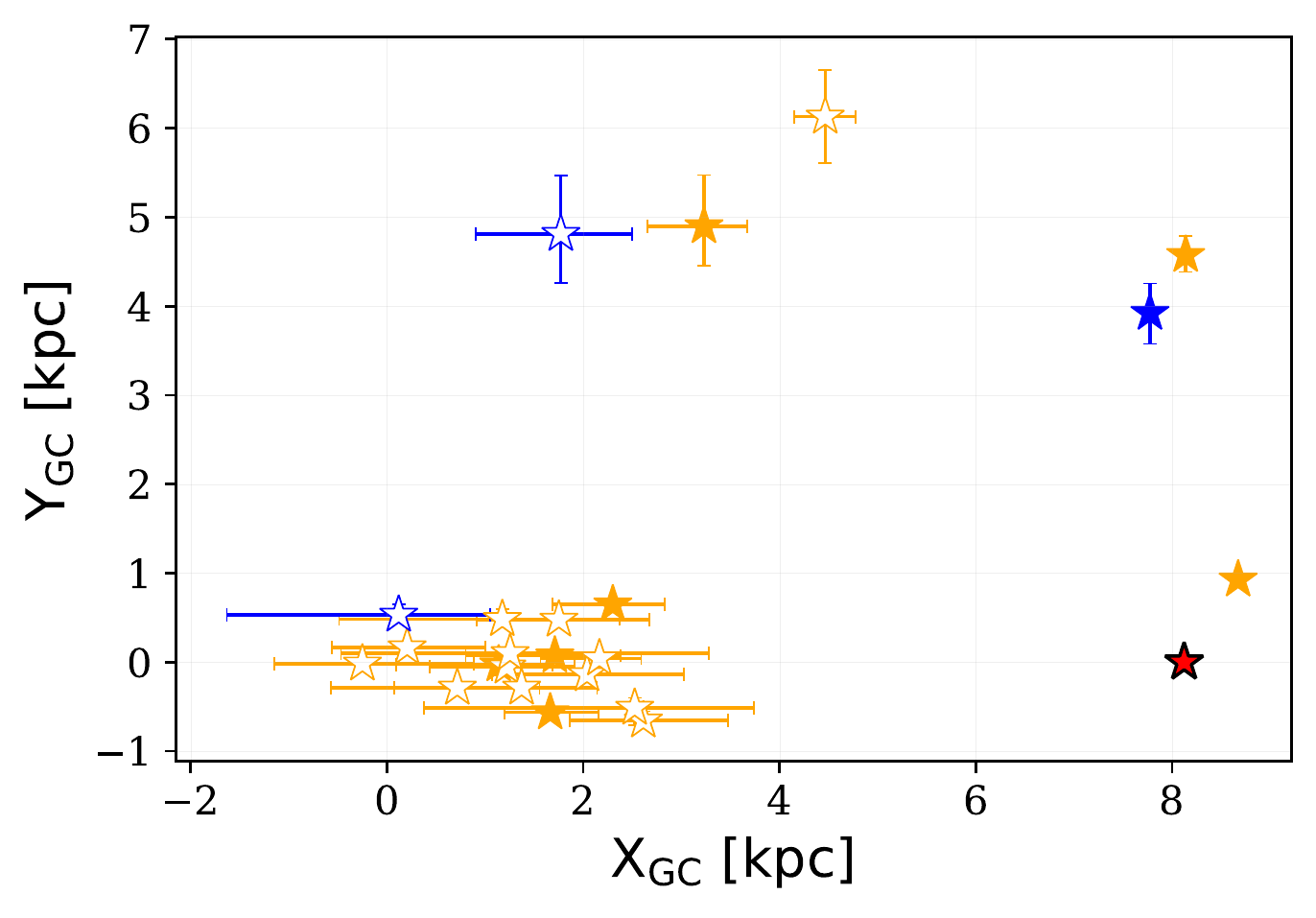}
    \includegraphics[width=\columnwidth]{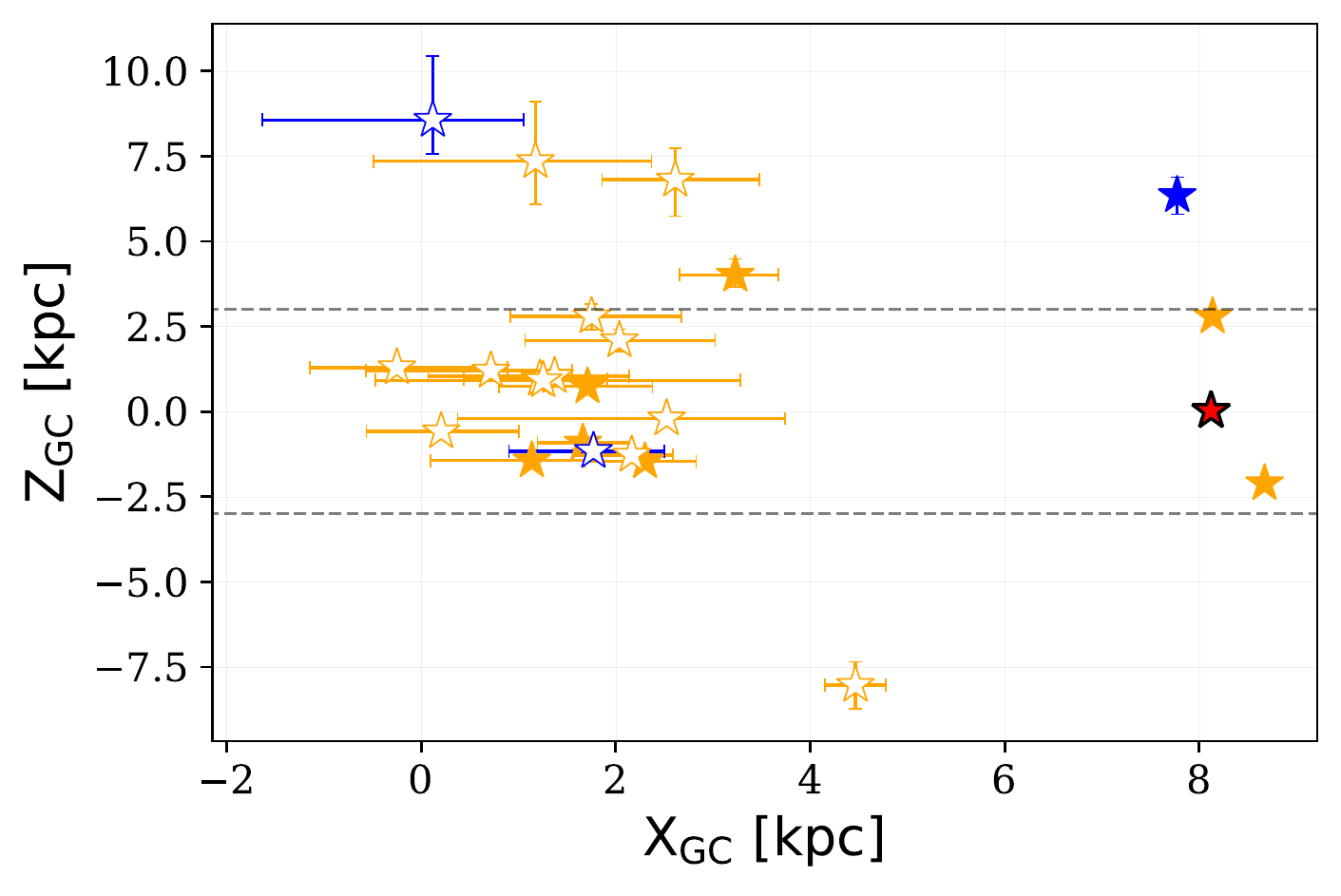}
    \caption{Spatial distribution of HiVel stars in the left handed Galactocentric coordinates. Left, distribution in the $X$-$Y$ plane. Right, distribution in the $X$-$Z$ plane. Dashed horizontal lines indicate the transition between the thick disk and the halo. The Sun is located at $(X,Y,Z)=(8.12,\,0.00,\,0.02)$~kpc. Markers and colors are the same as in Fig. \ref{fig:identifying_unbound_candidates}.}
    \label{fig:spatial_distribution}
\end{figure*}

Using the positions and velocities in the left-handed GC reference frame, Fig. \ref{fig:spatial_distribution} shows the HiVel stars spatial distribution. The left panel shows the spatial distribution in the $XY$-plane (the Galactic plane), and the right panel shows the spatial distribution in the $XZ$-plane. The horizontal dashed lines in the $XZ$-plane mark the boundaries between the thick disk and the halo region. Some HiVel stars are in the disk region, while other fall in the halo region.

\begin{figure}
    \centering
    \includegraphics[width=\columnwidth]{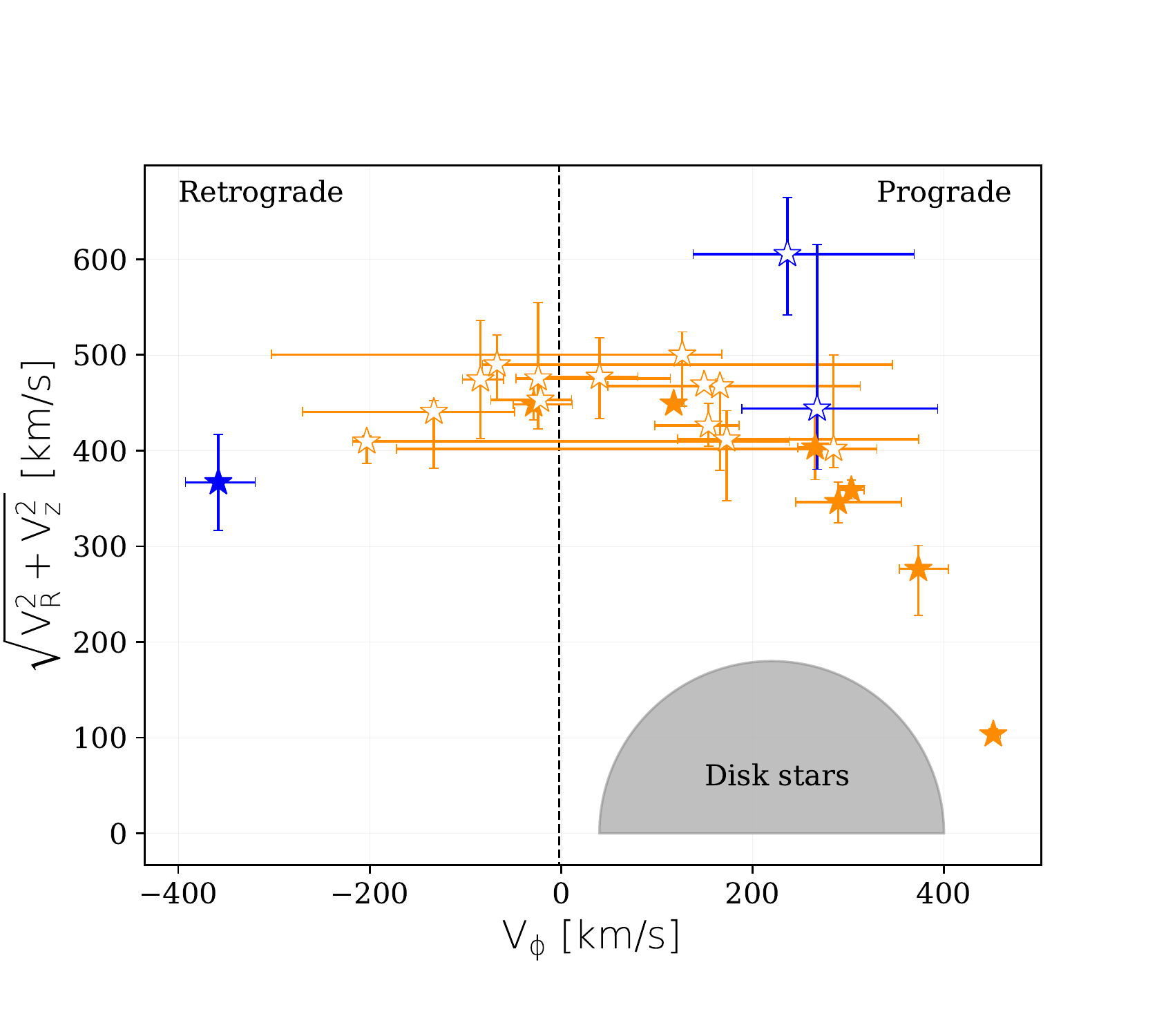}
    \caption{Toomre diagram of the HiVel sample. Markers and colors are the same as in Fig. \ref{fig:identifying_unbound_candidates}. The grey region represents stars with kinematic behavior typical of the disk. The dashed vertical line separates stars with retrograde and prograde motion.}
    \label{fig:toomre_diagram}
\end{figure}

Aiming to identify how these stars behave kinematically, we use the Toomre diagram in Galactocentric cylindrical coordinates. This diagram is widely used in the literature and allows stars to be classified into Galactic populations using information based on the velocity components, without being necessary to assume a gravitational potential for the Galaxy. On the $X$-axis and $Y$-axis of Fig. \ref{fig:toomre_diagram}, we represent the azimuthal velocity component $V_{\phi}$ and the $\sqrt{V_{R}^2 + V_{z}^2}$ component, respectively. Stars with a kinetic behavior typical of the disk occupy the gray region in the plot. We conclude that all the HiVel stars behave kinetically as halo stars. We also see that some stars display retrograde $(V_{\phi} < 0)$ motions, and other have prograde motions $(V_{\phi} > 0)$.

\subsection{Orbit integration}
\label{sec:orbit}

Using the full information in the phase space (positions and velocities in GC), we perform back-in-time orbit integrations considering 1\,000 MC realizations for each star. The aim for this is twofold: (i) to calculate the orbital parameters, in order to confirm the kinematics of these stars, and (ii) to put constraints on the possible places of origin through the analysis of the stars' trajectories. For the orbit integration, we used the Irrgang Model I gravitational potential \citep{Irrgang2013}, implemented in \texttt{galpy}. This potential is an updated version of classical \citet{Allen1991} model, and it is composed of a Plummer potential for the bulge, a Miyamoto-Nagai potential for the disk and a spherical potential for the Galactic halo. We choose this potential because recent results from Gaia DR2, using dynamical tracers like globular clusters and dwarf galaxies, is in agreement with this Model I parameters \citep{Irrgang2018}. Details about the model can be found in Appendix \ref{app:potential}. 

\begin{figure*}
    \centering
    \includegraphics[width=\columnwidth]{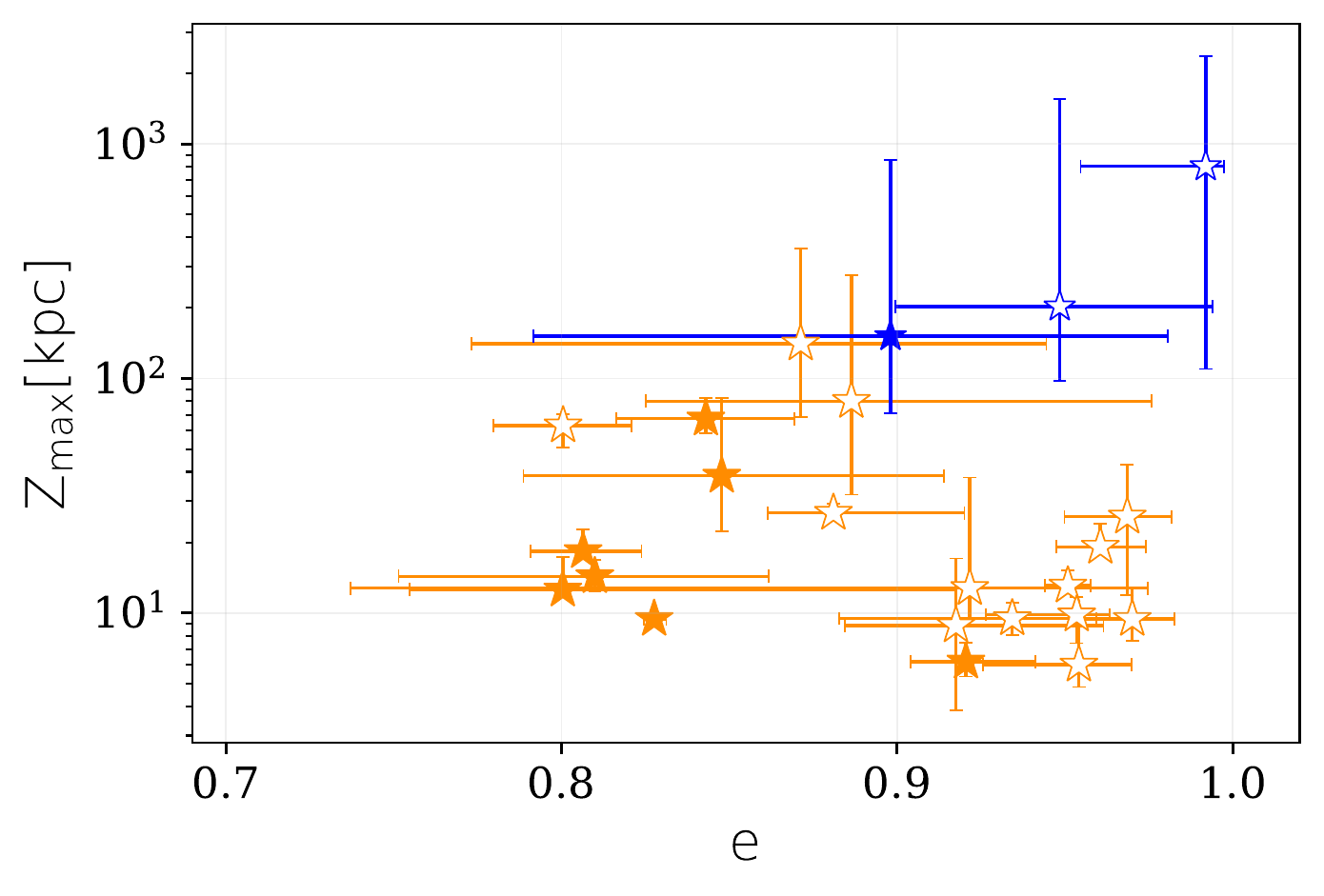}
    \includegraphics[width=\columnwidth]{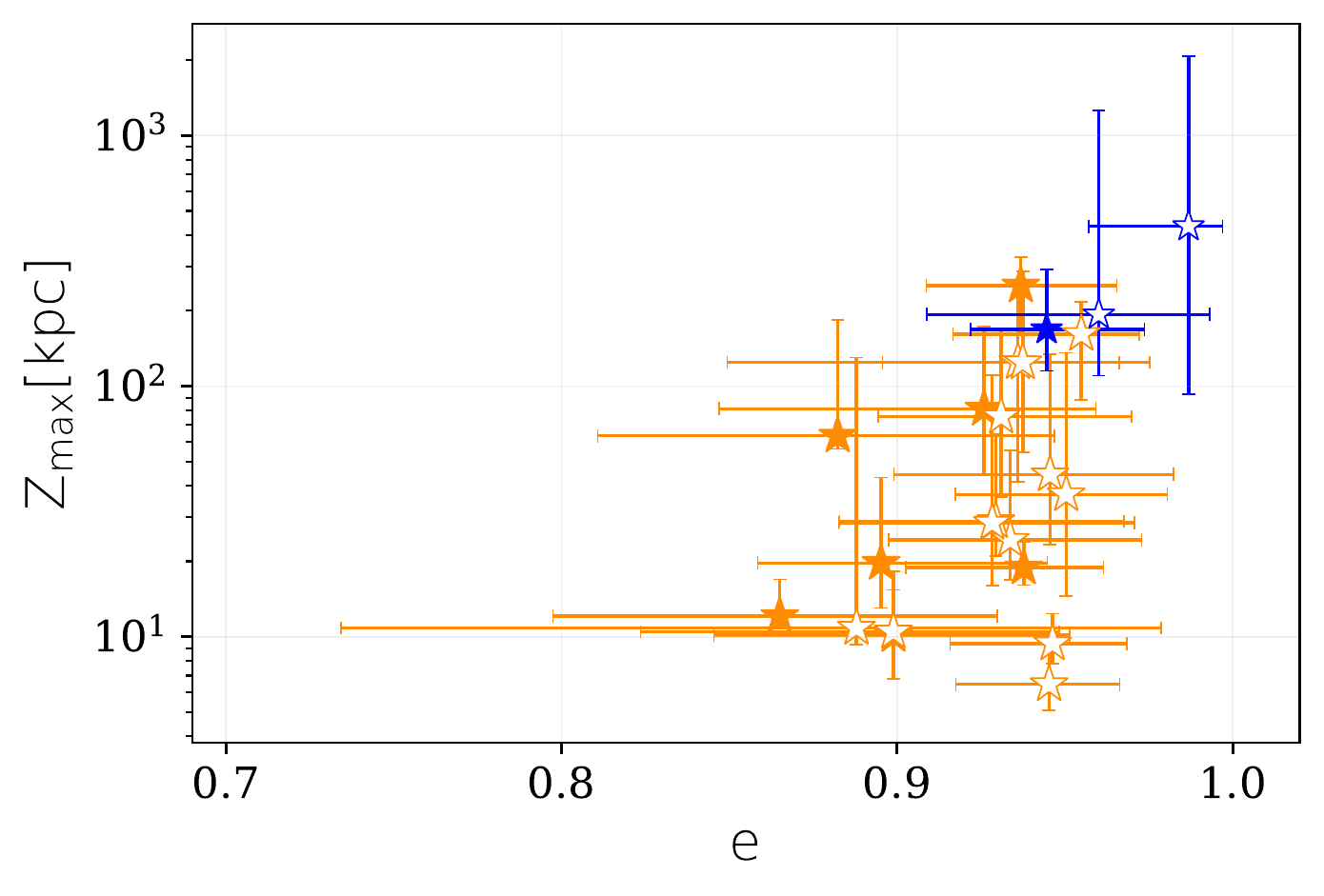}
    \caption{Maximum height over the Galactic plane ($Z_\mathrm{max}$) reached during orbit integration as a function of orbital eccentricity ($e$). Markers and colors are the same as in Fig. \ref{fig:identifying_unbound_candidates}. Left, without the LMC. Right, including the LMC.}
    \label{fig:zmax-ecc}
\end{figure*}

Table \ref{tab:Orbital parameters} shows the orbital parameters obtained from a total integration time of 10~Ga. The left panel of Fig. \ref{fig:zmax-ecc} shows the maximum distance above the Galactic plane ($Z_\mathrm{{max}}$) as a function of the orbital eccentricity ($e$). Since stars with similar orbits reside in specific regions of this plane, this kind of diagram was used by \citet{Boeche2013} to identify different populations of stars. We can see that the HiVel stars are highly eccentric ($e > 0.7$), and reach $Z_\mathrm{max} > 6~\mathrm{kpc}$, similar to the halo stars in the Galaxy.

\begin{table*}[ht]
\caption{Orbital parameters derived for the HiVel stars in the Irrgang Model I potential: perigalacticon and apogalacticon, orbital eccentricity, maximum height over the Galactic plane, and orbital energy. The last three rows correspond to the unbound candidates.}
\label{tab:Orbital parameters}
\footnotesize
\centering
\begin{tabular}{lccccc}
\hline\hline
APOGEE ID & $R_\mathrm{{peri}}$ & $R_\mathrm{{apo}}$ & $e$ & $Z_\mathrm{{max}}$ & $E$  \\
& (kpc) & (kpc) & & (kpc) & ($\mathrm{km^{2}\,s^{-2}}$) \\
\hline
2M18333156-3439135 &  $1.40_{-0.60}^{+0.28}$ &   $12.61_{-0.52}^{+4.87}$       & $0.80_{-0.05}^{+0.12}$  &  $12.57_{-0.51}^{+4.80}$       & $-161629_{-1602}^{+14663}$   \\
2M17183052+2300281 &  $4.74_{-0.65}^{+0.88}$ &   $57.48_{-22.68}^{+67.38}$     & $0.85_{-0.06}^{+0.07}$  &  $38.45_{-16.11}^{+44.24}$     & $-92582_{-21546}^{+32264}$   \\
2M00465509-0022516 &  $8.75_{-0.04}^{+0.06}$ &   $81.66_{-6.96}^{+9.65}$       & $0.81_{-0.02}^{+0.02}$  &  $18.34_{-0.99}^{+4.32}$       & $-77197_{-3610}^{+4520}$     \\
2M17223795-2451372 &  $0.63_{-0.28}^{+0.12}$ &   $14.19_{-4.42}^{+4.30}$       & $0.92_{-0.02}^{+0.02}$  &  $6.20_{-0.84}^{+1.28}$        & $-159900_{-21318}^{+13641}$  \\
2M18562350-2948361 &  $2.08_{-0.26}^{+0.23}$ &   $21.74_{-2.23}^{+2.40}$       & $0.83_{-0.00}^{+0.00}$  &  $9.43_{-0.17}^{+0.37}$        & $-136686_{-5149}^{+4940}$    \\
2M17472865+6118530 &  $8.48_{-0.15}^{+0.17}$ &   $99.59_{-17.02}^{+24.02}$     & $0.84_{-0.03}^{+0.03}$  &  $67.41_{-9.00}^{+15.30}$      & $-69195_{-7615}^{+8769}$     \\
2M18070909-3716087 &  $1.82_{-0.39}^{+0.43}$ &   $17.38_{-1.47}^{+1.85}$       & $0.81_{-0.06}^{+0.05}$  &  $14.33_{-1.93}^{+2.51}$       & $-146571_{-3655}^{+4415}$    \\
2M14473273-0018111 &  $7.03_{-0.57}^{+0.59}$ &   $63.56_{-11.19}^{+13.87}$     & $0.80_{-0.02}^{+0.02}$  &  $63.05_{-12.22}^{+7.45}$      & $-87709_{-8117}^{+8176}$     \\
2M17145903-2457509 &  $0.18_{-0.08}^{+0.11}$ &   $11.85_{-2.13}^{+2.93}$       & $0.97_{-0.02}^{+0.01}$  &  $9.42_{-1.82}^{+3.35}$        & $-170460_{-11162}^{+12666}$  \\
2M16323360-1200297 &  $1.73_{-0.59}^{+0.55}$ &   $27.77_{-0.71}^{+2.76}$       & $0.88_{-0.02}^{+0.04}$  &  $26.71_{-0.80}^{+2.56}$       & $-125178_{-1191}^{+4366}$    \\
2M17122912-2411516 &  $0.41_{-0.18}^{+0.45}$ &   $11.46_{-1.29}^{+4.19}$       & $0.93_{-0.05}^{+0.03}$  &  $9.48_{-1.46}^{+1.58}$        & $-168914_{-7569}^{+15243}$   \\
2M15191912+0202334 &  $4.90_{-1.71}^{+4.24}$ &   $80.47_{-47.83}^{+668.97}$    & $0.89_{-0.06}^{+0.09}$  &  $80.10_{-48.14}^{+195.29}$    & $-78481_{-38755}^{+67471}$   \\
2M17054467-2540270 &  $0.93_{-0.61}^{+0.73}$ &   $13.00_{-3.42}^{+26.04}$      & $0.92_{-0.18}^{+0.05}$  &  $12.75_{-3.21}^{+25.04}$      & $-161549_{-12858}^{+51354}$  \\
2M16344515-1900280 &  $0.48_{-0.19}^{+0.40}$ &   $28.87_{-5.71}^{+14.18}$      & $0.97_{-0.02}^{+0.01}$  &  $25.73_{-13.83}^{+17.08}$     & $-124089_{-10863}^{+18192}$  \\
2M22242563-0438021 &  $9.86_{-0.92}^{+0.92}$ &  $143.20_{-73.18}^{+232.11}$    & $0.87_{-0.10}^{+0.07}$  &  $140.93_{-72.56}^{+217.03}$   & $-54355_{-28938}^{+32468}$   \\
2M18051096-3001402 &  $0.21_{-0.09}^{+0.16}$ &    $8.35_{-1.79}^{+4.07}$       & $0.95_{-0.03}^{+0.02}$  &  $6.02_{-1.18}^{+3.19}$        & $-189533_{-13576}^{+21818}$  \\
2M18364421-3418367 &  $0.66_{-0.18}^{+0.17}$ &   $25.97_{-4.09}^{+3.64}$       & $0.95_{-0.01}^{+0.01}$  &  $13.08_{-0.43}^{+2.10}$       & $-129243_{-8428}^{+6150}$    \\
2M17065425-2606471 &  $0.46_{-0.14}^{+0.23}$ &   $21.23_{-0.35}^{+7.15}$       & $0.96_{-0.01}^{+0.01}$  &  $19.14_{-1.37}^{+4.94}$       & $-138320_{-796}^{+13488}$    \\
2M17191361-2407018 &  $0.97_{-0.56}^{+0.85}$ &   $22.85_{-5.60}^{+10.03}$      & $0.92_{-0.03}^{+0.04}$  &  $8.84_{-4.98}^{+8.25}$        & $-135364_{-14145}^{+17308}$  \\
2M17412026-3431349 &  $0.89_{-0.51}^{+0.44}$ &   $34.80_{-17.44}^{+14.51}$     & $0.95_{-0.03}^{+0.01}$  &  $9.84_{-2.41}^{+1.83}$        & $-115913_{-33836}^{+15721}$  \\
\hline
2M14503361+4921331 & $10.72_{-0.43}^{+0.48}$ &  $199.69_{-110.96}^{+941.34}$   & $0.90_{-0.11}^{+0.08}$  &  $151.62_{-80.66}^{+706.35}$   & $-40949_{-32571}^{+34658}$   \\
2M15180013+0209292 &  $6.78_{-1.60}^{+3.80}$ &  $252.06_{-154.06}^{+3238.42}$  & $0.95_{-0.05}^{+0.05}$  &  $202.81_{-105.19}^{+1355.99}$ & $-32646_{-37646}^{+82007}$   \\
2M19284379-0005176 &  $5.26_{-2.38}^{+0.44}$ & $1302.64_{-1180.38}^{+2946.40}$ & $0.99_{-0.04}^{+0.01}$  &  $803.96_{-694.01}^{+1559.30}$ & $-4113_{-57260}^{+82321}$ \\
\hline
\end{tabular}
\end{table*}

\begin{figure*}
    \centering
    \includegraphics[width=\columnwidth]{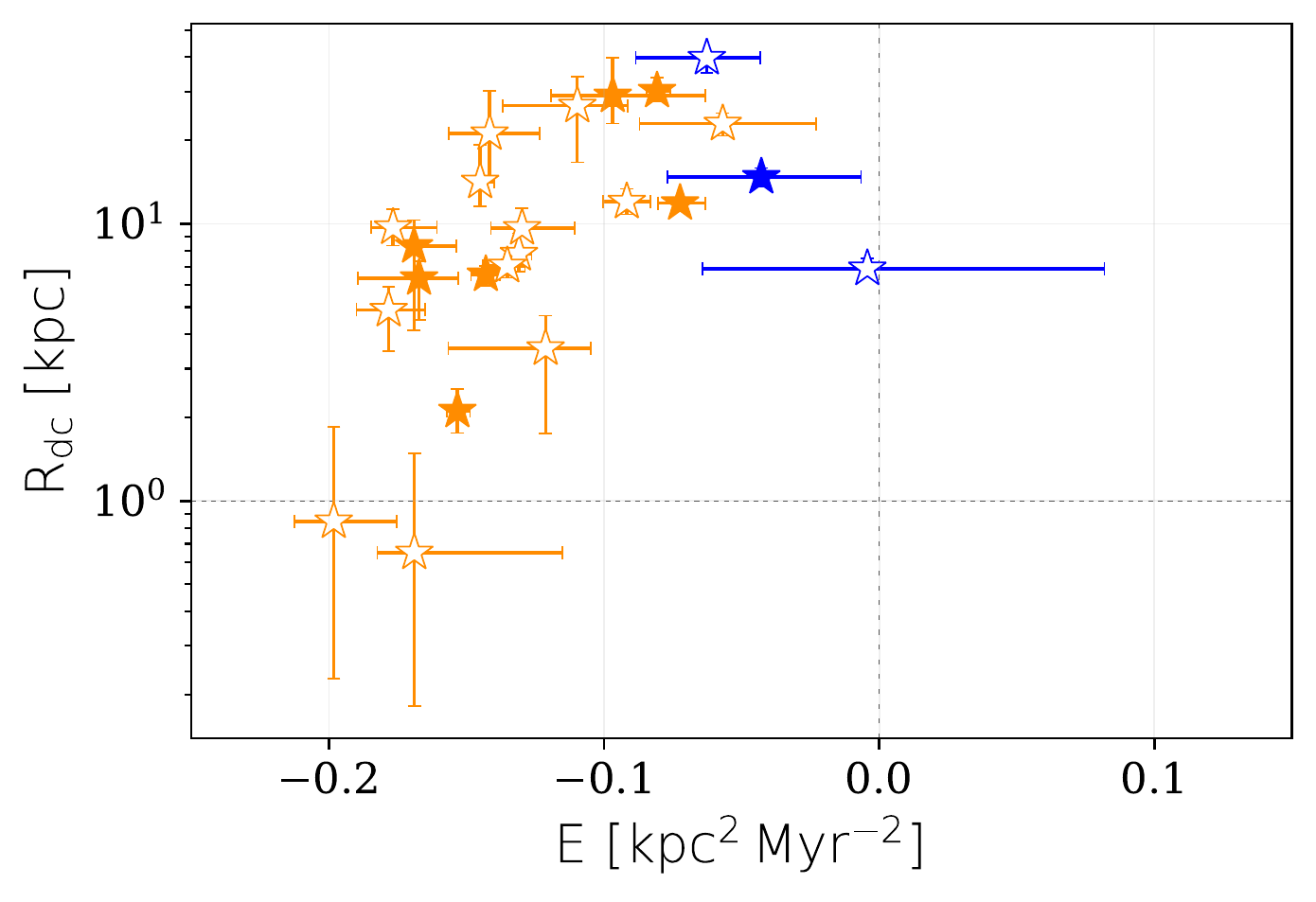}
    \includegraphics[width=\columnwidth]{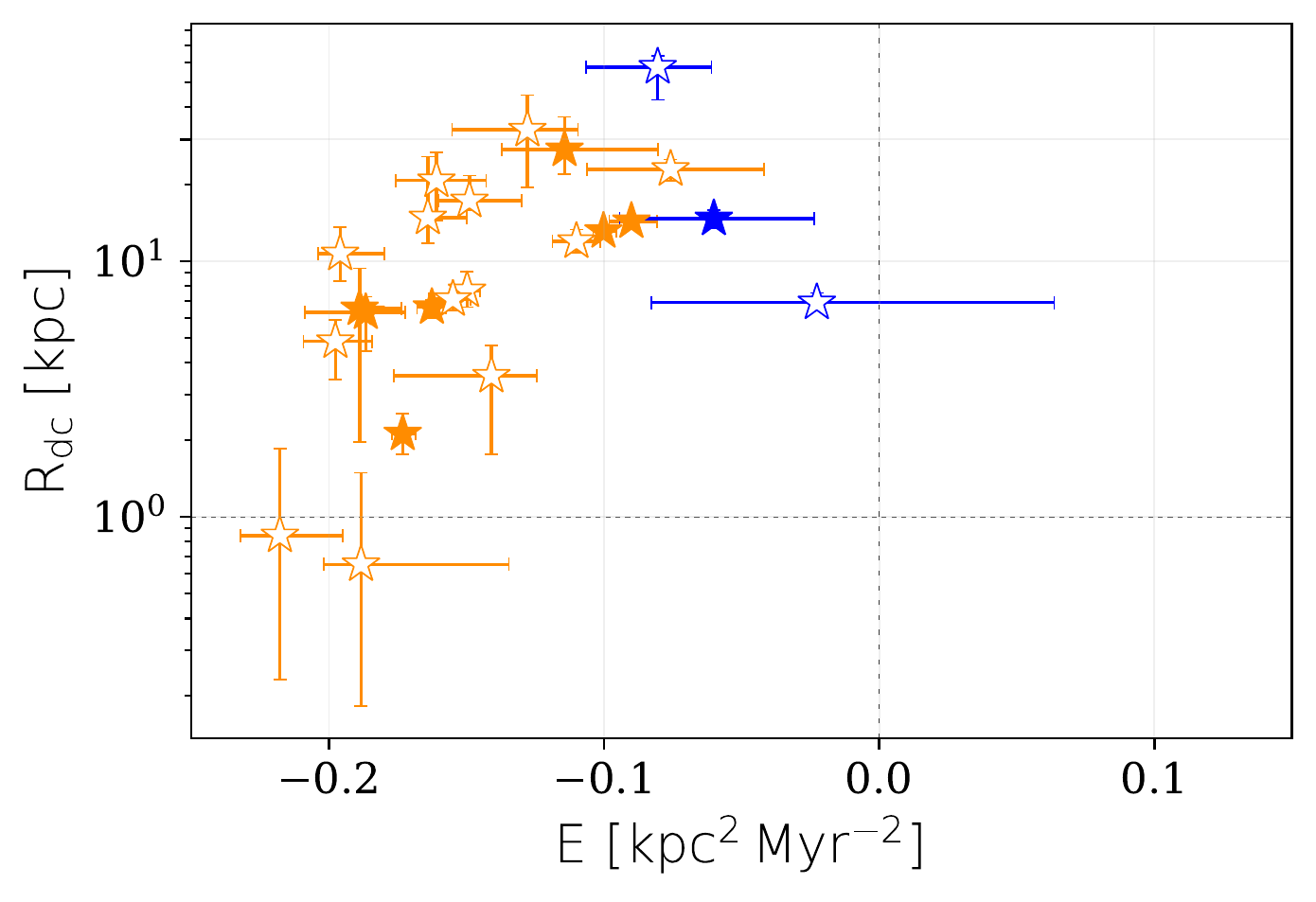}
    \caption{Distance from the Galactic center to the last intersection of the orbit with the Galactic plane ($R_\mathrm{dc}$), as a function of the orbital energy. The vertical defines the limit between bounded (negative energy) and unbounded (positive energy) orbits. The horizontal line corresponds to a Galactocentric distance of 1~kpc, for reference. Markers and colors are the same as in Fig. \ref{fig:identifying_unbound_candidates}. Left, without the LMC. Right, including the LMC.}
    \label{fig:disk_cross}
\end{figure*}

Aiming to put constraints on the spatial origin of the HiVel stars, we also compute the distance ($R_\mathrm{dc}$) to the Galactic center at the last intersection of the orbit with the disk plane. The left panel in Fig. \ref{fig:disk_cross} shows $R_\mathrm{dc}$ as a function of the orbital energy; the stars 2M17054467-2540270 and 2M18051096-3001402 have $R_\mathrm{dc} \lesssim 1~\mathrm{kpc}$, and within the uncertainties, they have a chance of being originated in the Galactic center. Such an origin for these sources needs to be analyzed in the context of chemical abundances that will be discussed below. The dashed vertical line in the figure divides the bound stars $(E < 0)$ from the unbound stars $(E > 0)$. All stars result to have bound orbits. 

We perform an additional analysis including in the integrations the gravitational effect of the Large Magellanic Cloud (LMC). This analysis is justified because the LMC is the most massive dwarf galaxy satellite of the Milky Way, and it might have a relevant gravitational effect on the orbital evolution of the HiVel stars. We use the \texttt{MovingObjectPotential} function implemented in \texttt{galpy}. For the LMC gravitational potential, we consider a Plummer profile with a mass of $M_\mathrm{LMC} = 1.8 \times 10^{11}~M_{\odot}$, taken from \citet{Shipp2021}. The scale radius, $b = 17~\mathrm{kpc}$, is calculated assuming an enclosed mass of $1.7 \times 10^{10}~M_{\odot}$ within $8.7~\mathrm{kpc}$ \citep{vanderMarel2014}. 

In the right panels of Figs. \ref{fig:zmax-ecc} and \ref{fig:disk_cross}, we plot the $Z_\mathrm{max}$ and $R_\mathrm{dc}$ as functions of the eccentricity and orbital energy, respectively, including the effect of the LMC. The heights over the Galactic plane are slightly affected and the orbits of some stars become more eccentric, while the orbital energy slightly decreases. However, the kinematic behavior continues to be similar to that of the halo stars.
For the orbit integrations taking into account the LMC, we calculate the minimum distance $d_\mathrm{LMC}$ between the star and the LMC during the integration. We find only one star (2M14503361+4921331) with a probability $p\left(d_\mathrm{LMC} < 5~\mathrm{kpc}\right) = 0.18$ of passing within 5 kpc of the LMC, while all the other HiVel stars have $p\left(d_\mathrm{LMC} < 5~\mathrm{kpc}\right) < 0.1$. It is worth noting that, in the above simulations, we do not take into account the dynamical friction force suffered by the LMC, which affects its orbit in time.

\section{Stellar parameters}
\label{sec:stellar_param}

\begin{figure*}
    \centering
    \includegraphics[width=\columnwidth]{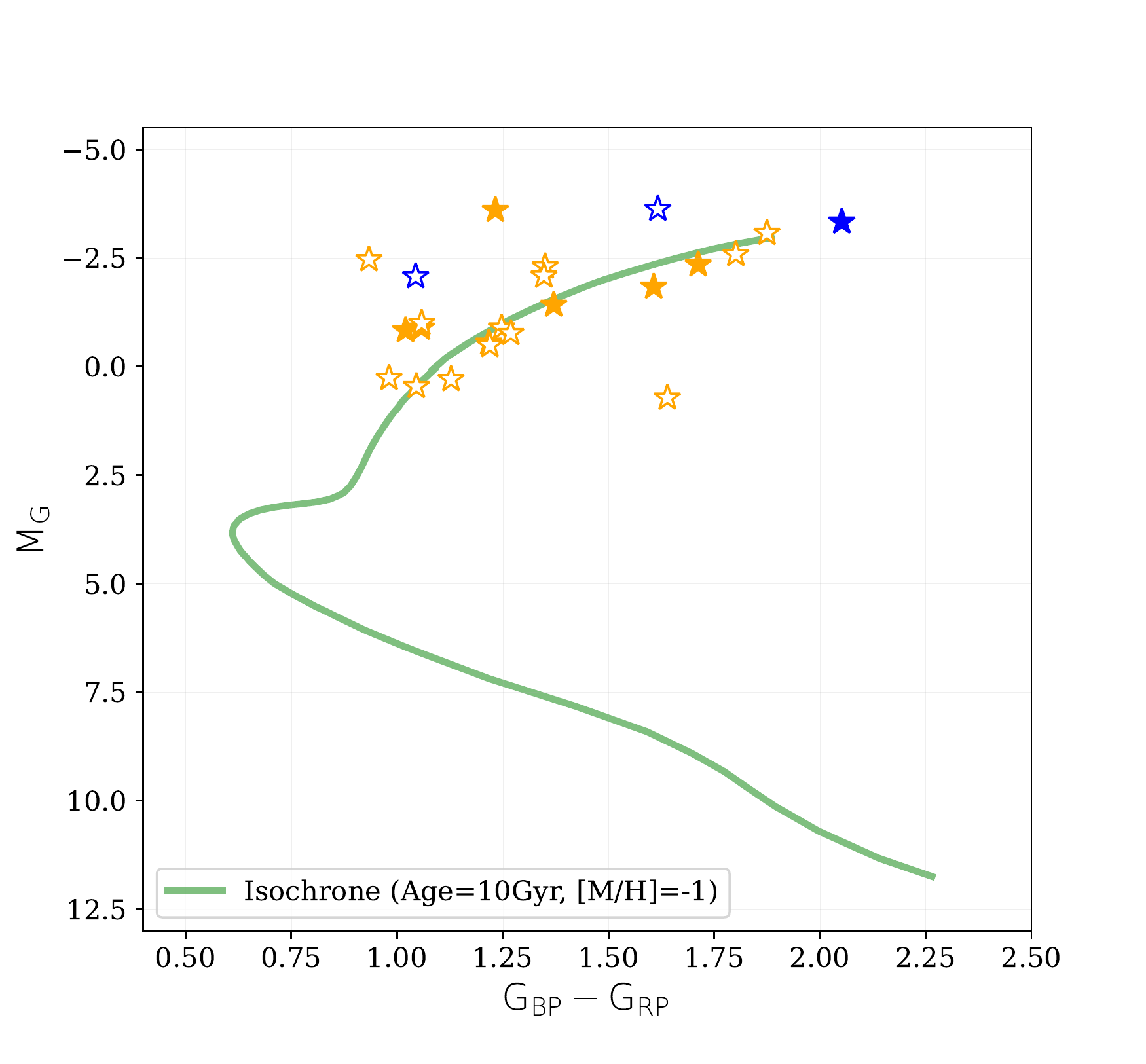}
    \includegraphics[width=\columnwidth]{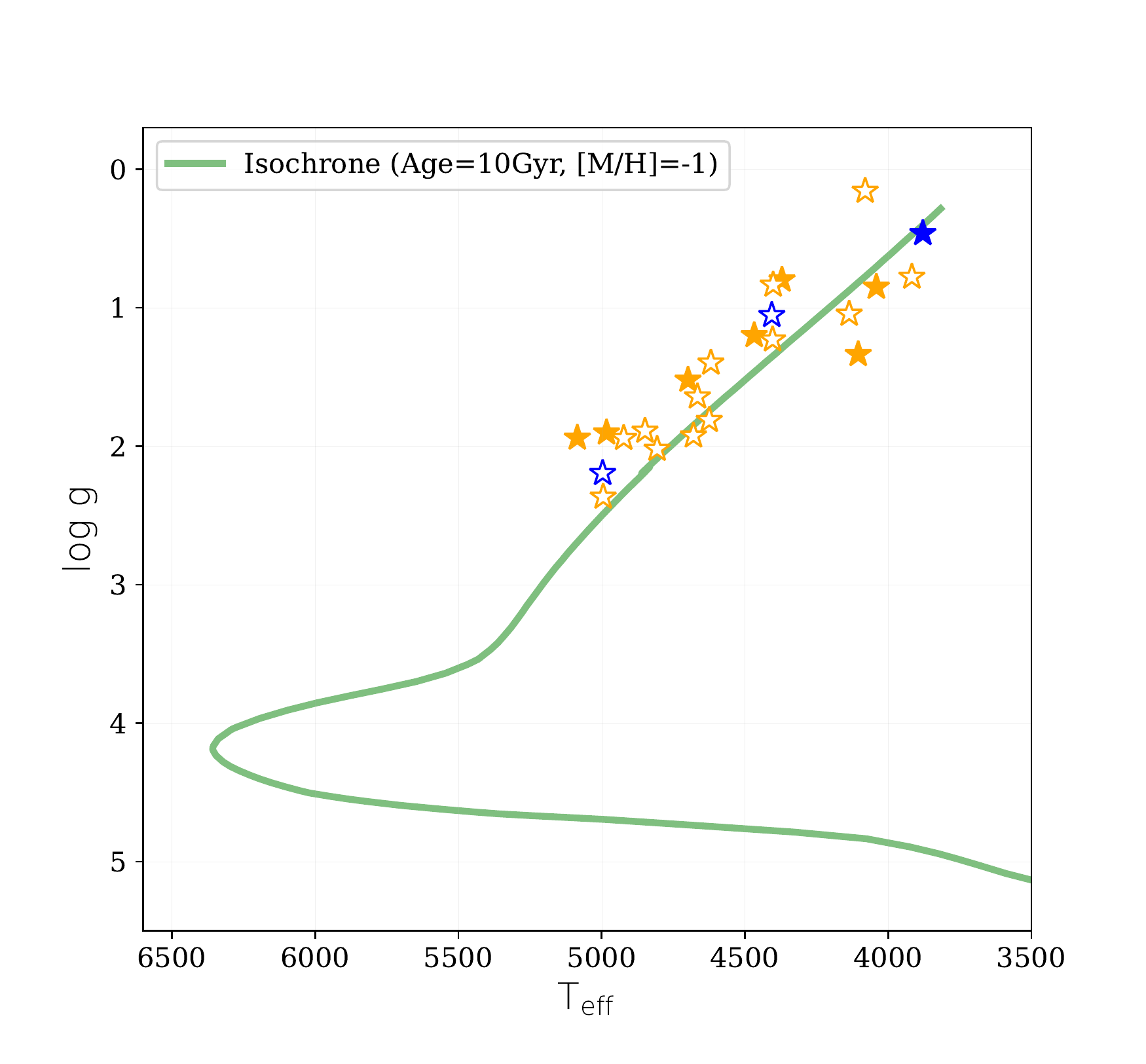}
    \caption{The H-R diagrams of the HiVel sample. Left panel shows the Gaia $G$ absolute magnitude as a function of the Gaia $BP - RP$ color. Right panel shows the surface gravity ($\log g$) as a function of the effective temperature ($T_\mathrm{eff}$). Markers and colors are the same as in Fig. \ref{fig:identifying_unbound_candidates}. Green lines indicate the stellar isochrones obtained from PARSEC.}
    \label{fig:color_magnitude_diagram}
\end{figure*}

Figure \ref{fig:color_magnitude_diagram} shows the Hertzsprung-Russell (H-R) diagrams for the HiVel sample. The lines in green are theoretical evolutionary paths or isochrones for $\mathrm{[Fe/H]} = 1$ and $10~\mathrm{Gyr}$ age, obtained from the PAdova and TRieste Stellar Evolution Code \citep[\texttt{PARSEC},][]{Bressan2012}. On the left panel we plot the absolute $G$ magnitudes against the $BP$-$RP$ colors from Gaia eDR3. Because all the stars in the HiVel sample have a 5-parameter astrometric solution, it is not necessary to apply the $G$-band photometry corrections \citep{Riello2021}. For our full sample, we carry out the dust extinction correction using the SFD2D dust map \citep{Schlafly2011}, through the \texttt{dustmaps} package \citep{Green2018}. In the right panel, we show the H-R diagram using the $T_\mathrm{eff}$ and $\log g$ from APOGEE DR17. All the HiVel stars are located in the Red Giant Branch (RGB).

\begin{table*}[ht]
\caption{Observed properties of the HiVel stars: absolute magnitudes from Gaia eDR3, effective temperature, metalicity and $\alpha$-elements abundance from APOGEE DR17. The last three rows correspond to the unbound candidates.}
\label{tab:Stellar_parameters}
\footnotesize
\centering
\begin{tabular}{lcccccc}
\hline\hline
APOGEE ID & $G$ & $BP$ & $RP$ & $T_\mathrm{eff}$ & [Fe/H] & [$\alpha$/Fe] \\
& (mag) & (mag) & (mag) & (K) & (dex) & (dex) \\
\hline
2M18333156-3439135 & 13.12 & 13.81 & 12.31 & 4467.51 $\,\pm$  8.70 & -1.21 $\pm$ 0.01 &  0.22 $\pm$ 0.01 \\
2M17183052+2300281 & 12.79 & 13.59 & 11.92 & 4104.49 $\,\pm$  5.91 & -0.64 $\pm$ 0.01 &  0.32 $\pm$ 0.01 \\
2M00465509-0022516 & 11.12 & 11.55 & 10.50 & 5084.93 $\,\pm$ 23.67 & -1.70 $\pm$ 0.02 &  0.18 $\pm$ 0.03 \\
2M17223795-2451372 & 12.63 & 13.86 & 11.54 & 4370.58 $\,\pm$  9.60 & -1.87 $\pm$ 0.01 &  0.23 $\pm$ 0.02 \\
2M18562350-2948361 & 13.72 & 14.34 & 12.95 & 4698.49 $\,\pm$ 12.87 & -1.22 $\pm$ 0.01 &  0.22 $\pm$ 0.01 \\
2M17472865+6118530 & 12.86 & 13.32 & 12.21 & 4982.99 $\,\pm$ 13.41 & -1.46 $\pm$ 0.01 &  0.26 $\pm$ 0.02 \\
2M18070909-3716087 & 12.08 & 13.01 & 11.13 & 4041.14 $\,\pm$  6.52 & -1.19 $\pm$ 0.01 &  0.19 $\pm$ 0.01 \\
2M14473273-0018111 & 15.29 & 15.75 & 14.65 & 4994.67 $\,\pm$ 29.81 & -1.12 $\pm$ 0.02 &  0.27 $\pm$ 0.02 \\
2M17145903-2457509 & 13.36 & 14.88 & 12.17 & 3917.18 $\,\pm$  5.74 & -1.01 $\pm$ 0.01 &  0.15 $\pm$ 0.01 \\
2M16323360-1200297 & 14.89 & 15.75 & 13.96 & 4665.54 $\,\pm$ 16.34 & -1.27 $\pm$ 0.02 &  0.21 $\pm$ 0.02 \\
2M17122912-2411516 & 16.41 & 17.23 & 15.51 & 4922.99 $\,\pm$ 40.24 & -1.52 $\pm$ 0.03 &  0.22 $\pm$ 0.03 \\
2M15191912+0202334 & 12.82 & 13.46 & 12.06 & 4618.68 $\,\pm$  9.45 & -1.14 $\pm$ 0.01 &  0.16 $\pm$ 0.01 \\
2M17054467-2540270 & 14.45 & 15.28 & 13.56 & 4401.37 $\,\pm$ 13.73 & -0.85 $\pm$ 0.02 & -0.07 $\pm$ 0.01 \\
2M16344515-1900280 & 15.19 & 16.29 & 14.16 & 4848.71 $\,\pm$ 17.92 & -1.22 $\pm$ 0.01 &  0.25 $\pm$ 0.02 \\
2M22242563-0438021\tablenotemark{\scriptsize a} & 13.29 & 13.97 & 12.52 & 4136.07 $\,\pm$ 10.20\tablenotemark{\scriptsize a} & -1.79 $\pm$ 0.02\tablenotemark{\scriptsize a} & -0.38 $\pm$ 0.02\tablenotemark{\scriptsize a} \\
2M18051096-3001402 & 13.10 & 14.65 & 11.89 & 4080.18 $\,\pm$  8.10 & -1.96 $\pm$ 0.01 &  0.10 $\pm$ 0.02 \\
2M18364421-3418367 & 14.52 & 15.08 & 13.81 & 4806.53 $\,\pm$ 15.54 & -1.22 $\pm$ 0.01 &  0.22 $\pm$ 0.02 \\
2M17065425-2606471 & 14.44 & 15.30 & 13.53 & 4624.50 $\,\pm$ 16.68 & -0.86 $\pm$ 0.02 &  0.11 $\pm$ 0.02 \\
2M17191361-2407018 & 17.15 & 18.73 & 15.95 & 4679.61 $\,\pm$ 14.01 & -0.80 $\pm$ 0.01 &  0.11 $\pm$ 0.01 \\
2M17412026-3431349 & 14.40 & 15.75 & 13.25 & 4403.45 $\,\pm$  8.74 & -1.30 $\pm$ 0.01 &  0.24 $\pm$ 0.01 \\
\hline
2M14503361+4921331 & 11.08 & 12.14 & 10.06 & 3878.89 $\,\pm$  5.78 & -1.15 $\pm$ 0.01 &  0.12 $\pm$ 0.01 \\
2M15180013+0209292 & 11.82 & 12.61 & 10.95 & 4406.14 $\,\pm$  8.15 & -1.12 $\pm$ 0.01 &  0.11 $\pm$ 0.01 \\
2M19284379-0005176 & 13.30 & 13.95 & 12.50 & 4996.48 $\,\pm$ 16.13 & -2.18 $\pm$ 0.01 &  0.26 $\pm$ 0.03 \\
\hline
\end{tabular}
\tablenotetext{a}{The temperature, metallicity and abundances provided by APOGEE DR17 for this star are most probably wrong and should be considered with caution. See Sect. \ref{sec:discussion}}
\end{table*}

\begin{table*}[ht]
\caption{Comparison of the $\log g$ values from APOGEE DR17 and the corresponding isochrone-derived masses and radii, with the values provided by the StarHorse2 catalog \citep{Anders2022}. The StarHorse2 radius has been derived from the corresponding $\log g$ and $M$ values. The two stars that do not have entries in the StarHorse2 catalog, have mass estimates in \citet{Anders2019} -- see text for details. The last three rows correspond to the unbound candidates.}
\label{tab:stellar_masses}
\footnotesize
\centering
\begin{tabular}{lccccccc}
\hline\hline
& \multicolumn{3}{c}{APOGEE DR17+isochrones} & & \multicolumn{3}{c}{StarHorse2+Gaia eDR3} \\
\cline{2-4} \cline{6-8}
APOGEE ID & $\log g$ & $M$ & $R$ & & $\log g$ & $M$ & $R$ \\
& & ($M_{\odot}$) & ($R_{\odot}$) & & & ($M_{\odot}$) & ($R_{\odot}$) \\
\hline
2M18333156-3439135 & 1.20 $\pm$ 0.04 & $1.04_{-0.25}^{+0.23}$ & $39.21_{-4.50}^{+4.54}$ & & $1.30_{-0.01}^{+0.03}$ & $0.87_{-0.08}^{+0.03}$ & 34.59 \\
2M17183052+2300281 & 1.34 $\pm$ 0.03 & $0.93_{-0.06}^{+0.07}$ & $40.29_{-1.81}^{+2.04}$ & & $1.16_{-0.36}^{+0.01}$ & $0.95_{-0.14}^{+0.03}$ & 42.48 \\
2M00465509-0022516 & 1.94 $\pm$ 0.06 & $1.13_{-0.17}^{+0.15}$ & $17.15_{-0.70}^{+0.84}$ & & $1.84_{-0.10}^{+0.09}$ & $0.80_{-0.03}^{+0.04}$ & 17,82 \\
2M17223795-2451372 & 0.80 $\pm$ 0.05 & $3.64_{-0.39}^{+0.49}$ & $90.46_{-7.68}^{+8.89}$ & & $0.88_{-0.42}^{+0.69}$ & $1.05_{-0.30}^{+3.40}$ & 61.64 \\
2M18562350-2948361 & 1.52 $\pm$ 0.04 & $0.78_{-0.02}^{+0.07}$ & $23.06_{-0.57}^{+0.75}$ & & $1.89_{-0.01}^{+0.03}$ & $0.90_{-0.07}^{+0.06}$ & 17.84 \\
2M17472865+6118530 & 1.90 $\pm$ 0.04 & $1.15_{-0.17}^{+0.20}$ & $18.52_{-1.27}^{+1.54}$ & & $1.79_{-0.05}^{+0.14}$ & $0.77_{-0.03}^{+0.07}$ & 18.52  \\
2M18070909-3716087 & 0.85 $\pm$ 0.04 & $1.47_{-0.59}^{+0.10}$ & $89.29_{-13.42}^{+3.19}$ & & $0.67_{-0.16}^{+0.20}$ & $0.91_{-0.31}^{+0.46}$ & 73.08 \\
2M14473273-0018111 & 2.37 $\pm$ 0.06 & $1.15_{-0.09}^{+0.08}$ & $10.71_{-0.72}^{+0.43}$ & & $2.52_{-0.14}^{+0.02}$ & $0.83_{-0.02}^{+0.04}$ & 8.30 \\
2M17145903-2457509 & 0.78 $\pm$ 0.03 & $0.93_{-0.09}^{+0.58}$ & $80.67_{-3.65}^{+11.94}$ & & $0.34_{-0.04}^{+0.26}$ & $1.09_{-0.34}^{+0.50}$ & 116,95 \\
2M16323360-1200297 & 1.64 $\pm$ 0.05 & $0.80_{-0.04}^{+0.10}$ & $23.14_{-0.53}^{+1.02}$ & & $1.83_{-0.10}^{+0.38}$ & $0.86_{-0.14}^{+0.01}$ & 18.69 \\
2M17122912-2411516 & 1.94 $\pm$ 0.09 & $1.31_{-0.08}^{+0.07}$ & $11.54_{-0.55}^{+0.24}$ & & $2.47_{-0.07}^{+0.32}$ & $0.88_{-0.04}^{+0.14}$ & 9.05 \\
2M15191912+0202334 & 1.40 $\pm$ 0.04 & $2.22_{-0.12}^{+0.08}$ & $46.12_{-1.82}^{+0.96}$ & & \nodata & \nodata & \nodata \\
2M17054467-2540270 & 0.84 $\pm$ 0.05 & $0.77_{-0.00}^{+0.01}$ & $24.86_{-0.38}^{+0.38}$ & & $1.68_{-0.19}^{+0.28}$ & $0.88_{-0.08}^{+0.47}$ & 22.47 \\
2M16344515-1900280 & 1.89 $\pm$ 0.05 & $2.78_{-0.05}^{+0.02}$ & $28.16_{-0.73}^{+0.32}$ & & $1.74_{-0.09}^{+0.43}$ & $0.81_{-0.08}^{+0.12}$ & 20.12 \\
2M22242563-0438021 & 1.05 $\pm$ 0.05 & $0.77_{-0.00}^{+0.00}$ & $40.83_{-0.15}^{+0.14}$ & & $1.16_{-0.11}^{+0.03}$ & $0.82_{-0.09}^{+0.01}$ & 39.46 \\
2M18051096-3001402 & 0.16 $\pm$ 0.05 & $0.87_{-0.00}^{+0.20}$ & $95.93_{-1.47}^{+10.28}$ & & $0.17_{-0.09}^{+0.07}$ & $0.59_{-0.00}^{+0.05}$ & 104.65 \\
2M18364421-3418367 & 2.02 $\pm$ 0.05 & $1.16_{-0.12}^{+0.09}$ & $17.49_{-1.30}^{+1.04}$ & & $2.12_{-0.38}^{+0.28}$ & $0.89_{-0.09}^{+0.24}$ & 13.61 \\
2M17065425-2606471 & 1.81 $\pm$ 0.05 & $0.88_{-0.09}^{+0.36}$ & $20.85_{-0.92}^{+3.39}$ & & $1.77_{-0.23}^{+0.01}$ & $0.90_{-0.11}^{+0.03}$ & 20.48 \\
2M17191361-2407018 & 1.92 $\pm$ 0.04 & $0.80_{-0.02}^{+0.03}$ & $17.03_{-0.20}^{+0.16}$ & & $2.75_{-0.09}^{+0.01}$ & $2.24_{-0.61}^{+0.32}$ & 10.46 \\
2M17412026-3431349 & 1.23 $\pm$ 0.04 & $4.28_{-0.25}^{+0.18}$ & $53.71_{-3.85}^{+3.83}$ & & $0.77_{-0.02}^{+0.13}$ & $0.75_{-0.03}^{+0.15}$ & 59.13 \\
\hline
2M14503361+4921331 & 0.46 $\pm$ 0.03 & $1.39_{-0.15}^{+0.13}$ & $132.28_{-7.35}^{+9.62}$ & & $0.33_{-0.00}^{+0.02}$ & $0.83_{-0.00}^{+0.01}$ & 103.24 \\
2M15180013+0209292 & 1.05 $\pm$ 0.03 & $3.17_{-0.26}^{+0.17}$ & $91.37_{-5.80}^{+4.01}$ & & \nodata & \nodata & \nodata \\
2M19284379-0005176 & 2.19 $\pm$ 0.06 & $1.46_{-0.15}^{+0.19}$ & $24.07_{-2.95}^{+1.92}$ & & $1.46_{-0.02}^{+0.03}$ & $0.77_{-0.03}^{+0.00}$ & 27.07 \\
\hline
\end{tabular}
\end{table*}

Table \ref{tab:Stellar_parameters} shows the stellar parameters for the HiVel sample, as determined from the APOGEE DR17 and Gaia eDR3 data. 
It is worth noting that the error of [$\alpha$/Fe] reported in this table comes directly from the APOGEE DR17 file, and it represents the dispersion of the abundance value determined by the \texttt{ASPCAP} over the different visits to the same star. This strategy to estimate of the uncertainty has been applied since DR16 \citep{Jonsson2020}, and differs from the empirical uncertainties derived in previous DRs by fitting the abundance scatter within stellar clusters \citep[e.g.][]{Holtzman2015, Jonsson2018}. Nevertheless, for individual species, typical abundance uncertainties may range up to 0.1 dex, depending on the temperature and metallicity of the star and on the S/N of the spectrum \citep{Jonsson2020}.
In most cases, the metallicities and $\alpha$-elements abundances reported in Table \ref{tab:Stellar_parameters} are typical of halo stars. A few peculiar cases will be discussed later in Sect. \ref{sec:discussion}.

\subsection{Masses and radii}

Using the stellar parameters $T_\mathrm{eff}$, $\log g$, and [Fe/H] from APOGEE DR17, the Gaia eDR3 parallaxes corrected by zero point bias\footnote{The correction is done using the \texttt{gaiadr3\_zeropoint Python} package, hosted at \url{https://gitlab.com/icc-ub/public/gaiadr3_zeropoint}.}, the $G$, $BP$ and $RP$ magnitudes corrected by extinction, and the 2MASS magnitudes in the $H$, $J$, $K$ bands, we infer the mass and radius of each star using the package \texttt{isochrones}. Among other features, this package uses the Mesa Isochrones \& Stellar Tracks models \citep[MIST,][]{Dotter2016, Choi2016, Paxton2011, Paxton2013, Paxton2015} to determine stellar properties based on arbitrary observables \citep{Morton2015}. The results are reported in Table \ref{tab:stellar_masses}.

We note that there are five stars for which the isochrone-inferred masses are larger than 2~$M_\odot$, that would not be typical of halo stars. To check the reliability of these masses, we search for an independent mass determination provided by the StarHorse2 catalog \citep{Anders2022}. The corresponding values are also reported in Table \ref{tab:stellar_masses}. We found that the StarHorse2 masses for three of these five stars is $\approx 1~M_\odot$, consistent with halo stars. For the other two stars (2M15191912+0202334 and 2M15180013+0209292), there is no mass determination in the StarHorse2 catalog, but they have a mass estimate in the original StarHorse catalog \citep{Anders2019} of $0.97^{+0.39}_{-0.16}~M_\odot$ ($27,40~R_\odot$) and $1.02^{+0.27}_{-0.18}~M_\odot$ ($63.62~R_\odot$), respectively. In view of this, we conclude that our isochrone-based mass determination for these five stars is most probably wrong. 

On the other hand, for the remaining stars in the HiVel sample, our isochrone-based masses are in agreement with those of StarHorse2. The only exception is the star  2M17191361-2407018, for which StarHorse2 provides a mass of 2.24~$M_\odot$, against our value of 0.80~$M_\odot$. Nevertheless, the mass estimate for this star in the original StarHorse catalog is of only $1.18^{+0.33}_{-0.20}~M_\odot$ ($12.03~R_\odot$), much closer to our value.

Assessing the sources of error that lead to estimate the wrong masses in some cases while not in others is beyond the scope of this work, although we believe that one possible source could be the bad quality values of $\log g$ provided by APOGEE. For example, if we apply the \texttt{isochrones} algorithm to the star 2M17412026-3431349, but we do not provide to the algorithm any information about the $\log g$ of the star, we obtain a mass estimate of $1.31~M_\odot$, quite different from the $4.28~M_\odot$ obtained when the algorithm is forced to use the $\log g$.

Concerning the radii, in general, the values that we estimate through the isochrones method are in agreement with the values derived from the $\log g$ and masses of StarHorse2. The largest discrepancies arise from the differences in $\log g$ and $M$ between the catalogs. In any case, we conclude that our HiVel sample is composed of low mass stars ($0.6\lesssim M\lesssim 1~M_\odot$), with big radii ($10\lesssim R\lesssim 100~R_\odot$), and low metallicities ($-2.2\lesssim \mathrm{[Fe/H]} \lesssim -0.6$), consistent with halo stars belonging to the RGB. 

\section{Chemical Patterns}
\label{sec:chemical}

The analysis presented in Sects. \ref{sec:kinematic} and \ref{sec:stellar_param} has been done without taking into account the chemical information of the HiVel stars. In this section we use the chemical abundance ratios provided by APOGEE DR17 to try to shed light on the origin of these stars. This procedure is called \textit{chemical tagging} \citep{Freeman2002}, and it was used, for example, by \citet{Hawkins2018} to analyze the origin of five HVS candidates in the Gaia DR2, reported by \citet{Marchetti2019}. This concept was also used by \citet{Reggiani2022} to characterize fifteen HVS candidates selected from \citet{Hattori2018A} and \citet{HerzogArbeitman2018}. In both cases, the chemical abundances ratios were obtained through high-resolution spectra. 

For our HiVel sample, the analysis will be restricted to the chemical abundances of O, Mg, Al, Si, Mn and Ni, that belong to the group of elements with very reliable measurements in APOGEE DR17. Except for the Mn abundances, available for only 17 stars, all the other elemental abundances are available for the entire HiVel sample. We discard the aluminum abundance for the star 2M19284379-0005176, that has the bitmask \texttt{AL\_FE\_FLAG} different from zero. As previously mentioned, the typical uncertainties in APOGEE abundances may range up to 0.1 dex, depending on temperature, metallicity and S/N. Following the results of \citet[][their figure 7]{Jonsson2020}, and taking into account the typical metallicities and temperatures of the HiVel stars, we may assume a mean uncertainty of $\sim0.05$~dex in the abundances of the six elements considered here.

\subsection{Alpha elements: Mg, Si, O} 

Alpha elements are a group of elements whose nucleus is composed of $\mathrm{^{4}He}$. According to chemical evolution models, the first $\alpha$-elements were synthesized through the $\alpha$ process inside very massive stars, initially composed of He and H. After a period of time in which the star burns all its fuel, necessary to maintain the hydrostatic equilibrium, the core of the star collapses (type II supernovae), and then the elements of the outer layers of the star, composed mostly of $\alpha$-elements and a lower fraction of iron-peak elements, are dispersed into the interstellar medium (ISM). Therefore, stars that formed from the material expelled by supernovae may be mostly enriched in $\alpha$-elements. Subsequently, with a more favorable environment for the formation of low-mass stars, the production of metal-enriched stars increased, mainly through type Ia supernovae, so that the ratio between the abundance of $\alpha$-elements and the metallicity can be used to study the environment in which the stars formed. 

\begin{figure*}
    \centering
    \includegraphics[width=\textwidth]{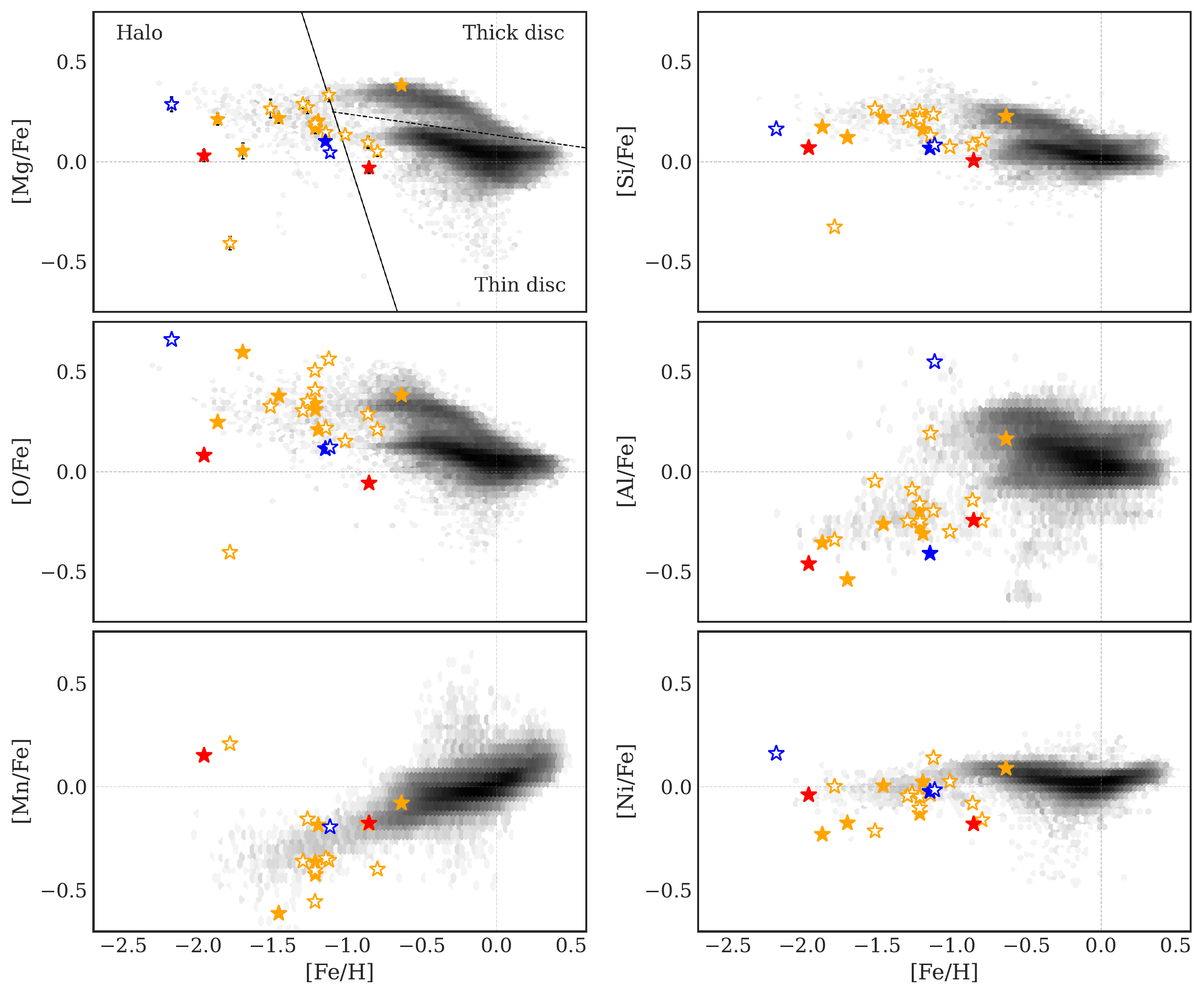}
    \caption{Abundances of individual elements as a function of metallicity. From top to bottom, left to right: $\alpha$-elements (Mg, Si, O), odd-$Z$ element (Al), and Fe-peak elements (Mn, Ni). Markers and colors are the same as in Fig. \ref{fig:identifying_unbound_candidates}, except for the red markers that correspond to the stars that might have an origin close to the Galactic center, from the kinematic point of view.
    The gray region represents the density distribution obtained from all the available APOGEE DR17 data. The solid and dashed lines in the top-left panel separate the halo stars from the disk stars, and the thin disc from the thick disc stars, respectively \citep{Lane2022}. The typical uncertainty in the abundances is $\sim0.05$ dex \citep{Jonsson2020}.}
    \label{fig:alpha_elements_dist}
\end{figure*}

The $[\alpha/\mathrm{Fe}]-\mathrm{[Fe/H]}$ for the stars in the Milky Way
segregates the stars into three large groups, with not very well defined boundaries between them, which correspond to the halo, thin disk, and thick disk, respectively. This is shown in the top-left panel of Fig. \ref{fig:alpha_elements_dist}, where the solid and dashed black lines have been adopted from \citet{Lane2022} as references. The gray regions in all panels of Fig. \ref{fig:alpha_elements_dist} represent the density distribution obtained from the APOGEE DR17 data. Using the reference lines in the top-left panel to distinguish between the populations, we can conclude that most of the HiVel stars are metal poor and are enriched in $\alpha$-elements, like the halo stars.  The unbound candidates (shown as blue stars) are clearly in the halo region, and some stars are in the boundary between the thin disk and the halo. The stars shown in red (2M18051096-3001402 and 2M17054467-2540270) are those that have a possible origin in the center of the Galaxy ($R_\mathrm{dc} < 1~\mathrm{kpc}$), and we can see that one of them is in the halo region while the other is in the boundary of the halo and the low-$\alpha$ thin disk. 

On the other hand, we find a star (2M17183052+2300281) that seems to belong to the thick disk, and another star (2M22242563-0438021) with metallicity $\mathrm{[Fe/H]} = -1.79$ that appears to be very poor in $\alpha$-elements. We will better discuss these two cases in Sect. \ref{sec:discussion}.

\subsection{Odd-Z element: Al}

Like the $\alpha$-elements, the odd-$Z$ elements are dispersed into the ISM mainly through type II supernovae. But unlike $\alpha$-elements, these elements are strongly dependent on the metallicity of the parent star \citep{Nomoto2013,Kobayashi2020A}. The $\mathrm{[Al/Fe]}-\mathrm{[Fe/H]}$ distribution follows a characteristic pattern, that increases from low metallicities to $\mathrm{[Fe/H]} \sim -1$ and then decreases to about $\mathrm{[Fe/H]} \sim 0$ \citep{Kobayashi2020A}. The beginning of the decline is associated with the contribution of type Ia supernovae, which produce lower amounts of Al compared to Fe. 

The middle-right panel of Fig. \ref{fig:alpha_elements_dist} shows the $\mathrm{[Al/Fe]}-\mathrm{[Fe/H]}$ distribution for the HiVel sample. We verify that most HiVel stars have sub-solar values of the [Al/Fe] ratios, which is also the case for the APOGEE halo stars in the figure, but one HiVel halo star appears slightly enriched in aluminum, and another one (2M15180013+0209292) is strongly enriched, with $\mathrm{[Al/Fe]} > +0.5$. This latter star is further discussed in Sect. \ref{sec:discussion}.

\subsection{Fe-peak elements: Mn, Ni}

The iron-peak elements $\left(21 \leq Z \leq 32 \right)$, in contrast to the $\alpha$-elements and the odd-$Z$ elements, are synthesized mostly via type Ia supernovae and, in less amount, via incomplete or complete Si-burning regions during core-collapse supernovae \citep{Kobayashi2020}. The abundance distributions of iron-peak elements as a function of metallicity do not follow the same trends as the $\alpha$ and odd-$Z$ elements. The bottom panels of Fig. \ref{fig:alpha_elements_dist} show the $\mathrm{[Mn/Fe]}$ and $\mathrm{[Ni/Fe]}$ abundances as a function of metallicity. As expected by observations and Galactic chemical models, the Mn abundances decrease towards lower metallicities, and most HiVel stars follow this trend. However, there are two stars with overabundance of Mn which would require further analysis.  

\section{Discussion}
\label{sec:discussion}

\subsection{Dynamical+chemical constraints}

The orbital parameters, the highly eccentric orbits, the fact of having stars with prograde motion and others with retrograde motion, and their location in the Toomre diagram, suggest that all HiVel stars have a kinematic behavior similar to halo stars. According to their location in the HR diagram, the estimated values for their masses, radii, and metallicities, the possibility of being OB runaway stars or hyper-runaway stars can be ruled out. 

Taking into account that the potential energy, considering the Irrgang Model I potential, is less than zero for all the HiVel stars, and that it decreases even more when including the LMC, we may discard our unbound candidates as truly unbound stars. Moreover, these stars follow a chemical pattern similar to the halo stars, thus also ruling out an origin in the Galactic center or the Galaxy disk. The two bound stars with probable origin in the Galactic center ($R_\mathrm{dc} < 1~\mathrm{kpc}$) may also be discarded because these stars follow a chemical pattern typical of halo stars.

From the orbital analysis, an origin in the LMC for the HiVel stars may also be ruled out. The only star that is likely to have a dynamical origin in the LMC (probability of 18\%), is located, from the chemical point of view, in a region where the LMC chemistry overlaps with the chemical abundances of the largest dwarf galaxy that merged with the Milky Way, as shown in figure 5 of \citet{Hasselquist2021}. Therefore, it is difficult to identify an origin in the LMC when we only consider O, Mg, Al, Si, and Mn abundances. 

\subsection{In situ or accreted stars}

\citet{Nissen2010} proposed that halo stars with metallicities between $-1.8$ and $-0.4$ can be divided into a high-$\alpha$ and a low-$\alpha$ populations. The first one would correspond to in-situ stars (i.e. stars that formed in the Milky Way), while the second population would represent stars accreted during mergers over the Galaxy's evolution.
\citet{Hayes2018} arrived to a similar conclusion, but analyzing the Mg abundances of the APOGEE DR13 halo stars, showing that they can be divided into a high-Mg and a low-Mg populations, the latter related to the accreted stars.

\begin{figure*}
    \centering
    \includegraphics[width=\columnwidth]{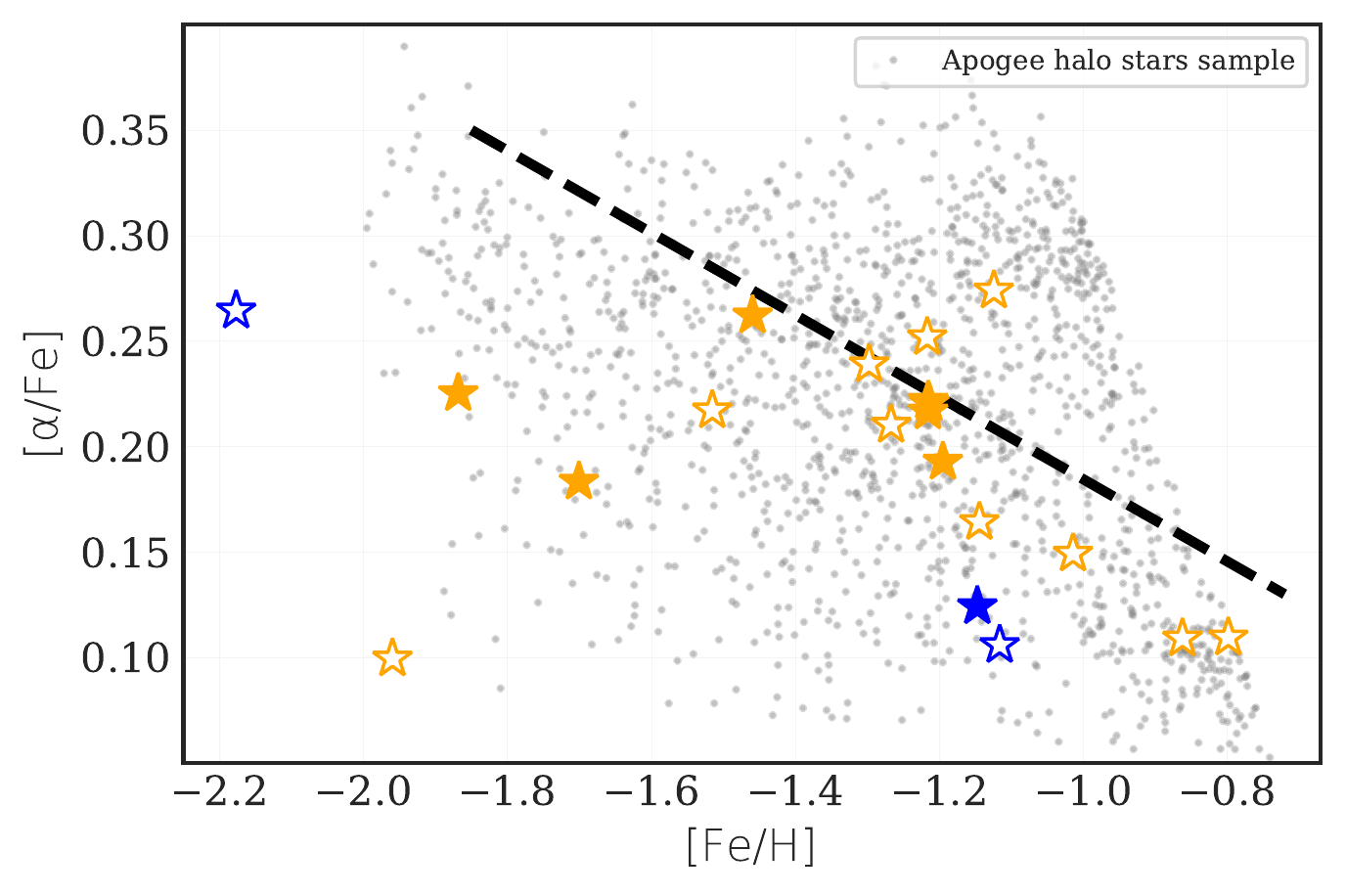}
    \includegraphics[width=\columnwidth]{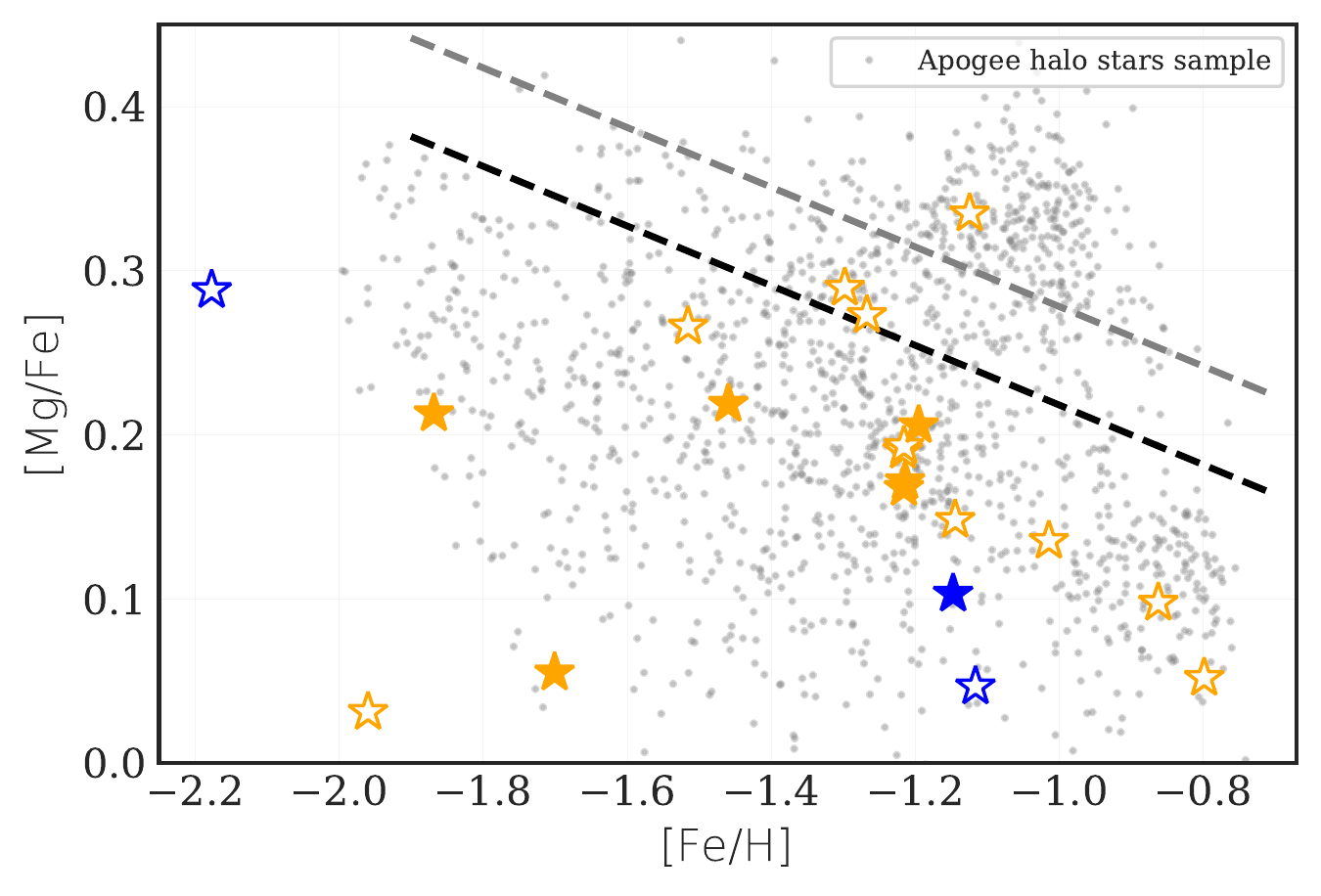}
    \caption{Halo stars in the $[\alpha/\mathrm{Fe}]$ vs. $\mathrm{[Fe/H]}$ space (left), and the $[\mathrm{Mg/Fe}]$ vs. $\mathrm{[Fe/H]}$ space (right). Markers and colors for the HiVel sample are the same as in Fig. \ref{fig:identifying_unbound_candidates}.
    Gray dots are all the halo stars from APOGEE DR17. The black dashed lines separate the high-$\alpha$ or high-Mg (in-situ) population from the low-$\alpha$ or low-Mg (accreted) population, as defined by \citet{Nissen2010} (left) and \citet{Hayes2018} (right). The gray dashed line in the right panel is the corrected limit accounting for the systematic differences in Mg abundances between APOGEE DR13 and DR17. Typical uncertainties in [Mg/Fe] are below 0.05 dex \citep{Jonsson2020}.}
    \label{fig:alpha-feh}
\end{figure*}

Figure \ref{fig:alpha-feh} shows the distribution of $\alpha$-elements abundances (left panel) and Mg abundances (right panel) against metallicities for the HiVel sample (color symbols) and for the APOGEE DR17 halo stars (gray dots). The latter have been selected according to the boundaries shown in Fig. \ref{fig:alpha_elements_dist}.
The black dashed lines in both panels represent the limits between the in-situ population (located in the top-right part of the plot) and the accreted population (located in the bottom-left part). For the $\alpha$ elements distribution, this limit has been defined from the \citet{Nissen2010} sample of halo stars. For the Mg abundances, the limit has been defined from the APOGEE DR13 stars considered in \citet{Hayes2018}. It is worth noting that, in both cases, most of the HiVel sample falls in the accreted population region (i.e. low-$\alpha$ or low-Mg). 

We verify that there is a systematic difference between the Mg abundances of APOGEE DR13 and those in DR17. The latter are, on average, 0.06 dex higher than the former, with a dispersion of $\sigma=\pm0.05$~dex. This implies that the black dashed line in the right panel of Fig. \ref{fig:alpha-feh} should be shifted up to the position of the gray dashed line, in order to establish the appropriate limit between low-Mg and high-Mg for DR17 data. In view of this, we may conclude that most of the HiVel stars in our sample belong to a low-$\alpha$ halo population of probably accreted stars. According to \citet{Hawkins2015} and \citet{Belokurov2022}, this result would also be supported by the abundances of aluminum with sub-solar values observed in Fig. \ref{fig:alpha_elements_dist}. 

In Appendix \ref{app:lz_action_energy_space}, we also present additional dynamical arguments supporting the hypothesis of an accreted origin of most HiVel stars. 

\subsection{Peculiar cases}

From the analysis of the chemical abundances, we find three stars that do not follow the expected chemical pattern of halo stars. We discuss these three peculiar stars in the following.

\paragraph{2M22242563-0438021 (HE 2221-0453)}
This star has been classified in the literature as a Carbon-Enhanced Metal-Poor star (CEMP), based on high-resolution optical spectroscopic analysis \citep{Aoki2007,Yoon2016}. In particular, \citet{Aoki2007} provide radial velocity measurement for this star that is in good agreement with APOGEE's value. However, these authors also provide very different results for the stellar parameters, metallicity, and all the measured abundances compared to the values estimated by APOGEE DR17. The fact of being a CEMP star may explain the big differences that are evident in Table \ref{tab:cemp_properties}. With a true carbon abundance ratio $\mathrm{[C/Fe]} = 1.83$, for example, this star falls outside the grid of models that are used by \texttt{ASPCAP}. Therefore, although \texttt{ASPCAP} is able to find a good fit to the APOGEE spectrum of this star, that fit corresponds to a local minimum within the models grid and does not represent the actual solution. We have verified by manual synthesis that both Aoki et al's solution and \texttt{ASPCAP} solution provide good fits to the APOGEE observed spectrum, with similar $\chi^2$ residuals. But the \texttt{ASPCAP} solution is abnormally poor in $\alpha$-elements, as well as abnormally enriched in Mn, being totally outside the expected trend for halo stars. Thus, there is no doubt that Aoki et al's solution is the correct one. On the other hand, the radial velocity of this star is well determined by APOGEE, so its classification in the HiVel sample is still valid, turning this star a rare, if not the first, example of a high-velocity CEMP star. \citet{Aoki2007} provide a barium abundance of $[\mathrm{Ba/Fe}] = 1.75$, leading to classify this star as a CEMP-s, which would imply to be a binary star. However, since europium abundance has not been determined, it cannot be confirmed as a truly CEMP-s. A discussion on the possible binary character of this star would depend on precise measurements of radial velocity variations, that do not exist in the literature.

\begin{table}[ht]
\caption{Comparison of stellar parameters, chemical abundances [X/Fe], and radial velocities of the star 2M22242563-0438021 (HE 2221-0453), as given by \citet{Aoki2007} and by APOGEE DR17. All abundances are in dex.}
\label{tab:cemp_properties}
\centering
\begin{tabular}{ccc}
\hline\hline
& Aoki et al. (2007) & APOGEE DR17 \\
\hline
$T_\mathrm{eff}$ (K) & 4400 & 4136 \\
$\log g$ & 0.4 & 1.04 \\
$\mathrm{[Fe/H]}$ & -2.2 & -1.79 \\
Mg & 0.8 & -0.4 \\
Ca & 0.82 & -0.56 \\
C & 1.83 & 0.53 \\
N & 0.84 & 0.08 \\
Ti & 0.54 & -0.68 \\
$v_\mathrm{rad}$ (km\,s$^{-1}$) & -189.9 & -180.2 \\
\hline
\end{tabular}
\end{table}

\paragraph{2M17183052+2300281} This star is the most metal-rich in our sample, and belongs to the thick disk region (high-$\alpha$ sequence) in the $\mathrm{[Mg/Fe]} - \mathrm{[Fe/H]}$ space. It follows well the abundance pattern of the thick disk stars. \citet{Mackereth2019}, combining data from APOGEE DR14, Gaia DR2, and orbital eccentricity, show a large sample of highly eccentric stars located in the thick disk region. \citet{Belokurov2020} argued that these stars belong to a special population of stars dubbed as the \textit{Splash population}, formed in the Galaxy (i.e. in-situ stars). An important parameter to probe if this star belongs to this population would be its age, since it is believed that such stars are younger than the stars belonging to the satellite galaxies that merged with our Galaxy. However, age determinations for individual stars belonging to the RGB are quite uncertain and difficult to obtain.

\paragraph{2M15180013+0209292} This star is one of the unbound candidates. It is highly enriched in aluminum and is also enriched in nitrogen, $\mathrm{[N/Fe]} \sim 1$. These features fit well to the population of stars reported by \citet{FernandezTrincado2020}, who based on the kinematic and chemical analysis of 29 stars, proposed that such stars were probably accreted from globular clusters by the Galaxy during the mergers experienced by the Milky Way in the past. 

\section{Conclusions}
\label{sec:conclusion}

In this work, we have analyzed a sample of 23 high velocity stars that we identified in the APOGEE DR17 catalog. The Galactocentric velocities of these stars were derived combining the Gaia eDR3 proper motions, the \citet{BailerJones2021} distances, and the APOGEE DR17 radial velocities.
Our conclusions can be summarized as follows:
\begin{itemize}
    \item We found 3 stars that are unbounded to the MWPotential2014 galactic potential, but they are bounded to the Irrgang Model I potential.
    Twenty other stars display velocities greater than 450~km\,s$^{-1}$, but they are all bounded to the Galaxy.
    Including the gravitational effect from the LMC tends to make the stars even more bounded to the Milky Way.
    \item From the kinematic point of view, all the stars are typical halo stars.
    Two stars in the sample passed close to the Galactic center ($<0.9$~kpc) in the past, but not close enough to allow to invoke the Hills mechanism as the source of their high velocity.
    \item Stellar parameters provided by APOGEE spectra and Gaia photometry indicate that the stars belong to the red giant branch ($0.6\lesssim M\lesssim 1~M_\odot$; $10\lesssim R\lesssim 100~R_\odot$).
    Stars are metal poor ($-2.2\lesssim \mathrm{[Fe/H]} \lesssim -0.6$) and show abundances that, in most cases, are compatible with low-$\alpha$ halo stars (accreted stars).
    \item One star shows unusually low abundances of all the $\alpha$-elements, while is enriched in Mn. This star resulted to be a CEMP star, previously identified in the literature, whose abundances are not properly determined by the APOGEE automatic reduction pipeline. It is a peculiar example of a high velocity ($v_\mathrm{GC}=482~\mathrm{km\,s}^{-1}$) CEMP star. 
    \item The most metal rich star in the sample ($\mathrm{[Fe/H]} \simeq -0.6$) appears to belong to the thick disk.
    \item One of the fastest stars in the sample ($v_\mathrm{GC}=546~\mathrm{km\,s}^{-1}$) is unusually enriched in Al, Ni and O. This star would deserve further analysis through high resolution spectroscopy.
    \item We did not identify any confirmed hypervelocity star in the APOGEE DR17, but we found some stars that are borderline. Classification of such stars as HVS is sensitive to the adopted distances.
    \item Taking into account the uncertainties in the estimated Galactocentric velocities, the classification of HVS stars seems to be not too sensitive to the assumed potential for the Galaxy.
    \item The fact that most stars in the sample follow an abundance pattern typical of accreted stars seems compatible with the idea that their high velocities originated in strong dynamical interactions during mergers of dwarf galaxies with the Milky Way.
\end{itemize}

\vspace{2cm}

This work has been partially financed by the Coordena\c{c}\~ao de Aperfei\c{c}oamento de Pessoal de N\'{\i}vel Superior – Brasil (CAPES) – Finance Code 001. FR and CBP wish to thank financial support from the Brazilian National Council of Research - CNPq.

Funding for the Sloan Digital Sky Survey IV has been provided by the Alfred P. Sloan Foundation, the U.S. Department of Energy Office of Science, and the Participating Institutions. SDSS acknowledges support and resources from the Center for High-Performance Computing at the University of Utah. The SDSS web site is www.sdss.org.

SDSS is managed by the Astrophysical Research Consortium for the Participating Institutions of the SDSS Collaboration including the Brazilian Participation Group, the Carnegie Institution for Science, Carnegie Mellon University, Center for Astrophysics | Harvard \& Smithsonian (CfA), the Chilean Participation Group, the French Participation Group, Instituto de Astrofísica de Canarias, The Johns Hopkins University, Kavli Institute for the Physics and Mathematics of the Universe (IPMU) / University of Tokyo, the Korean Participation Group, Lawrence Berkeley National Laboratory, Leibniz Institut für Astrophysik Potsdam (AIP), Max-Planck-Institut für Astronomie (MPIA Heidelberg), Max-Planck-Institut für Astrophysik (MPA Garching), Max-Planck-Institut für Extraterrestrische Physik (MPE), National Astronomical Observatories of China, New Mexico State University, New York University, University of Notre Dame, Observatório Nacional / MCTI, The Ohio State University, Pennsylvania State University, Shanghai Astronomical Observatory, United Kingdom Participation Group, Universidad Nacional Autónoma de México, University of Arizona, University of Colorado Boulder, University of Oxford, University of Portsmouth, University of Utah, University of Virginia, University of Washington, University of Wisconsin, Vanderbilt University, and Yale University.

\appendix

\section{Distances}
\label{app:distances}

All our analysis of the HiVel stars is based on the photogeometric distances estimated by \citet{BailerJones2021}.
Since the velocity of a star is most sensitive to its distance, it is important to compare the adopted distances to other estimates available in the literature. We consider here the StarHorse2 catalog \citep{Anders2022} and the AstroNN catalog \citep{Leung2019}, as well as the distances obtained as the inverse of the Gaia parallax. 

\begin{figure}
    \centering
    \includegraphics[width=0.8\columnwidth]{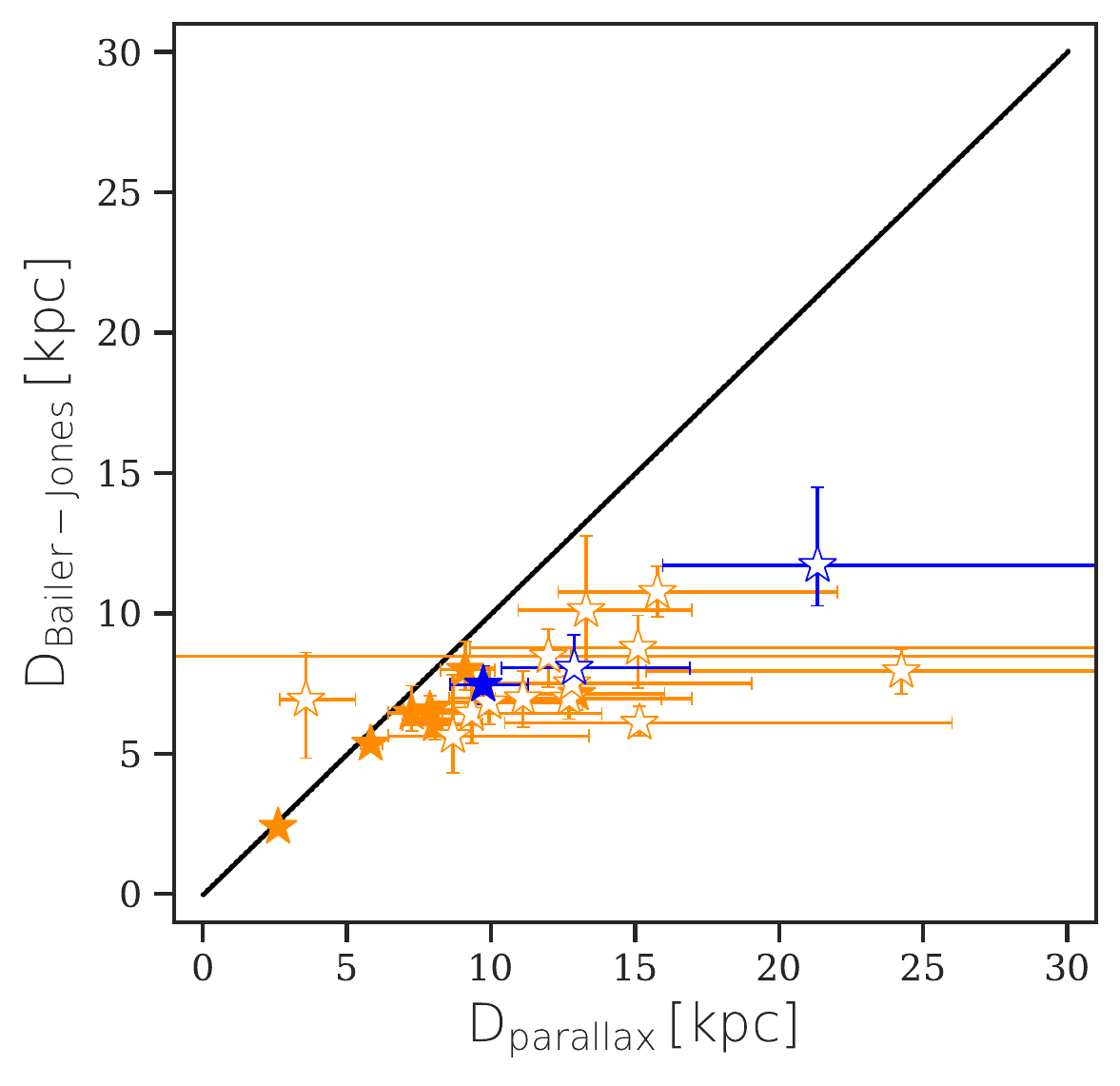}
    \includegraphics[width=0.8\columnwidth]{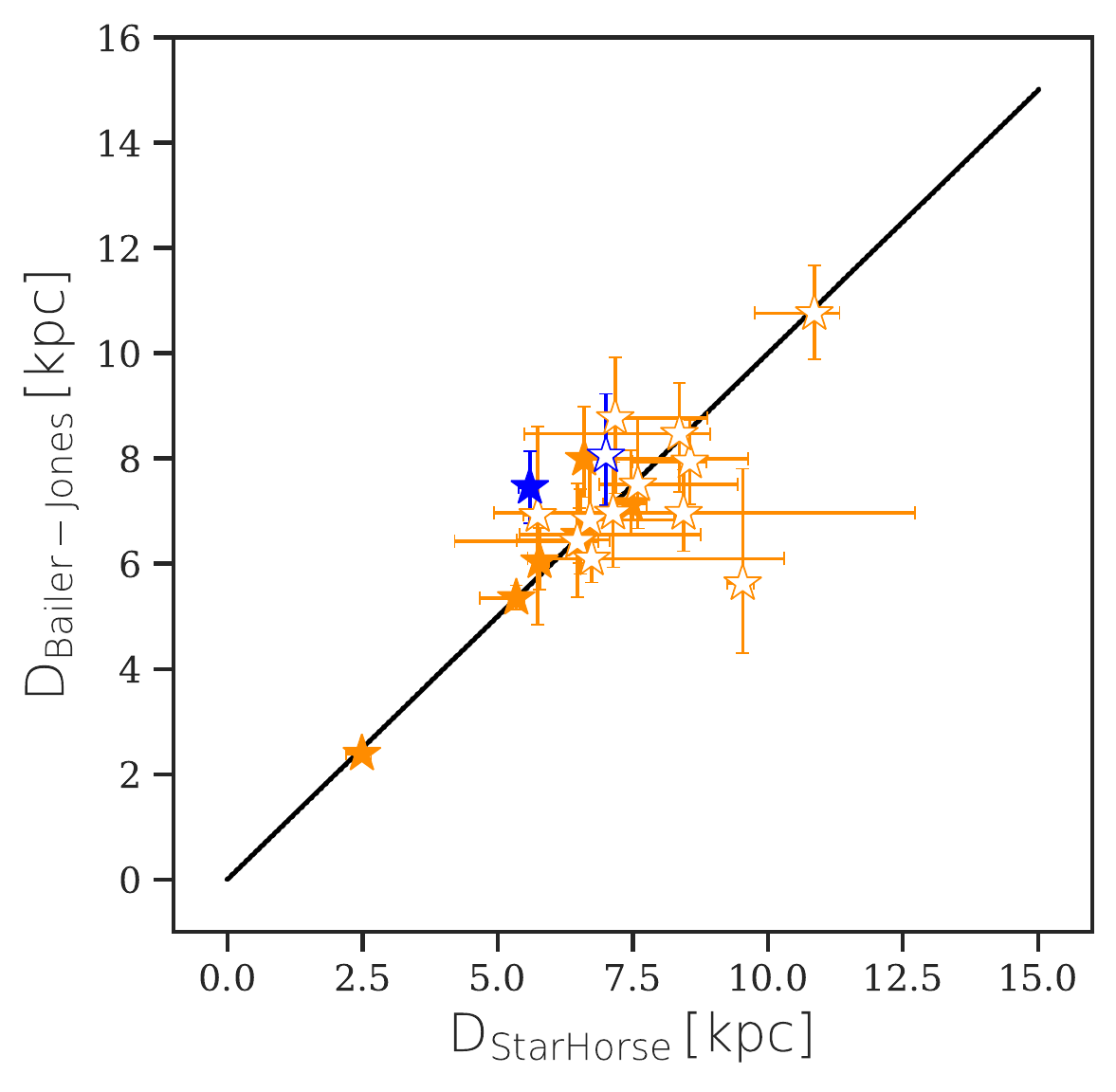}
    \includegraphics[width=0.8\columnwidth]{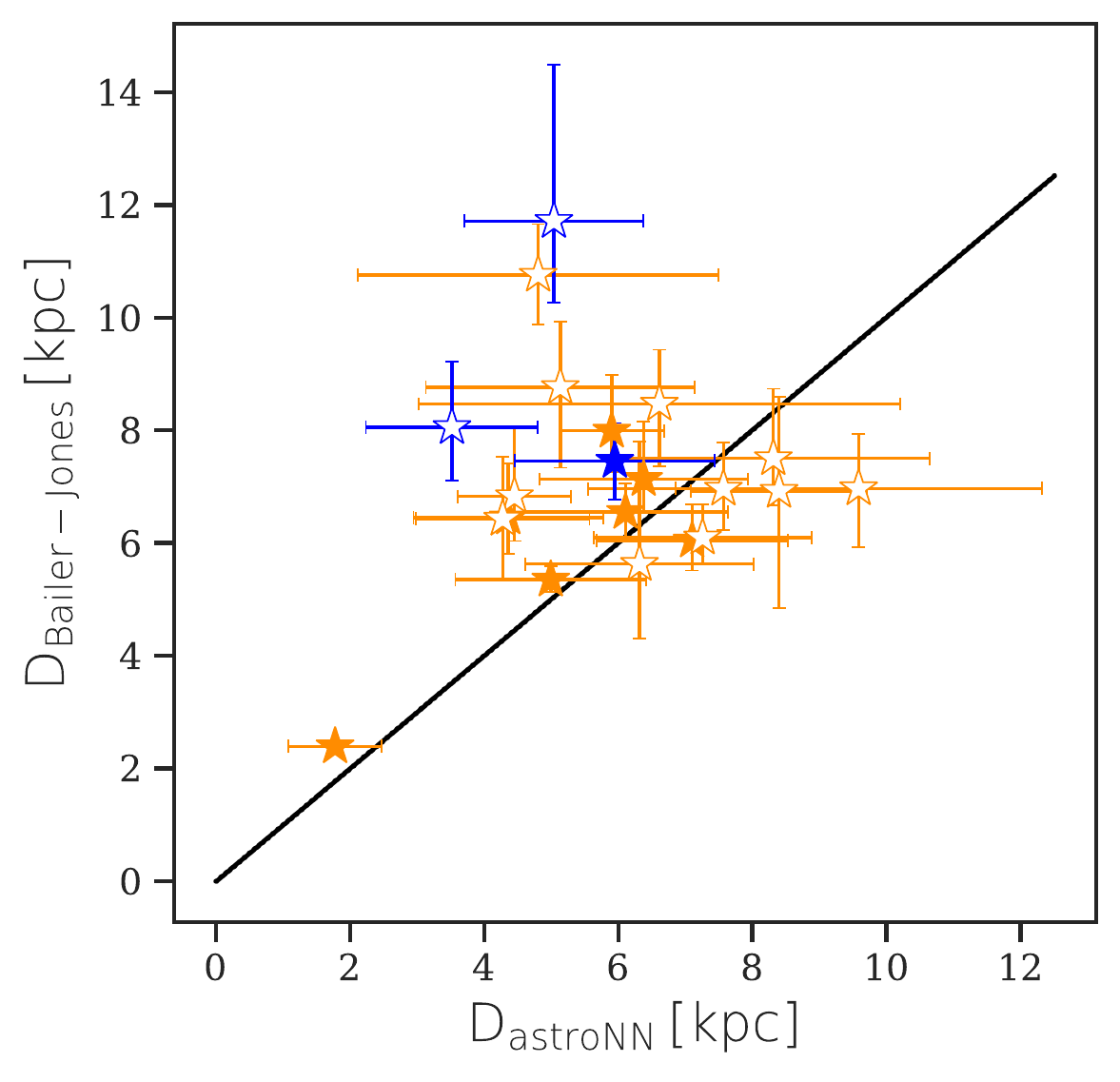}
    \caption{Comparison between the photogeometric distances of \citet{BailerJones2021} used in this work (vertical axis) and other distance estimates for the HiVel stars: Gaia parallaxes (top panel), StarHorse2 (middle panel), and AstroNN (bottom panel). Markers and colors are the same as in Fig. \ref{fig:identifying_unbound_candidates}.}
    \label{fig:distances}
\end{figure}

The Bailer-Jones catalog uses information on paralaxes and colors from the Gaia eDR3 to estimate the distances. For nearby sources, the distance is computed directly from the parallax, but for distant sources, the authors used a Bayesian inference method that provides the posterior distribution of distances depending on the given priors. The priors are taken from a synthetic mock catalog of the eDR3 \citep{Rybizki2020}, which gives the positions, distances, magnitudes, colors, and extinctions of 1.5 billion individual stars. This mock catalog covers well the population of RGB halo stars, and it is expected to provide good distance estimates of our HiVel stars.

The StarHorse2 catalog also uses a Bayesian inference to estimate distances, but based on a different set of parameters and priors \citep{Queiroz2018}. The catalog combines information on stellar parameters (temperature, surface gravity, metallicity, etc.) from other catalogs, including APOGEE  \citep{Queiroz2020}. In particular, it assumes Gaussian priors for the age and matallicity which are broad enough to accommodate most or all of the recent distributions found in the literature. Metallicity priors, for example, follow a distribution with $-0.6\pm0.5$ for the thick disk, and $-1.6\pm0.5$ for the halo, which encompass well the metallicity range of our HiVel stars.

Finally, the AstroNN catalog combines two deep neural networks to estimate the distances. One network is trained is trained using Gaia magnitudes and colors, and the other is trained using normalized continuum spectra obtained from APOGEE and the pseudo-luminosity obtained from the 2MASS $K_S$ band. This training set is also expected to encompass well the stars in our HiVel sample.

The comparisons between the catalogs are shown in Fig. \ref{fig:distances}. It is worth noting that some HiVel stars do not have distance estimates in either StarHorse2 or AstroNN.

Gaia paralaxes provide, in general, much larger distances than Bailer-Jones, but this is not surprising since the latter is precisely an unbiased set of the former. If we use the Gaia paralaxes directly, many HiVel stars will be misidentified as hypervelocity stars. 

The situation is more consistent when comparing to StarHorse2 distances that, in general, show a good agreement. The only exception is 2M17412026-3431349, that has a StarHorse2 distance of about twice its Bailer-Jones distance, although with such StarHorse2 distance it still remains bounded to the Galaxy. On the other hand, star 2M22242563-0438021 shows a difference in distance of only 0.1 kpc between the two catalogs, but this is enough to turn it an unbound candidate when assuming the StarHorse2 distance. We recall that this is the peculiar CEMP star in our sample.

The comparison to AstroNN distances shows bigger dispersion. In general, AstroNN provides much smaller distances for the HiVel stars than Bailer-Jones. The three candidates that we found to be unbounded to the MWPotential2014 become bounded according to the AstroNN distances, and they will not be classified as HiVel stars either. Actually, many stars in our HiVel sample show Galactocentric velocities smaller than 450~$\mathrm{km\,s}^{-1}$ when assuming the AstroNN distances.

\section{Model I Gravitational potential}
\label{app:potential}

The Model I gravitational potential described in \citet{Irrgang2013} is a revised version of the \citet{Allen1991} potential. The bulge is modelled as a Plummer potential:
\begin{equation}
    \Phi_{b}(R, z) = - \frac{G M_{b}}{\sqrt{R^{2} + b_{b}^{2} + z^{2}}}.
\end{equation}
where $(R, z)$ are the components of the cylindrical coordinate system. The disk is axisymmetric and modelled by a Miyamoto-Nagai potential \citep{Miyamoto1975}:
\begin{equation}
    \Phi_{d}(R, z) = - \frac{G M_{d}}{\sqrt{R^{2} + \left(a_{d} + \sqrt{b_{d}^{2} + z^{2}}\right)^{2}}}.
\end{equation}
Finally, the galactic halo is modelled by a spherical potential given by:
\begin{equation}
    \Phi_h(R, z)=\begin{cases}
        \displaystyle\frac{G M_h}{a_h}\left[ \frac{1}{\gamma - 1} \ln{\left(\frac{1 + \left( r/a_h \right)^{\gamma - 1}}{1 + \beta^{\gamma - 1}} \right)} \right. \\
        \qquad\qquad\qquad\quad\displaystyle - \left. \frac{\beta^{\gamma - 1}}{1 + \beta^{\gamma - 1}}\right],
        \quad\mathrm{if}~r<\Lambda,\\ \\
        \displaystyle-\frac{G M_h}{r}\frac{\beta^\gamma}{1 + \beta^{\gamma - 1}}, \qquad\qquad\qquad\mathrm{if}~r\geq\Lambda .
    \end{cases}
\end{equation}
where $\beta=\Lambda/a_h$, $M_b$, $M_d$ and $M_h$ are the masses of the components in Galactic mass ($M_\mathrm{Gal}=2.325\times 10^7~M_{\odot}$), $b_b$, $a_d$, $b_d$ and $a_h$ are the scale lengths of the components, and  $r=\sqrt{R^2 + z^2}$. The values of these parameters are given in Table \ref{tab:model I parameters}.

\begin{table}[h]
\caption{Irrgang Model I parameters}
\label{tab:model I parameters}
\centering
\begin{tabular}{cc}
\hline\hline
$M_b\:(M_\mathrm{Gal})$ & $409 \pm 63$ \\
$M_d\:(M_\mathrm{Gal})$ & $2856_{-202}^{+376}$ \\
$M_h\:(M_\mathrm{Gal})$ & $1018^{+27933}_{-603}$ \\
$b_b\:\mathrm{(kpc)}$ & $0.23 \pm 0.03$ \\
$a_d\:\mathrm{(kpc)}$ & $4.22^{+0.53}_{-0.99}$ \\ 
$b_d\:\mathrm{(kpc)}$ & $0.292_{+0.020}^{-0.025}$ \\
$a_h\:\mathrm{(kpc)}$ & $2.562^{+25.963}_{-1.419}$ \\
$\Lambda\:\mathrm{(kpc)}$ & $200^{+0}_{-83}$ \\
$\gamma$ & $2$ \\
\hline
\end{tabular}
\end{table}

\section{Action diamond space}
\label{app:lz_action_energy_space}
\begin{figure}
    \centering
    \includegraphics[width=0.9\columnwidth]{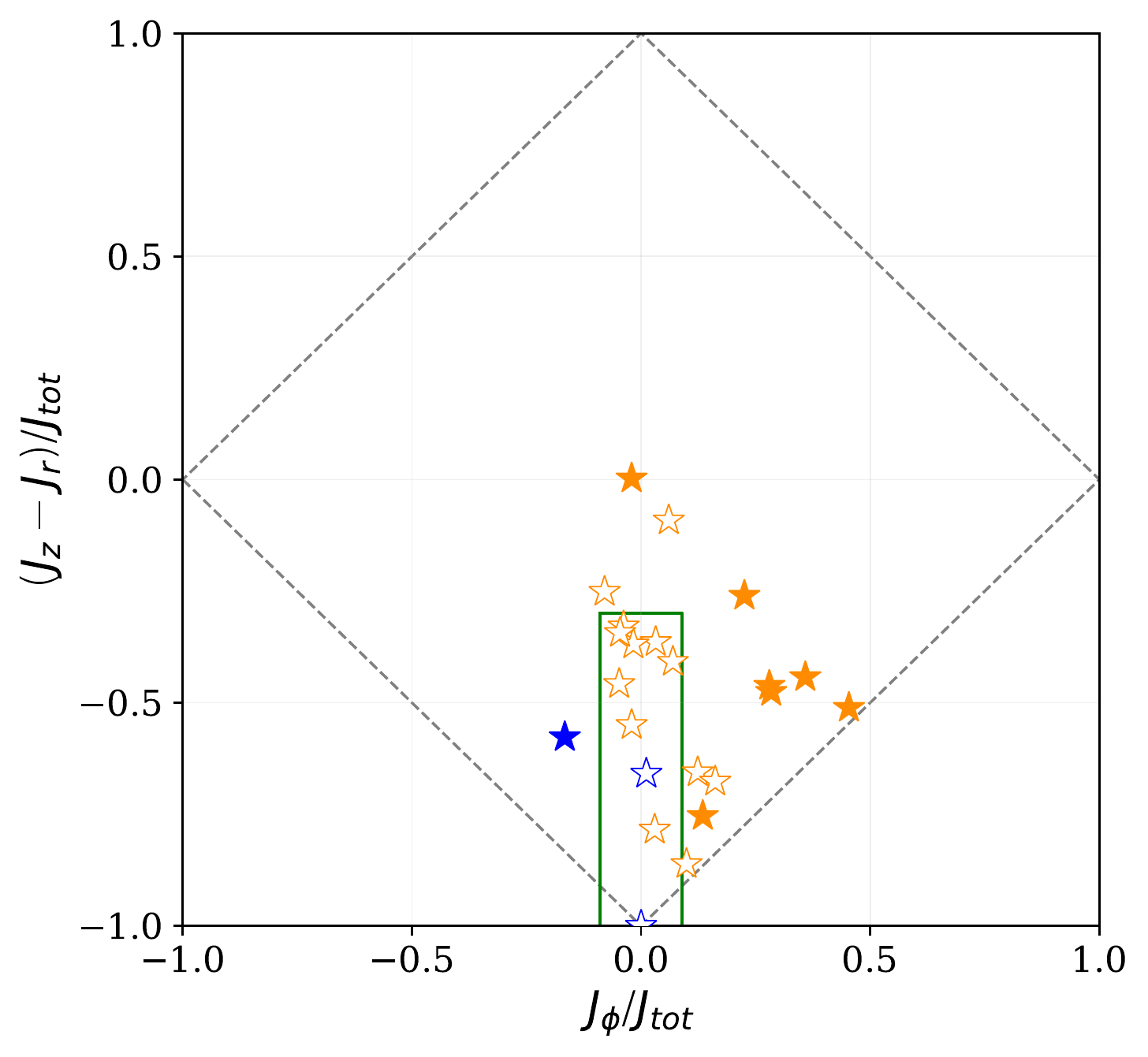}
    \caption{Distribution of the HiVel sample in the action diamond space. Markers and colors are the same as in Fig. \ref{fig:identifying_unbound_candidates}. The green box represents the locations of the Gaia-Sausage remnant.}
    \label{fig:action_map}
\end{figure}

Figure \ref{fig:action_map} shows the place occupied by the HiVel stars in the action diamond space. The actions are calculated using \texttt{galpy} considering the Stackel approximation \citep{Binney2012} and the Irrgang Model I potential. Most of the stars lie in the region of highly radial orbits, as expected. The green box, adopted from \citet{Monty2020}, is the region that would be occupied by stars belonging to the Gaia-Sausage-Enceladus dwarf galaxy, that merged with the Milky Way. Several HiVel stars fall in this region, reinforcing their possible origin as accreted stars. 

\bibliography{manuscript-hvs}{}

\begin{thebibliography}{}
\expandafter\ifx\csname natexlab\endcsname\relax\def\natexlab#1{#1}\fi
\providecommand{\url}[1]{\href{#1}{#1}}
\providecommand{\dodoi}[1]{doi:~\href{http://doi.org/#1}{\nolinkurl{#1}}}
\providecommand{\doeprint}[1]{\href{http://ascl.net/#1}{\nolinkurl{http://ascl.net/#1}}}
\providecommand{\doarXiv}[1]{\href{https://arxiv.org/abs/#1}{\nolinkurl{https://arxiv.org/abs/#1}}}

\bibitem[{{Abadi} {et~al.}(2009){Abadi}, {Navarro}, \& {Steinmetz}}]{Abadi2009}
{Abadi}, M.~G., {Navarro}, J.~F., \& {Steinmetz}, M. 2009, \apjl, 691, L63,
  \dodoi{10.1088/0004-637X/691/2/L63}

\bibitem[{{Abdurro'uf} {et~al.}(2022){Abdurro'uf}, {Accetta}, {Aerts}, {Silva
  Aguirre}, {Ahumada}, {Ajgaonkar}, {Filiz Ak}, {Alam}, {Allende Prieto},
  {Almeida}, {Anders}, {Anderson}, {Andrews}, {Anguiano}, {Aquino-Ort{\'\i}z},
  {Arag{\'o}n-Salamanca}, {Argudo-Fern{\'a}ndez}, {Ata}, {Aubert},
  {Avila-Reese}, {Badenes}, {Barb{\'a}}, {Barger}, {Barrera-Ballesteros},
  {Beaton}, {Beers}, {Belfiore}, {Bender}, {Bernardi}, {Bershady}, {Beutler},
  {Bidin}, {Bird}, {Bizyaev}, {Blanc}, {Blanton}, {Boardman}, {Bolton},
  {Boquien}, {Borissova}, {Bovy}, {Brandt}, {Brown}, {Brownstein}, {Brusa},
  {Buchner}, {Bundy}, {Burchett}, {Bureau}, {Burgasser}, {Cabang}, {Campbell},
  {Cappellari}, {Carlberg}, {Wanderley}, {Carrera}, {Cash}, {Chen}, {Chen},
  {Cherinka}, {Chiappini}, {Choi}, {Chojnowski}, {Chung}, {Clerc}, {Cohen},
  {Comerford}, {Comparat}, {da Costa}, {Covey}, {Crane}, {Cruz-Gonzalez},
  {Culhane}, {Cunha}, {Dai}, {Damke}, {Darling}, {Davidson}, {Davies},
  {Dawson}, {De Lee}, {Diamond-Stanic}, {Cano-D{\'\i}az}, {S{\'a}nchez},
  {Donor}, {Duckworth}, {Dwelly}, {Eisenstein}, {Elsworth}, {Emsellem},
  {Eracleous}, {Escoffier}, {Fan}, {Farr}, {Feng}, {Fern{\'a}ndez-Trincado},
  {Feuillet}, {Filipp}, {Fillingham}, {Frinchaboy}, {Fromenteau}, {Galbany},
  {Garc{\'\i}a}, {Garc{\'\i}a-Hern{\'a}ndez}, {Ge}, {Geisler}, {Gelfand},
  {G{\'e}ron}, {Gibson}, {Goddy}, {Godoy-Rivera}, {Grabowski}, {Green},
  {Greener}, {Grier}, {Griffith}, {Guo}, {Guy}, {Hadjara}, {Harding},
  {Hasselquist}, {Hayes}, {Hearty}, {Hern{\'a}ndez}, {Hill}, {Hogg},
  {Holtzman}, {Horta}, {Hsieh}, {Hsu}, {Hsu}, {Huber}, {Huertas-Company},
  {Hutchinson}, {Hwang}, {Ibarra-Medel}, {Chitham}, {Ilha}, {Imig}, {Jaekle},
  {Jayasinghe}, {Ji}, {Johnson}, {Jones}, {J{\"o}nsson}, {Katkov}, {Khalatyan},
  {Kinemuchi}, {Kisku}, {Knapen}, {Kneib}, {Kollmeier}, {Kong}, {Kounkel},
  {Kreckel}, {Krishnarao}, {Lacerna}, {Lane}, {Langgin}, {Lavender}, {Law},
  {Lazarz}, {Leung}, {Leung}, {Lewis}, {Li}, {Li}, {Lian}, {Liang}, {Lin},
  {Lin}, {Lin}, {Lintott}, {Long}, {Longa-Pe{\~n}a}, {L{\'o}pez-Cob{\'a}},
  {Lu}, {Lundgren}, {Luo}, {Mackereth}, {de la Macorra}, {Mahadevan},
  {Majewski}, {Manchado}, {Mandeville}, {Maraston}, {Margalef-Bentabol},
  {Masseron}, {Masters}, {Mathur}, {McDermid}, {Mckay}, {Merloni},
  {Merrifield}, {Meszaros}, {Miglio}, {Di Mille}, {Minniti}, {Minsley},
  {Monachesi}, {Moon}, {Mosser}, {Mulchaey}, {Muna}, {Mu{\~n}oz}, {Myers},
  {Myers}, {Nadathur}, {Nair}, {Nandra}, {Neumann}, {Newman}, {Nidever},
  {Nikakhtar}, {Nitschelm}, {O'Connell}, {Garma-Oehmichen}, {Luan Souza de
  Oliveira}, {Olney}, {Oravetz}, {Ortigoza-Urdaneta}, {Osorio}, {Otter},
  {Pace}, {Padilla}, {Pan}, {Pan}, {Parikh}, {Parker}, {Peirani}, {Pe{\~n}a
  Ram{\'\i}rez}, {Penny}, {Percival}, {Perez-Fournon}, {Pinsonneault},
  {Poidevin}, {Poovelil}, {Price-Whelan}, {B{\'a}rbara de Andrade Queiroz},
  {Raddick}, {Ray}, {Rembold}, {Riddle}, {Riffel}, {Riffel}, {Rix}, {Robin},
  {Rodr{\'\i}guez-Puebla}, {Roman-Lopes}, {Rom{\'a}n-Z{\'u}{\~n}iga}, {Rose},
  {Ross}, {Rossi}, {Rubin}, {Salvato}, {S{\'a}nchez}, {S{\'a}nchez-Gallego},
  {Sanderson}, {Santana Rojas}, {Sarceno}, {Sarmiento}, {Sayres}, {Sazonova},
  {Schaefer}, {Schiavon}, {Schlegel}, {Schneider}, {Schultheis}, {Schwope},
  {Serenelli}, {Serna}, {Shao}, {Shapiro}, {Sharma}, {Shen}, {Shetrone}, {Shu},
  {Simon}, {Skrutskie}, {Smethurst}, {Smith}, {Sobeck}, {Spoo}, {Sprague},
  {Stark}, {Stassun}, {Steinmetz}, {Stello}, {Stone-Martinez},
  {Storchi-Bergmann}, {Stringfellow}, {Stutz}, {Su}, {Taghizadeh-Popp},
  {Talbot}, {Tayar}, {Telles}, {Teske}, {Thakar}, {Theissen}, {Tkachenko},
  {Thomas}, {Tojeiro}, {Hernandez Toledo}, {Troup}, {Trump}, {Trussler},
  {Turner}, {Tuttle}, {Unda-Sanzana}, {V{\'a}zquez-Mata}, {Valentini},
  {Valenzuela}, {Vargas-Gonz{\'a}lez}, {Vargas-Maga{\~n}a}, {Alfaro},
  {Villanova}, {Vincenzo}, {Wake}, {Warfield}, {Washington}, {Weaver},
  {Weijmans}, {Weinberg}, {Weiss}, {Westfall}, {Wild}, {Wilde}, {Wilson},
  {Wilson}, {Wilson}, {Wolf}, {Wood-Vasey}, {Yan}, {Zamora}, {Zasowski},
  {Zhang}, {Zhao}, {Zheng}, {Zheng}, \& {Zhu}}]{Abdurrouf2022}
{Abdurro'uf}, {Accetta}, K., {Aerts}, C., {et~al.} 2022, \apjs, 259, 35,
  \dodoi{10.3847/1538-4365/ac4414}

\bibitem[{{Allen} \& {Santillan}(1991)}]{Allen1991}
{Allen}, C., \& {Santillan}, A. 1991, rmxaa, 22, 255

\bibitem[{{Allende Prieto} {et~al.}(2006){Allende Prieto}, {Beers}, {Wilhelm},
  {Newberg}, {Rockosi}, {Yanny}, \& {Lee}}]{AllendePrieto2006}
{Allende Prieto}, C., {Beers}, T.~C., {Wilhelm}, R., {et~al.} 2006, \apj, 636,
  804, \dodoi{10.1086/498131}

\bibitem[{{Anders} {et~al.}(2019){Anders}, {Khalatyan}, {Chiappini}, {Queiroz},
  {Santiago}, {Jordi}, {Girardi}, {Brown}, {Matijevi{\v{c}}}, {Monari},
  {Cantat-Gaudin}, {Weiler}, {Khan}, {Miglio}, {Carrillo}, {Romero-G{\'o}mez},
  {Minchev}, {de Jong}, {Antoja}, {Ramos}, {Steinmetz}, \& {Enke}}]{Anders2019}
{Anders}, F., {Khalatyan}, A., {Chiappini}, C., {et~al.} 2019, \aap, 628, A94,
  \dodoi{10.1051/0004-6361/201935765}

\bibitem[{{Anders} {et~al.}(2022){Anders}, {Khalatyan}, {Queiroz}, {Chiappini},
  {Ard{\`e}vol}, {Casamiquela}, {Figueras}, {Jim{\'e}nez-Arranz}, {Jordi},
  {Mongui{\'o}}, {Romero-G{\'o}mez}, {Altamirano}, {Antoja}, {Assaad},
  {Cantat-Gaudin}, {Castro-Ginard}, {Enke}, {Girardi}, {Guiglion}, {Khan},
  {Luri}, {Miglio}, {Minchev}, {Ramos}, {Santiago}, \&
  {Steinmetz}}]{Anders2022}
{Anders}, F., {Khalatyan}, A., {Queiroz}, A.~B.~A., {et~al.} 2022, \aap, 658,
  A91, \dodoi{10.1051/0004-6361/202142369}

\bibitem[{{Aoki} {et~al.}(2007){Aoki}, {Beers}, {Christlieb}, {Norris}, {Ryan},
  \& {Tsangarides}}]{Aoki2007}
{Aoki}, W., {Beers}, T.~C., {Christlieb}, N., {et~al.} 2007, \apj, 655, 492,
  \dodoi{10.1086/509817}

\bibitem[{{Astropy Collaboration} {et~al.}(2013){Astropy Collaboration},
  {Robitaille}, {Tollerud}, {Greenfield}, {Droettboom}, {Bray}, {Aldcroft},
  {Davis}, {Ginsburg}, {Price-Whelan}, {Kerzendorf}, {Conley}, {Crighton},
  {Barbary}, {Muna}, {Ferguson}, {Grollier}, {Parikh}, {Nair}, {Unther},
  {Deil}, {Woillez}, {Conseil}, {Kramer}, {Turner}, {Singer}, {Fox}, {Weaver},
  {Zabalza}, {Edwards}, {Azalee Bostroem}, {Burke}, {Casey}, {Crawford},
  {Dencheva}, {Ely}, {Jenness}, {Labrie}, {Lim}, {Pierfederici}, {Pontzen},
  {Ptak}, {Refsdal}, {Servillat}, \& {Streicher}}]{astropy:2013}
{Astropy Collaboration}, {Robitaille}, T.~P., {Tollerud}, E.~J., {et~al.} 2013,
  \aap, 558, A33, \dodoi{10.1051/0004-6361/201322068}

\bibitem[{{Astropy Collaboration} {et~al.}(2018){Astropy Collaboration},
  {Price-Whelan}, {Sip{\H{o}}cz}, {G{\"u}nther}, {Lim}, {Crawford}, {Conseil},
  {Shupe}, {Craig}, {Dencheva}, {Ginsburg}, {Vand erPlas}, {Bradley},
  {P{\'e}rez-Su{\'a}rez}, {de Val-Borro}, {Aldcroft}, {Cruz}, {Robitaille},
  {Tollerud}, {Ardelean}, {Babej}, {Bach}, {Bachetti}, {Bakanov}, {Bamford},
  {Barentsen}, {Barmby}, {Baumbach}, {Berry}, {Biscani}, {Boquien}, {Bostroem},
  {Bouma}, {Brammer}, {Bray}, {Breytenbach}, {Buddelmeijer}, {Burke},
  {Calderone}, {Cano Rodr{\'\i}guez}, {Cara}, {Cardoso}, {Cheedella}, {Copin},
  {Corrales}, {Crichton}, {D'Avella}, {Deil}, {Depagne}, {Dietrich}, {Donath},
  {Droettboom}, {Earl}, {Erben}, {Fabbro}, {Ferreira}, {Finethy}, {Fox},
  {Garrison}, {Gibbons}, {Goldstein}, {Gommers}, {Greco}, {Greenfield},
  {Groener}, {Grollier}, {Hagen}, {Hirst}, {Homeier}, {Horton}, {Hosseinzadeh},
  {Hu}, {Hunkeler}, {Ivezi{\'c}}, {Jain}, {Jenness}, {Kanarek}, {Kendrew},
  {Kern}, {Kerzendorf}, {Khvalko}, {King}, {Kirkby}, {Kulkarni}, {Kumar},
  {Lee}, {Lenz}, {Littlefair}, {Ma}, {Macleod}, {Mastropietro}, {McCully},
  {Montagnac}, {Morris}, {Mueller}, {Mumford}, {Muna}, {Murphy}, {Nelson},
  {Nguyen}, {Ninan}, {N{\"o}the}, {Ogaz}, {Oh}, {Parejko}, {Parley}, {Pascual},
  {Patil}, {Patil}, {Plunkett}, {Prochaska}, {Rastogi}, {Reddy Janga},
  {Sabater}, {Sakurikar}, {Seifert}, {Sherbert}, {Sherwood-Taylor}, {Shih},
  {Sick}, {Silbiger}, {Singanamalla}, {Singer}, {Sladen}, {Sooley},
  {Sornarajah}, {Streicher}, {Teuben}, {Thomas}, {Tremblay}, {Turner},
  {Terr{\'o}n}, {van Kerkwijk}, {de la Vega}, {Watkins}, {Weaver}, {Whitmore},
  {Woillez}, {Zabalza}, \& {Astropy Contributors}}]{astropy:2018}
{Astropy Collaboration}, {Price-Whelan}, A.~M., {Sip{\H{o}}cz}, B.~M., {et~al.}
  2018, \aj, 156, 123, \dodoi{10.3847/1538-3881/aabc4f}

\bibitem[{{Bailer-Jones} {et~al.}(2021){Bailer-Jones}, {Rybizki}, {Fouesneau},
  {Demleitner}, \& {Andrae}}]{BailerJones2021}
{Bailer-Jones}, C.~A.~L., {Rybizki}, J., {Fouesneau}, M., {Demleitner}, M., \&
  {Andrae}, R. 2021, \aj, 161, 147, \dodoi{10.3847/1538-3881/abd806}

\bibitem[{{Beaton} {et~al.}(2021){Beaton}, {Oelkers}, {Hayes}, {Covey},
  {Chojnowski}, {De Lee}, {Sobeck}, {Majewski}, {Cohen},
  {Fern{\'a}ndez-Trincado}, {Longa-Pe{\~n}a}, {O'Connell}, {Santana},
  {Stringfellow}, {Zasowski}, {Aerts}, {Anguiano}, {Bender}, {Ca{\~n}as},
  {Cunha}, {Donor}, {Fleming}, {Frinchaboy}, {Feuillet}, {Harding},
  {Hasselquist}, {Holtzman}, {Johnson}, {Kollmeier}, {Kounkel}, {Mahadevan},
  {Price-Whelan}, {Rojas-Arriagada}, {Rom{\'a}n-Z{\'u}{\~n}iga}, {Schlafly},
  {Schultheis}, {Shetrone}, {Simon}, {Stassun}, {Stutz}, {Tayar}, {Teske},
  {Tkachenko}, {Troup}, {Albareti}, {Bizyaev}, {Bovy}, {Burgasser}, {Comparat},
  {Downes}, {Geisler}, {Inno}, {Manchado}, {Ness}, {Pinsonneault}, {Prada},
  {Roman-Lopes}, {Simonian}, {Smith}, {Yan}, \& {Zamora}}]{Beaton2021}
{Beaton}, R.~L., {Oelkers}, R.~J., {Hayes}, C.~R., {et~al.} 2021, \aj, 162,
  302, \dodoi{10.3847/1538-3881/ac260c}

\bibitem[{{Belokurov} {et~al.}(2018){Belokurov}, {Erkal}, {Evans}, {Koposov},
  \& {Deason}}]{Belokurov2018}
{Belokurov}, V., {Erkal}, D., {Evans}, N.~W., {Koposov}, S.~E., \& {Deason},
  A.~J. 2018, \mnras, 478, 611, \dodoi{10.1093/mnras/sty982}

\bibitem[{{Belokurov} \& {Kravtsov}(2022)}]{Belokurov2022}
{Belokurov}, V., \& {Kravtsov}, A. 2022, arXiv e-prints, arXiv:2203.04980.
\newblock \doarXiv{2203.04980}

\bibitem[{{Belokurov} {et~al.}(2020){Belokurov}, {Sanders}, {Fattahi}, {Smith},
  {Deason}, {Evans}, \& {Grand}}]{Belokurov2020}
{Belokurov}, V., {Sanders}, J.~L., {Fattahi}, A., {et~al.} 2020, \mnras, 494,
  3880, \dodoi{10.1093/mnras/staa876}

\bibitem[{{Bennett} \& {Bovy}(2019)}]{Bennett2019}
{Bennett}, M., \& {Bovy}, J. 2019, \mnras, 482, 1417,
  \dodoi{10.1093/mnras/sty2813}

\bibitem[{{Binney}(2012)}]{Binney2012}
{Binney}, J. 2012, \mnras, 426, 1324, \dodoi{10.1111/j.1365-2966.2012.21757.x}

\bibitem[{{Blaauw}(1961)}]{Blaauw1961}
{Blaauw}, A. 1961, \bain, 15, 265

\bibitem[{{Blanton} {et~al.}(2017){Blanton}, {Bershady}, {Abolfathi},
  {Albareti}, {Allende Prieto}, {Almeida}, {Alonso-Garc{\'\i}a}, {Anders},
  {Anderson}, {Andrews}, {Aquino-Ort{\'\i}z}, {Arag{\'o}n-Salamanca},
  {Argudo-Fern{\'a}ndez}, {Armengaud}, {Aubourg}, {Avila-Reese}, {Badenes},
  {Bailey}, {Barger}, {Barrera-Ballesteros}, {Bartosz}, {Bates}, {Baumgarten},
  {Bautista}, {Beaton}, {Beers}, {Belfiore}, {Bender}, {Berlind}, {Bernardi},
  {Beutler}, {Bird}, {Bizyaev}, {Blanc}, {Blomqvist}, {Bolton}, {Boquien},
  {Borissova}, {van den Bosch}, {Bovy}, {Brandt}, {Brinkmann}, {Brownstein},
  {Bundy}, {Burgasser}, {Burtin}, {Busca}, {Cappellari}, {Delgado Carigi},
  {Carlberg}, {Carnero Rosell}, {Carrera}, {Chanover}, {Cherinka}, {Cheung},
  {G{\'o}mez Maqueo Chew}, {Chiappini}, {Choi}, {Chojnowski}, {Chuang},
  {Chung}, {Cirolini}, {Clerc}, {Cohen}, {Comparat}, {da Costa}, {Cousinou},
  {Covey}, {Crane}, {Croft}, {Cruz-Gonzalez}, {Garrido Cuadra}, {Cunha},
  {Damke}, {Darling}, {Davies}, {Dawson}, {de la Macorra}, {Dell'Agli}, {De
  Lee}, {Delubac}, {Di Mille}, {Diamond-Stanic}, {Cano-D{\'\i}az}, {Donor},
  {Downes}, {Drory}, {du Mas des Bourboux}, {Duckworth}, {Dwelly}, {Dyer},
  {Ebelke}, {Eigenbrot}, {Eisenstein}, {Emsellem}, {Eracleous}, {Escoffier},
  {Evans}, {Fan}, {Fern{\'a}ndez-Alvar}, {Fernandez-Trincado}, {Feuillet},
  {Finoguenov}, {Fleming}, {Font-Ribera}, {Fredrickson}, {Freischlad},
  {Frinchaboy}, {Fuentes}, {Galbany}, {Garcia-Dias},
  {Garc{\'\i}a-Hern{\'a}ndez}, {Gaulme}, {Geisler}, {Gelfand},
  {Gil-Mar{\'\i}n}, {Gillespie}, {Goddard}, {Gonzalez-Perez}, {Grabowski},
  {Green}, {Grier}, {Gunn}, {Guo}, {Guy}, {Hagen}, {Hahn}, {Hall}, {Harding},
  {Hasselquist}, {Hawley}, {Hearty}, {Gonzalez Hern{\'a}ndez}, {Ho}, {Hogg},
  {Holley-Bockelmann}, {Holtzman}, {Holzer}, {Huehnerhoff}, {Hutchinson},
  {Hwang}, {Ibarra-Medel}, {da Silva Ilha}, {Ivans}, {Ivory}, {Jackson},
  {Jensen}, {Johnson}, {Jones}, {J{\"o}nsson}, {Jullo}, {Kamble}, {Kinemuchi},
  {Kirkby}, {Kitaura}, {Klaene}, {Knapp}, {Kneib}, {Kollmeier}, {Lacerna},
  {Lane}, {Lang}, {Law}, {Lazarz}, {Lee}, {Le Goff}, {Liang}, {Li}, {Li},
  {Lian}, {Lima}, {Lin}, {Lin}, {Bertran de Lis}, {Liu}, {de Icaza Lizaola},
  {Long}, {Lucatello}, {Lundgren}, {MacDonald}, {Deconto Machado}, {MacLeod},
  {Mahadevan}, {Geimba Maia}, {Maiolino}, {Majewski}, {Malanushenko},
  {Malanushenko}, {Manchado}, {Mao}, {Maraston}, {Marques-Chaves}, {Masseron},
  {Masters}, {McBride}, {McDermid}, {McGrath}, {McGreer}, {Medina Pe{\~n}a},
  {Melendez}, {Merloni}, {Merrifield}, {Meszaros}, {Meza}, {Minchev},
  {Minniti}, {Miyaji}, {More}, {Mulchaey}, {M{\"u}ller-S{\'a}nchez}, {Muna},
  {Munoz}, {Myers}, {Nair}, {Nandra}, {Correa do Nascimento}, {Negrete},
  {Ness}, {Newman}, {Nichol}, {Nidever}, {Nitschelm}, {Ntelis}, {O'Connell},
  {Oelkers}, {Oravetz}, {Oravetz}, {Pace}, {Padilla}, {Palanque-Delabrouille},
  {Alonso Palicio}, {Pan}, {Parejko}, {Parikh}, {P{\^a}ris}, {Park}, {Patten},
  {Peirani}, {Pellejero-Ibanez}, {Penny}, {Percival}, {Perez-Fournon},
  {Petitjean}, {Pieri}, {Pinsonneault}, {Pisani}, {Poleski}, {Prada},
  {Prakash}, {Queiroz}, {Raddick}, {Raichoor}, {Barboza Rembold}, {Richstein},
  {Riffel}, {Riffel}, {Rix}, {Robin}, {Rockosi}, {Rodr{\'\i}guez-Torres},
  {Roman-Lopes}, {Rom{\'a}n-Z{\'u}{\~n}iga}, {Rosado}, {Ross}, {Rossi}, {Ruan},
  {Ruggeri}, {Rykoff}, {Salazar-Albornoz}, {Salvato}, {S{\'a}nchez}, {Aguado},
  {S{\'a}nchez-Gallego}, {Santana}, {Santiago}, {Sayres}, {Schiavon}, {da Silva
  Schimoia}, {Schlafly}, {Schlegel}, {Schneider}, {Schultheis}, {Schuster},
  {Schwope}, {Seo}, {Shao}, {Shen}, {Shetrone}, {Shull}, {Simon}, {Skinner},
  {Skrutskie}, {Slosar}, {Smith}, {Sobeck}, {Sobreira}, {Somers}, {Souto},
  {Stark}, {Stassun}, {Stauffer}, {Steinmetz}, {Storchi-Bergmann},
  {Streblyanska}, {Stringfellow}, {Su{\'a}rez}, {Sun}, {Suzuki}, {Szigeti},
  {Taghizadeh-Popp}, {Tang}, {Tao}, {Tayar}, {Tembe}, {Teske}, {Thakar},
  {Thomas}, {Thompson}, {Tinker}, {Tissera}, {Tojeiro}, {Hernandez Toledo}, {de
  la Torre}, {Tremonti}, {Troup}, {Valenzuela}, {Martinez Valpuesta},
  {Vargas-Gonz{\'a}lez}, {Vargas-Maga{\~n}a}, {Vazquez}, {Villanova}, {Vivek},
  {Vogt}, {Wake}, {Walterbos}, {Wang}, {Weaver}, {Weijmans}, {Weinberg},
  {Westfall}, {Whelan}, {Wild}, {Wilson}, {Wood-Vasey}, {Wylezalek}, {Xiao},
  {Yan}, {Yang}, {Ybarra}, {Y{\`e}che}, {Zakamska}, {Zamora}, {Zarrouk},
  {Zasowski}, {Zhang}, {Zhao}, {Zheng}, {Zheng}, {Zhou}, {Zhou}, {Zhu},
  {Zoccali}, \& {Zou}}]{Blanton2017}
{Blanton}, M.~R., {Bershady}, M.~A., {Abolfathi}, B., {et~al.} 2017, \aj, 154,
  28, \dodoi{10.3847/1538-3881/aa7567}

\bibitem[{{Boeche} {et~al.}(2013){Boeche}, {Chiappini}, {Minchev}, {Williams},
  {Steinmetz}, {Sharma}, {Kordopatis}, {Bland-Hawthorn}, {Bienaym{\'e}},
  {Gibson}, {Gilmore}, {Grebel}, {Helmi}, {Munari}, {Navarro}, {Parker},
  {Reid}, {Seabroke}, {Siebert}, {Siviero}, {Watson}, {Wyse}, \&
  {Zwitter}}]{Boeche2013}
{Boeche}, C., {Chiappini}, C., {Minchev}, I., {et~al.} 2013, \aap, 553, A19,
  \dodoi{10.1051/0004-6361/201219607}

\bibitem[{{Boubert} \& {Evans}(2016)}]{Boubert2016}
{Boubert}, D., \& {Evans}, N.~W. 2016, \apjl, 825, L6,
  \dodoi{10.3847/2041-8205/825/1/L6}

\bibitem[{{Boubert} {et~al.}(2018){Boubert}, {Guillochon}, {Hawkins},
  {Ginsburg}, {Evans}, \& {Strader}}]{Boubert2018}
{Boubert}, D., {Guillochon}, J., {Hawkins}, K., {et~al.} 2018, \mnras, 479,
  2789, \dodoi{10.1093/mnras/sty1601}

\bibitem[{{Bovy}(2015)}]{Bovy2015}
{Bovy}, J. 2015, \apjs, 216, 29, \dodoi{10.1088/0067-0049/216/2/29}

\bibitem[{{Bowen} \& {Vaughan}(1973)}]{Bowen1973}
{Bowen}, I.~S., \& {Vaughan}, A.~H., J. 1973, \ao, 12, 1430,
  \dodoi{10.1364/AO.12.001430}

\bibitem[{{Bressan} {et~al.}(2012){Bressan}, {Marigo}, {Girardi}, {Salasnich},
  {Dal Cero}, {Rubele}, \& {Nanni}}]{Bressan2012}
{Bressan}, A., {Marigo}, P., {Girardi}, L., {et~al.} 2012, \mnras, 427, 127,
  \dodoi{10.1111/j.1365-2966.2012.21948.x}

\bibitem[{{Brown}(2021)}]{Brown2021}
{Brown}, A. G.~A. 2021, arXiv e-prints, arXiv:2102.11712.
\newblock \doarXiv{2102.11712}

\bibitem[{{Brown} {et~al.}(2014){Brown}, {Geller}, \& {Kenyon}}]{Brown2014}
{Brown}, W.~R., {Geller}, M.~J., \& {Kenyon}, S.~J. 2014, \apj, 787, 89,
  \dodoi{10.1088/0004-637X/787/1/89}

\bibitem[{{Brown} {et~al.}(2005){Brown}, {Geller}, {Kenyon}, \&
  {Kurtz}}]{Brown2005}
{Brown}, W.~R., {Geller}, M.~J., {Kenyon}, S.~J., \& {Kurtz}, M.~J. 2005,
  \apjl, 622, L33, \dodoi{10.1086/429378}

\bibitem[{{Brown} {et~al.}(2018){Brown}, {Lattanzi}, {Kenyon}, \&
  {Geller}}]{Brown2018}
{Brown}, W.~R., {Lattanzi}, M.~G., {Kenyon}, S.~J., \& {Geller}, M.~J. 2018,
  \apj, 866, 39, \dodoi{10.3847/1538-4357/aadb8e}

\bibitem[{{Capuzzo-Dolcetta} \& {Fragione}(2015)}]{CapuzzoDolcetta2015}
{Capuzzo-Dolcetta}, R., \& {Fragione}, G. 2015, \mnras, 454, 2677,
  \dodoi{10.1093/mnras/stv2123}

\bibitem[{{Choi} {et~al.}(2016){Choi}, {Dotter}, {Conroy}, {Cantiello},
  {Paxton}, \& {Johnson}}]{Choi2016}
{Choi}, J., {Dotter}, A., {Conroy}, C., {et~al.} 2016, \apj, 823, 102,
  \dodoi{10.3847/0004-637X/823/2/102}

\bibitem[{{Cui} {et~al.}(2012){Cui}, {Zhao}, {Chu}, {Li}, {Li}, {Zhang}, {Su},
  {Yao}, {Wang}, {Xing}, {Li}, {Zhu}, {Wang}, {Gu}, {Luo}, {Xu}, {Zhang},
  {Liu}, {Zhang}, {Yang}, {Cao}, {Chen}, {Chen}, {Chen}, {Chen}, {Chu}, {Feng},
  {Gong}, {Hou}, {Hu}, {Hu}, {Hu}, {Jia}, {Jiang}, {Jiang}, {Jiang}, {Jin},
  {Li}, {Li}, {Li}, {Liu}, {Liu}, {Lu}, {Mao}, {Men}, {Qi}, {Qi}, {Shi},
  {Tang}, {Tao}, {Wang}, {Wang}, {Wang}, {Wang}, {Wang}, {Wang}, {Wang},
  {Wang}, {Wang}, {Wang}, {Wang}, {Wang}, {Xu}, {Xu}, {Yang}, {Yu}, {Yuan},
  {Yuan}, {Zhai}, {Zhang}, {Zhang}, {Zhang}, {Zhao}, {Zhou}, {Zhou}, {Zhu}, \&
  {Zou}}]{Cui2012}
{Cui}, X.-Q., {Zhao}, Y.-H., {Chu}, Y.-Q., {et~al.} 2012, Research in Astronomy
  and Astrophysics, 12, 1197, \dodoi{10.1088/1674-4527/12/9/003}

\bibitem[{{Cunha} {et~al.}(2017){Cunha}, {Smith}, {Hasselquist}, {Souto},
  {Shetrone}, {Allende Prieto}, {Bizyaev}, {Frinchaboy},
  {Garc{\'\i}a-Hern{\'a}ndez}, {Holtzman}, {Johnson}, {J{\H{o}}nsson},
  {Majewski}, {M{\'e}sz{\'a}ros}, {Nidever}, {Pinsonneault}, {Schiavon},
  {Sobeck}, {Skrutskie}, {Zamora}, {Zasowski}, \&
  {Fern{\'a}ndez-Trincado}}]{Cunha2017}
{Cunha}, K., {Smith}, V.~V., {Hasselquist}, S., {et~al.} 2017, \apj, 844, 145,
  \dodoi{10.3847/1538-4357/aa7beb}

\bibitem[{{Di Matteo} {et~al.}(2019){Di Matteo}, {Haywood}, {Lehnert}, {Katz},
  {Khoperskov}, {Snaith}, {G{\'o}mez}, \& {Robichon}}]{DiMatteo2019}
{Di Matteo}, P., {Haywood}, M., {Lehnert}, M.~D., {et~al.} 2019, \aap, 632, A4,
  \dodoi{10.1051/0004-6361/201834929}

\bibitem[{{Dotter}(2016)}]{Dotter2016}
{Dotter}, A. 2016, \apjs, 222, 8, \dodoi{10.3847/0067-0049/222/1/8}

\bibitem[{{Drimmel} \& {Poggio}(2018)}]{Drimmel2018}
{Drimmel}, R., \& {Poggio}, E. 2018, Research Notes of the American
  Astronomical Society, 2, 210, \dodoi{10.3847/2515-5172/aaef8b}

\bibitem[{{Eisenstein} {et~al.}(2011){Eisenstein}, {Weinberg}, {Agol},
  {Aihara}, {Allende Prieto}, {Anderson}, {Arns}, {Aubourg}, {Bailey},
  {Balbinot}, {Barkhouser}, {Beers}, {Berlind}, {Bickerton}, {Bizyaev},
  {Blanton}, {Bochanski}, {Bolton}, {Bosman}, {Bovy}, {Brandt}, {Breslauer},
  {Brewington}, {Brinkmann}, {Brown}, {Brownstein}, {Burger}, {Busca},
  {Campbell}, {Cargile}, {Carithers}, {Carlberg}, {Carr}, {Chang}, {Chen},
  {Chiappini}, {Comparat}, {Connolly}, {Cortes}, {Croft}, {Cunha}, {da Costa},
  {Davenport}, {Dawson}, {De Lee}, {Porto de Mello}, {de Simoni}, {Dean},
  {Dhital}, {Ealet}, {Ebelke}, {Edmondson}, {Eiting}, {Escoffier}, {Esposito},
  {Evans}, {Fan}, {Femen{\'\i}a Castell{\'a}}, {Dutra Ferreira}, {Fitzgerald},
  {Fleming}, {Font-Ribera}, {Ford}, {Frinchaboy}, {Garc{\'\i}a P{\'e}rez},
  {Gaudi}, {Ge}, {Ghezzi}, {Gillespie}, {Gilmore}, {Girardi}, {Gott}, {Gould},
  {Grebel}, {Gunn}, {Hamilton}, {Harding}, {Harris}, {Hawley}, {Hearty},
  {Hennawi}, {Gonz{\'a}lez Hern{\'a}ndez}, {Ho}, {Hogg}, {Holtzman},
  {Honscheid}, {Inada}, {Ivans}, {Jiang}, {Jiang}, {Johnson}, {Jordan},
  {Jordan}, {Kauffmann}, {Kazin}, {Kirkby}, {Klaene}, {Knapp}, {Kneib},
  {Kochanek}, {Koesterke}, {Kollmeier}, {Kron}, {Lampeitl}, {Lang}, {Lawler},
  {Le Goff}, {Lee}, {Lee}, {Leisenring}, {Lin}, {Liu}, {Long}, {Loomis},
  {Lucatello}, {Lundgren}, {Lupton}, {Ma}, {Ma}, {MacDonald}, {Mack},
  {Mahadevan}, {Maia}, {Majewski}, {Makler}, {Malanushenko}, {Malanushenko},
  {Mandelbaum}, {Maraston}, {Margala}, {Maseman}, {Masters}, {McBride},
  {McDonald}, {McGreer}, {McMahon}, {Mena Requejo}, {M{\'e}nard},
  {Miralda-Escud{\'e}}, {Morrison}, {Mullally}, {Muna}, {Murayama}, {Myers},
  {Naugle}, {Neto}, {Nguyen}, {Nichol}, {Nidever}, {O'Connell}, {Ogando},
  {Olmstead}, {Oravetz}, {Padmanabhan}, {Paegert}, {Palanque-Delabrouille},
  {Pan}, {Pandey}, {Parejko}, {P{\^a}ris}, {Pellegrini}, {Pepper}, {Percival},
  {Petitjean}, {Pfaffenberger}, {Pforr}, {Phleps}, {Pichon}, {Pieri}, {Prada},
  {Price-Whelan}, {Raddick}, {Ramos}, {Reid}, {Reyle}, {Rich}, {Richards},
  {Rieke}, {Rieke}, {Rix}, {Robin}, {Rocha-Pinto}, {Rockosi}, {Roe},
  {Rollinde}, {Ross}, {Ross}, {Rossetto}, {S{\'a}nchez}, {Santiago}, {Sayres},
  {Schiavon}, {Schlegel}, {Schlesinger}, {Schmidt}, {Schneider}, {Sellgren},
  {Shelden}, {Sheldon}, {Shetrone}, {Shu}, {Silverman}, {Simmerer}, {Simmons},
  {Sivarani}, {Skrutskie}, {Slosar}, {Smee}, {Smith}, {Snedden}, {Stassun},
  {Steele}, {Steinmetz}, {Stockett}, {Stollberg}, {Strauss}, {Szalay},
  {Tanaka}, {Thakar}, {Thomas}, {Tinker}, {Tofflemire}, {Tojeiro}, {Tremonti},
  {Vargas Maga{\~n}a}, {Verde}, {Vogt}, {Wake}, {Wan}, {Wang}, {Weaver},
  {White}, {White}, {Wilson}, {Wisniewski}, {Wood-Vasey}, {Yanny}, {Yasuda},
  {Y{\`e}che}, {York}, {Young}, {Zasowski}, {Zehavi}, \&
  {Zhao}}]{Eisenstein2011}
{Eisenstein}, D.~J., {Weinberg}, D.~H., {Agol}, E., {et~al.} 2011, \aj, 142,
  72, \dodoi{10.1088/0004-6256/142/3/72}

\bibitem[{{Fern{\'a}ndez-Trincado} {et~al.}(2020){Fern{\'a}ndez-Trincado},
  {Beers}, {Minniti}, {Tang}, {Villanova}, {Geisler}, {P{\'e}rez-Villegas}, \&
  {Vieira}}]{FernandezTrincado2020}
{Fern{\'a}ndez-Trincado}, J.~G., {Beers}, T.~C., {Minniti}, D., {et~al.} 2020,
  \aap, 643, L4, \dodoi{10.1051/0004-6361/202039207}

\bibitem[{{Freeman} \& {Bland-Hawthorn}(2002)}]{Freeman2002}
{Freeman}, K., \& {Bland-Hawthorn}, J. 2002, \araa, 40, 487,
  \dodoi{10.1146/annurev.astro.40.060401.093840}

\bibitem[{{Gaia Collaboration} {et~al.}(2016){Gaia Collaboration}, {Prusti},
  {de Bruijne}, {Brown}, {Vallenari}, {Babusiaux}, {Bailer-Jones}, {Bastian},
  {Biermann}, {Evans}, {Eyer}, {Jansen}, {Jordi}, {Klioner}, {Lammers},
  {Lindegren}, {Luri}, {Mignard}, {Milligan}, {Panem}, {Poinsignon},
  {Pourbaix}, {Randich}, {Sarri}, {Sartoretti}, {Siddiqui}, {Soubiran},
  {Valette}, {van Leeuwen}, {Walton}, {Aerts}, {Arenou}, {Cropper}, {Drimmel},
  {H{\o}g}, {Katz}, {Lattanzi}, {O'Mullane}, {Grebel}, {Holland}, {Huc},
  {Passot}, {Bramante}, {Cacciari}, {Casta{\~n}eda}, {Chaoul}, {Cheek}, {De
  Angeli}, {Fabricius}, {Guerra}, {Hern{\'a}ndez}, {Jean-Antoine-Piccolo},
  {Masana}, {Messineo}, {Mowlavi}, {Nienartowicz}, {Ord{\'o}{\~n}ez-Blanco},
  {Panuzzo}, {Portell}, {Richards}, {Riello}, {Seabroke}, {Tanga},
  {Th{\'e}venin}, {Torra}, {Els}, {Gracia-Abril}, {Comoretto},
  {Garcia-Reinaldos}, {Lock}, {Mercier}, {Altmann}, {Andrae}, {Astraatmadja},
  {Bellas-Velidis}, {Benson}, {Berthier}, {Blomme}, {Busso}, {Carry},
  {Cellino}, {Clementini}, {Cowell}, {Creevey}, {Cuypers}, {Davidson}, {De
  Ridder}, {de Torres}, {Delchambre}, {Dell'Oro}, {Ducourant}, {Fr{\'e}mat},
  {Garc{\'\i}a-Torres}, {Gosset}, {Halbwachs}, {Hambly}, {Harrison}, {Hauser},
  {Hestroffer}, {Hodgkin}, {Huckle}, {Hutton}, {Jasniewicz}, {Jordan},
  {Kontizas}, {Korn}, {Lanzafame}, {Manteiga}, {Moitinho}, {Muinonen},
  {Osinde}, {Pancino}, {Pauwels}, {Petit}, {Recio-Blanco}, {Robin}, {Sarro},
  {Siopis}, {Smith}, {Smith}, {Sozzetti}, {Thuillot}, {van Reeven}, {Viala},
  {Abbas}, {Abreu Aramburu}, {Accart}, {Aguado}, {Allan}, {Allasia},
  {Altavilla}, {{\'A}lvarez}, {Alves}, {Anderson}, {Andrei}, {Anglada Varela},
  {Antiche}, {Antoja}, {Ant{\'o}n}, {Arcay}, {Atzei}, {Ayache}, {Bach},
  {Baker}, {Balaguer-N{\'u}{\~n}ez}, {Barache}, {Barata}, {Barbier}, {Barblan},
  {Baroni}, {Barrado y Navascu{\'e}s}, {Barros}, {Barstow}, {Becciani},
  {Bellazzini}, {Bellei}, {Bello Garc{\'\i}a}, {Belokurov}, {Bendjoya},
  {Berihuete}, {Bianchi}, {Bienaym{\'e}}, {Billebaud}, {Blagorodnova},
  {Blanco-Cuaresma}, {Boch}, {Bombrun}, {Borrachero}, {Bouquillon}, {Bourda},
  {Bouy}, {Bragaglia}, {Breddels}, {Brouillet}, {Br{\"u}semeister},
  {Bucciarelli}, {Budnik}, {Burgess}, {Burgon}, {Burlacu}, {Busonero}, {Buzzi},
  {Caffau}, {Cambras}, {Campbell}, {Cancelliere}, {Cantat-Gaudin}, {Carlucci},
  {Carrasco}, {Castellani}, {Charlot}, {Charnas}, {Charvet}, {Chassat},
  {Chiavassa}, {Clotet}, {Cocozza}, {Collins}, {Collins}, {Costigan}, {Crifo},
  {Cross}, {Crosta}, {Crowley}, {Dafonte}, {Damerdji}, {Dapergolas}, {David},
  {David}, {De Cat}, {de Felice}, {de Laverny}, {De Luise}, {De March}, {de
  Martino}, {de Souza}, {Debosscher}, {del Pozo}, {Delbo}, {Delgado},
  {Delgado}, {di Marco}, {Di Matteo}, {Diakite}, {Distefano}, {Dolding}, {Dos
  Anjos}, {Drazinos}, {Dur{\'a}n}, {Dzigan}, {Ecale}, {Edvardsson}, {Enke},
  {Erdmann}, {Escolar}, {Espina}, {Evans}, {Eynard Bontemps}, {Fabre},
  {Fabrizio}, {Faigler}, {Falc{\~a}o}, {Farr{\`a}s Casas}, {Faye}, {Federici},
  {Fedorets}, {Fern{\'a}ndez-Hern{\'a}ndez}, {Fernique}, {Fienga}, {Figueras},
  {Filippi}, {Findeisen}, {Fonti}, {Fouesneau}, {Fraile}, {Fraser}, {Fuchs},
  {Furnell}, {Gai}, {Galleti}, {Galluccio}, {Garabato}, {Garc{\'\i}a-Sedano},
  {Gar{\'e}}, {Garofalo}, {Garralda}, {Gavras}, {Gerssen}, {Geyer}, {Gilmore},
  {Girona}, {Giuffrida}, {Gomes}, {Gonz{\'a}lez-Marcos},
  {Gonz{\'a}lez-N{\'u}{\~n}ez}, {Gonz{\'a}lez-Vidal}, {Granvik}, {Guerrier},
  {Guillout}, {Guiraud}, {G{\'u}rpide}, {Guti{\'e}rrez-S{\'a}nchez}, {Guy},
  {Haigron}, {Hatzidimitriou}, {Haywood}, {Heiter}, {Helmi}, {Hobbs},
  {Hofmann}, {Holl}, {Holland}, {Hunt}, {Hypki}, {Icardi}, {Irwin}, {Jevardat
  de Fombelle}, {Jofr{\'e}}, {Jonker}, {Jorissen}, {Julbe}, {Karampelas},
  {Kochoska}, {Kohley}, {Kolenberg}, {Kontizas}, {Koposov}, {Kordopatis},
  {Koubsky}, {Kowalczyk}, {Krone-Martins}, {Kudryashova}, {Kull}, {Bachchan},
  {Lacoste-Seris}, {Lanza}, {Lavigne}, {Le Poncin-Lafitte}, {Lebreton},
  {Lebzelter}, {Leccia}, {Leclerc}, {Lecoeur-Taibi}, {Lemaitre}, {Lenhardt},
  {Leroux}, {Liao}, {Licata}, {Lindstr{\o}m}, {Lister}, {Livanou}, {Lobel},
  {L{\"o}ffler}, {L{\'o}pez}, {Lopez-Lozano}, {Lorenz}, {Loureiro},
  {MacDonald}, {Magalh{\~a}es Fernandes}, {Managau}, {Mann}, {Mantelet},
  {Marchal}, {Marchant}, {Marconi}, {Marie}, {Marinoni}, {Marrese},
  {Marschalk{\'o}}, {Marshall}, {Mart{\'\i}n-Fleitas}, {Martino}, {Mary},
  {Matijevi{\v{c}}}, {Mazeh}, {McMillan}, {Messina}, {Mestre}, {Michalik},
  {Millar}, {Miranda}, {Molina}, {Molinaro}, {Molinaro}, {Moln{\'a}r},
  {Moniez}, {Montegriffo}, {Monteiro}, {Mor}, {Mora}, {Morbidelli}, {Morel},
  {Morgenthaler}, {Morley}, {Morris}, {Mulone}, {Muraveva}, {Musella},
  {Narbonne}, {Nelemans}, {Nicastro}, {Noval}, {Ord{\'e}novic},
  {Ordieres-Mer{\'e}}, {Osborne}, {Pagani}, {Pagano}, {Pailler}, {Palacin},
  {Palaversa}, {Parsons}, {Paulsen}, {Pecoraro}, {Pedrosa}, {Pentik{\"a}inen},
  {Pereira}, {Pichon}, {Piersimoni}, {Pineau}, {Plachy}, {Plum}, {Poujoulet},
  {Pr{\v{s}}a}, {Pulone}, {Ragaini}, {Rago}, {Rambaux}, {Ramos-Lerate},
  {Ranalli}, {Rauw}, {Read}, {Regibo}, {Renk}, {Reyl{\'e}}, {Ribeiro},
  {Rimoldini}, {Ripepi}, {Riva}, {Rixon}, {Roelens}, {Romero-G{\'o}mez},
  {Rowell}, {Royer}, {Rudolph}, {Ruiz-Dern}, {Sadowski}, {Sagrist{\`a}
  Sell{\'e}s}, {Sahlmann}, {Salgado}, {Salguero}, {Sarasso}, {Savietto},
  {Schnorhk}, {Schultheis}, {Sciacca}, {Segol}, {Segovia}, {Segransan},
  {Serpell}, {Shih}, {Smareglia}, {Smart}, {Smith}, {Solano}, {Solitro},
  {Sordo}, {Soria Nieto}, {Souchay}, {Spagna}, {Spoto}, {Stampa}, {Steele},
  {Steidelm{\"u}ller}, {Stephenson}, {Stoev}, {Suess}, {S{\"u}veges}, {Surdej},
  {Szabados}, {Szegedi-Elek}, {Tapiador}, {Taris}, {Tauran}, {Taylor},
  {Teixeira}, {Terrett}, {Tingley}, {Trager}, {Turon}, {Ulla}, {Utrilla},
  {Valentini}, {van Elteren}, {Van Hemelryck}, {van Leeuwen}, {Varadi},
  {Vecchiato}, {Veljanoski}, {Via}, {Vicente}, {Vogt}, {Voss}, {Votruba},
  {Voutsinas}, {Walmsley}, {Weiler}, {Weingrill}, {Werner}, {Wevers},
  {Whitehead}, {Wyrzykowski}, {Yoldas}, {{\v{Z}}erjal}, {Zucker}, {Zurbach},
  {Zwitter}, {Alecu}, {Allen}, {Allende Prieto}, {Amorim},
  {Anglada-Escud{\'e}}, {Arsenijevic}, {Azaz}, {Balm}, {Beck}, {Bernstein},
  {Bigot}, {Bijaoui}, {Blasco}, {Bonfigli}, {Bono}, {Boudreault}, {Bressan},
  {Brown}, {Brunet}, {Bunclark}, {Buonanno}, {Butkevich}, {Carret}, {Carrion},
  {Chemin}, {Ch{\'e}reau}, {Corcione}, {Darmigny}, {de Boer}, {de Teodoro}, {de
  Zeeuw}, {Delle Luche}, {Domingues}, {Dubath}, {Fodor}, {Fr{\'e}zouls},
  {Fries}, {Fustes}, {Fyfe}, {Gallardo}, {Gallegos}, {Gardiol}, {Gebran},
  {Gomboc}, {G{\'o}mez}, {Grux}, {Gueguen}, {Heyrovsky}, {Hoar}, {Iannicola},
  {Isasi Parache}, {Janotto}, {Joliet}, {Jonckheere}, {Keil}, {Kim},
  {Klagyivik}, {Klar}, {Knude}, {Kochukhov}, {Kolka}, {Kos}, {Kutka}, {Lainey},
  {LeBouquin}, {Liu}, {Loreggia}, {Makarov}, {Marseille}, {Martayan},
  {Martinez-Rubi}, {Massart}, {Meynadier}, {Mignot}, {Munari}, {Nguyen},
  {Nordlander}, {Ocvirk}, {O'Flaherty}, {Olias Sanz}, {Ortiz}, {Osorio},
  {Oszkiewicz}, {Ouzounis}, {Palmer}, {Park}, {Pasquato}, {Peltzer}, {Peralta},
  {P{\'e}turaud}, {Pieniluoma}, {Pigozzi}, {Poels}, {Prat}, {Prod'homme},
  {Raison}, {Rebordao}, {Risquez}, {Rocca-Volmerange}, {Rosen}, {Ruiz-Fuertes},
  {Russo}, {Sembay}, {Serraller Vizcaino}, {Short}, {Siebert}, {Silva},
  {Sinachopoulos}, {Slezak}, {Soffel}, {Sosnowska}, {Strai{\v{z}}ys}, {ter
  Linden}, {Terrell}, {Theil}, {Tiede}, {Troisi}, {Tsalmantza}, {Tur},
  {Vaccari}, {Vachier}, {Valles}, {Van Hamme}, {Veltz}, {Virtanen}, {Wallut},
  {Wichmann}, {Wilkinson}, {Ziaeepour}, \& {Zschocke}}]{GaiaCol2016}
{Gaia Collaboration}, {Prusti}, T., {de Bruijne}, J.~H.~J., {et~al.} 2016,
  \aap, 595, A1, \dodoi{10.1051/0004-6361/201629272}

\bibitem[{{Gaia Collaboration} {et~al.}(2018){Gaia Collaboration}, {Brown},
  {Vallenari}, {Prusti}, {de Bruijne}, {Babusiaux}, {Bailer-Jones}, {Biermann},
  {Evans}, {Eyer}, {Jansen}, {Jordi}, {Klioner}, {Lammers}, {Lindegren},
  {Luri}, {Mignard}, {Panem}, {Pourbaix}, {Randich}, {Sartoretti}, {Siddiqui},
  {Soubiran}, {van Leeuwen}, {Walton}, {Arenou}, {Bastian}, {Cropper},
  {Drimmel}, {Katz}, {Lattanzi}, {Bakker}, {Cacciari}, {Casta{\~n}eda},
  {Chaoul}, {Cheek}, {De Angeli}, {Fabricius}, {Guerra}, {Holl}, {Masana},
  {Messineo}, {Mowlavi}, {Nienartowicz}, {Panuzzo}, {Portell}, {Riello},
  {Seabroke}, {Tanga}, {Th{\'e}venin}, {Gracia-Abril}, {Comoretto},
  {Garcia-Reinaldos}, {Teyssier}, {Altmann}, {Andrae}, {Audard},
  {Bellas-Velidis}, {Benson}, {Berthier}, {Blomme}, {Burgess}, {Busso},
  {Carry}, {Cellino}, {Clementini}, {Clotet}, {Creevey}, {Davidson}, {De
  Ridder}, {Delchambre}, {Dell'Oro}, {Ducourant},
  {Fern{\'a}ndez-Hern{\'a}ndez}, {Fouesneau}, {Fr{\'e}mat}, {Galluccio},
  {Garc{\'\i}a-Torres}, {Gonz{\'a}lez-N{\'u}{\~n}ez}, {Gonz{\'a}lez-Vidal},
  {Gosset}, {Guy}, {Halbwachs}, {Hambly}, {Harrison}, {Hern{\'a}ndez},
  {Hestroffer}, {Hodgkin}, {Hutton}, {Jasniewicz}, {Jean-Antoine-Piccolo},
  {Jordan}, {Korn}, {Krone-Martins}, {Lanzafame}, {Lebzelter}, {L{\"o}ffler},
  {Manteiga}, {Marrese}, {Mart{\'\i}n-Fleitas}, {Moitinho}, {Mora}, {Muinonen},
  {Osinde}, {Pancino}, {Pauwels}, {Petit}, {Recio-Blanco}, {Richards},
  {Rimoldini}, {Robin}, {Sarro}, {Siopis}, {Smith}, {Sozzetti}, {S{\"u}veges},
  {Torra}, {van Reeven}, {Abbas}, {Abreu Aramburu}, {Accart}, {Aerts},
  {Altavilla}, {{\'A}lvarez}, {Alvarez}, {Alves}, {Anderson}, {Andrei},
  {Anglada Varela}, {Antiche}, {Antoja}, {Arcay}, {Astraatmadja}, {Bach},
  {Baker}, {Balaguer-N{\'u}{\~n}ez}, {Balm}, {Barache}, {Barata}, {Barbato},
  {Barblan}, {Barklem}, {Barrado}, {Barros}, {Barstow}, {Bartholom{\'e}
  Mu{\~n}oz}, {Bassilana}, {Becciani}, {Bellazzini}, {Berihuete}, {Bertone},
  {Bianchi}, {Bienaym{\'e}}, {Blanco-Cuaresma}, {Boch}, {Boeche}, {Bombrun},
  {Borrachero}, {Bossini}, {Bouquillon}, {Bourda}, {Bragaglia}, {Bramante},
  {Breddels}, {Bressan}, {Brouillet}, {Br{\"u}semeister}, {Brugaletta},
  {Bucciarelli}, {Burlacu}, {Busonero}, {Butkevich}, {Buzzi}, {Caffau},
  {Cancelliere}, {Cannizzaro}, {Cantat-Gaudin}, {Carballo}, {Carlucci},
  {Carrasco}, {Casamiquela}, {Castellani}, {Castro-Ginard}, {Charlot},
  {Chemin}, {Chiavassa}, {Cocozza}, {Costigan}, {Cowell}, {Crifo}, {Crosta},
  {Crowley}, {Cuypers}, {Dafonte}, {Damerdji}, {Dapergolas}, {David}, {David},
  {de Laverny}, {De Luise}, {De March}, {de Martino}, {de Souza}, {de Torres},
  {Debosscher}, {del Pozo}, {Delbo}, {Delgado}, {Delgado}, {Di Matteo},
  {Diakite}, {Diener}, {Distefano}, {Dolding}, {Drazinos}, {Dur{\'a}n},
  {Edvardsson}, {Enke}, {Eriksson}, {Esquej}, {Eynard Bontemps}, {Fabre},
  {Fabrizio}, {Faigler}, {Falc{\~a}o}, {Farr{\`a}s Casas}, {Federici},
  {Fedorets}, {Fernique}, {Figueras}, {Filippi}, {Findeisen}, {Fonti},
  {Fraile}, {Fraser}, {Fr{\'e}zouls}, {Gai}, {Galleti}, {Garabato},
  {Garc{\'\i}a-Sedano}, {Garofalo}, {Garralda}, {Gavel}, {Gavras}, {Gerssen},
  {Geyer}, {Giacobbe}, {Gilmore}, {Girona}, {Giuffrida}, {Glass}, {Gomes},
  {Granvik}, {Gueguen}, {Guerrier}, {Guiraud}, {Guti{\'e}rrez-S{\'a}nchez},
  {Haigron}, {Hatzidimitriou}, {Hauser}, {Haywood}, {Heiter}, {Helmi}, {Heu},
  {Hilger}, {Hobbs}, {Hofmann}, {Holland}, {Huckle}, {Hypki}, {Icardi},
  {Jan{\ss}en}, {Jevardat de Fombelle}, {Jonker}, {Juh{\'a}sz}, {Julbe},
  {Karampelas}, {Kewley}, {Klar}, {Kochoska}, {Kohley}, {Kolenberg},
  {Kontizas}, {Kontizas}, {Koposov}, {Kordopatis}, {Kostrzewa-Rutkowska},
  {Koubsky}, {Lambert}, {Lanza}, {Lasne}, {Lavigne}, {Le Fustec}, {Le
  Poncin-Lafitte}, {Lebreton}, {Leccia}, {Leclerc}, {Lecoeur-Taibi},
  {Lenhardt}, {Leroux}, {Liao}, {Licata}, {Lindstr{\o}m}, {Lister}, {Livanou},
  {Lobel}, {L{\'o}pez}, {Managau}, {Mann}, {Mantelet}, {Marchal}, {Marchant},
  {Marconi}, {Marinoni}, {Marschalk{\'o}}, {Marshall}, {Martino}, {Marton},
  {Mary}, {Massari}, {Matijevi{\v{c}}}, {Mazeh}, {McMillan}, {Messina},
  {Michalik}, {Millar}, {Molina}, {Molinaro}, {Moln{\'a}r}, {Montegriffo},
  {Mor}, {Morbidelli}, {Morel}, {Morris}, {Mulone}, {Muraveva}, {Musella},
  {Nelemans}, {Nicastro}, {Noval}, {O'Mullane}, {Ord{\'e}novic},
  {Ord{\'o}{\~n}ez-Blanco}, {Osborne}, {Pagani}, {Pagano}, {Pailler},
  {Palacin}, {Palaversa}, {Panahi}, {Pawlak}, {Piersimoni}, {Pineau}, {Plachy},
  {Plum}, {Poggio}, {Poujoulet}, {Pr{\v{s}}a}, {Pulone}, {Racero}, {Ragaini},
  {Rambaux}, {Ramos-Lerate}, {Regibo}, {Reyl{\'e}}, {Riclet}, {Ripepi}, {Riva},
  {Rivard}, {Rixon}, {Roegiers}, {Roelens}, {Romero-G{\'o}mez}, {Rowell},
  {Royer}, {Ruiz-Dern}, {Sadowski}, {Sagrist{\`a} Sell{\'e}s}, {Sahlmann},
  {Salgado}, {Salguero}, {Sanna}, {Santana-Ros}, {Sarasso}, {Savietto},
  {Schultheis}, {Sciacca}, {Segol}, {Segovia}, {S{\'e}gransan}, {Shih},
  {Siltala}, {Silva}, {Smart}, {Smith}, {Solano}, {Solitro}, {Sordo}, {Soria
  Nieto}, {Souchay}, {Spagna}, {Spoto}, {Stampa}, {Steele},
  {Steidelm{\"u}ller}, {Stephenson}, {Stoev}, {Suess}, {Surdej}, {Szabados},
  {Szegedi-Elek}, {Tapiador}, {Taris}, {Tauran}, {Taylor}, {Teixeira},
  {Terrett}, {Teyssandier}, {Thuillot}, {Titarenko}, {Torra Clotet}, {Turon},
  {Ulla}, {Utrilla}, {Uzzi}, {Vaillant}, {Valentini}, {Valette}, {van Elteren},
  {Van Hemelryck}, {van Leeuwen}, {Vaschetto}, {Vecchiato}, {Veljanoski},
  {Viala}, {Vicente}, {Vogt}, {von Essen}, {Voss}, {Votruba}, {Voutsinas},
  {Walmsley}, {Weiler}, {Wertz}, {Wevers}, {Wyrzykowski}, {Yoldas},
  {{\v{Z}}erjal}, {Ziaeepour}, {Zorec}, {Zschocke}, {Zucker}, {Zurbach}, \&
  {Zwitter}}]{GaiaCol2018}
{Gaia Collaboration}, {Brown}, A.~G.~A., {Vallenari}, A., {et~al.} 2018, \aap,
  616, A1, \dodoi{10.1051/0004-6361/201833051}

\bibitem[{{Gaia Collaboration} {et~al.}(2021){Gaia Collaboration}, {Brown},
  {Vallenari}, {Prusti}, {de Bruijne}, {Babusiaux}, {Biermann}, {Creevey},
  {Evans}, {Eyer}, {Hutton}, {Jansen}, {Jordi}, {Klioner}, {Lammers},
  {Lindegren}, {Luri}, {Mignard}, {Panem}, {Pourbaix}, {Randich}, {Sartoretti},
  {Soubiran}, {Walton}, {Arenou}, {Bailer-Jones}, {Bastian}, {Cropper},
  {Drimmel}, {Katz}, {Lattanzi}, {van Leeuwen}, {Bakker}, {Cacciari},
  {Casta{\~n}eda}, {De Angeli}, {Ducourant}, {Fabricius}, {Fouesneau},
  {Fr{\'e}mat}, {Guerra}, {Guerrier}, {Guiraud}, {Jean-Antoine Piccolo},
  {Masana}, {Messineo}, {Mowlavi}, {Nicolas}, {Nienartowicz}, {Pailler},
  {Panuzzo}, {Riclet}, {Roux}, {Seabroke}, {Sordo}, {Tanga}, {Th{\'e}venin},
  {Gracia-Abril}, {Portell}, {Teyssier}, {Altmann}, {Andrae}, {Bellas-Velidis},
  {Benson}, {Berthier}, {Blomme}, {Brugaletta}, {Burgess}, {Busso}, {Carry},
  {Cellino}, {Cheek}, {Clementini}, {Damerdji}, {Davidson}, {Delchambre},
  {Dell'Oro}, {Fern{\'a}ndez-Hern{\'a}ndez}, {Galluccio}, {Garc{\'\i}a-Lario},
  {Garcia-Reinaldos}, {Gonz{\'a}lez-N{\'u}{\~n}ez}, {Gosset}, {Haigron},
  {Halbwachs}, {Hambly}, {Harrison}, {Hatzidimitriou}, {Heiter},
  {Hern{\'a}ndez}, {Hestroffer}, {Hodgkin}, {Holl}, {Jan{\ss}en}, {Jevardat de
  Fombelle}, {Jordan}, {Krone-Martins}, {Lanzafame}, {L{\"o}ffler}, {Lorca},
  {Manteiga}, {Marchal}, {Marrese}, {Moitinho}, {Mora}, {Muinonen}, {Osborne},
  {Pancino}, {Pauwels}, {Petit}, {Recio-Blanco}, {Richards}, {Riello},
  {Rimoldini}, {Robin}, {Roegiers}, {Rybizki}, {Sarro}, {Siopis}, {Smith},
  {Sozzetti}, {Ulla}, {Utrilla}, {van Leeuwen}, {van Reeven}, {Abbas}, {Abreu
  Aramburu}, {Accart}, {Aerts}, {Aguado}, {Ajaj}, {Altavilla}, {{\'A}lvarez},
  {{\'A}lvarez Cid-Fuentes}, {Alves}, {Anderson}, {Anglada Varela}, {Antoja},
  {Audard}, {Baines}, {Baker}, {Balaguer-N{\'u}{\~n}ez}, {Balbinot}, {Balog},
  {Barache}, {Barbato}, {Barros}, {Barstow}, {Bartolom{\'e}}, {Bassilana},
  {Bauchet}, {Baudesson-Stella}, {Becciani}, {Bellazzini}, {Bernet}, {Bertone},
  {Bianchi}, {Blanco-Cuaresma}, {Boch}, {Bombrun}, {Bossini}, {Bouquillon},
  {Bragaglia}, {Bramante}, {Breedt}, {Bressan}, {Brouillet}, {Bucciarelli},
  {Burlacu}, {Busonero}, {Butkevich}, {Buzzi}, {Caffau}, {Cancelliere},
  {C{\'a}novas}, {Cantat-Gaudin}, {Carballo}, {Carlucci}, {Carnerero},
  {Carrasco}, {Casamiquela}, {Castellani}, {Castro-Ginard}, {Castro Sampol},
  {Chaoul}, {Charlot}, {Chemin}, {Chiavassa}, {Cioni}, {Comoretto}, {Cooper},
  {Cornez}, {Cowell}, {Crifo}, {Crosta}, {Crowley}, {Dafonte}, {Dapergolas},
  {David}, {David}, {de Laverny}, {De Luise}, {De March}, {De Ridder}, {de
  Souza}, {de Teodoro}, {de Torres}, {del Peloso}, {del Pozo}, {Delbo},
  {Delgado}, {Delgado}, {Delisle}, {Di Matteo}, {Diakite}, {Diener},
  {Distefano}, {Dolding}, {Eappachen}, {Edvardsson}, {Enke}, {Esquej}, {Fabre},
  {Fabrizio}, {Faigler}, {Fedorets}, {Fernique}, {Fienga}, {Figueras},
  {Fouron}, {Fragkoudi}, {Fraile}, {Franke}, {Gai}, {Garabato},
  {Garcia-Gutierrez}, {Garc{\'\i}a-Torres}, {Garofalo}, {Gavras}, {Gerlach},
  {Geyer}, {Giacobbe}, {Gilmore}, {Girona}, {Giuffrida}, {Gomel}, {Gomez},
  {Gonzalez-Santamaria}, {Gonz{\'a}lez-Vidal}, {Granvik},
  {Guti{\'e}rrez-S{\'a}nchez}, {Guy}, {Hauser}, {Haywood}, {Helmi}, {Hidalgo},
  {Hilger}, {H{\l}adczuk}, {Hobbs}, {Holland}, {Huckle}, {Jasniewicz},
  {Jonker}, {Juaristi Campillo}, {Julbe}, {Karbevska}, {Kervella}, {Khanna},
  {Kochoska}, {Kontizas}, {Kordopatis}, {Korn}, {Kostrzewa-Rutkowska},
  {Kruszy{\'n}ska}, {Lambert}, {Lanza}, {Lasne}, {Le Campion}, {Le Fustec},
  {Lebreton}, {Lebzelter}, {Leccia}, {Leclerc}, {Lecoeur-Taibi}, {Liao},
  {Licata}, {Lindstr{\o}m}, {Lister}, {Livanou}, {Lobel}, {Madrero Pardo},
  {Managau}, {Mann}, {Marchant}, {Marconi}, {Marcos Santos}, {Marinoni},
  {Marocco}, {Marshall}, {Martin Polo}, {Mart{\'\i}n-Fleitas}, {Masip},
  {Massari}, {Mastrobuono-Battisti}, {Mazeh}, {McMillan}, {Messina},
  {Michalik}, {Millar}, {Mints}, {Molina}, {Molinaro}, {Moln{\'a}r},
  {Montegriffo}, {Mor}, {Morbidelli}, {Morel}, {Morris}, {Mulone}, {Munoz},
  {Muraveva}, {Murphy}, {Musella}, {Noval}, {Ord{\'e}novic}, {Orr{\`u}},
  {Osinde}, {Pagani}, {Pagano}, {Palaversa}, {Palicio}, {Panahi}, {Pawlak},
  {Pe{\~n}alosa Esteller}, {Penttil{\"a}}, {Piersimoni}, {Pineau}, {Plachy},
  {Plum}, {Poggio}, {Poretti}, {Poujoulet}, {Pr{\v{s}}a}, {Pulone}, {Racero},
  {Ragaini}, {Rainer}, {Raiteri}, {Rambaux}, {Ramos}, {Ramos-Lerate}, {Re
  Fiorentin}, {Regibo}, {Reyl{\'e}}, {Ripepi}, {Riva}, {Rixon}, {Robichon},
  {Robin}, {Roelens}, {Rohrbasser}, {Romero-G{\'o}mez}, {Rowell}, {Royer},
  {Rybicki}, {Sadowski}, {Sagrist{\`a} Sell{\'e}s}, {Sahlmann}, {Salgado},
  {Salguero}, {Samaras}, {Sanchez Gimenez}, {Sanna}, {Santove{\~n}a},
  {Sarasso}, {Schultheis}, {Sciacca}, {Segol}, {Segovia}, {S{\'e}gransan},
  {Semeux}, {Shahaf}, {Siddiqui}, {Siebert}, {Siltala}, {Slezak}, {Smart},
  {Solano}, {Solitro}, {Souami}, {Souchay}, {Spagna}, {Spoto}, {Steele},
  {Steidelm{\"u}ller}, {Stephenson}, {S{\"u}veges}, {Szabados}, {Szegedi-Elek},
  {Taris}, {Tauran}, {Taylor}, {Teixeira}, {Thuillot}, {Tonello}, {Torra},
  {Torra}, {Turon}, {Unger}, {Vaillant}, {van Dillen}, {Vanel}, {Vecchiato},
  {Viala}, {Vicente}, {Voutsinas}, {Weiler}, {Wevers}, {Wyrzykowski}, {Yoldas},
  {Yvard}, {Zhao}, {Zorec}, {Zucker}, {Zurbach}, \& {Zwitter}}]{GaiaCol2021}
---. 2021, \aap, 649, A1, \dodoi{10.1051/0004-6361/202039657}

\bibitem[{{Garc{\'{\i}}a P{\'e}rez} {et~al.}(2016){Garc{\'{\i}}a P{\'e}rez},
  {Allende Prieto}, {Holtzman}, {Shetrone}, {M{\'e}sz{\'a}ros}, {Bizyaev},
  {Carrera}, {Cunha}, {Garc{\'{\i}}a-Hern{\'a}ndez}, {Johnson}, {Majewski},
  {Nidever}, {Schiavon}, {Shane}, {Smith}, {Sobeck}, {Troup}, {Zamora},
  {Weinberg}, {Bovy}, {Eisenstein}, {Feuillet}, {Frinchaboy}, {Hayden},
  {Hearty}, {Nguyen}, {O'Connell}, {Pinsonneault}, {Wilson}, \&
  {Zasowski}}]{GarciaPerez2016}
{Garc{\'{\i}}a P{\'e}rez}, A.~E., {Allende Prieto}, C., {Holtzman}, J.~A.,
  {et~al.} 2016, \aj, 151, 144, \dodoi{10.3847/0004-6256/151/6/144}

\bibitem[{{Gnedin} {et~al.}(2005){Gnedin}, {Gould}, {Miralda-Escud{\'e}}, \&
  {Zentner}}]{Gnedin2005}
{Gnedin}, O.~Y., {Gould}, A., {Miralda-Escud{\'e}}, J., \& {Zentner}, A.~R.
  2005, \apj, 634, 344, \dodoi{10.1086/496958}

\bibitem[{{Gravity Collaboration} {et~al.}(2018){Gravity Collaboration},
  {Abuter}, {Amorim}, {Anugu}, {Baub{\"o}ck}, {Benisty}, {Berger}, {Blind},
  {Bonnet}, {Brandner}, {Buron}, {Collin}, {Chapron}, {Cl{\'e}net}, {Coud{\'e}
  Du Foresto}, {de Zeeuw}, {Deen}, {Delplancke-Str{\"o}bele}, {Dembet},
  {Dexter}, {Duvert}, {Eckart}, {Eisenhauer}, {Finger}, {F{\"o}rster
  Schreiber}, {F{\'e}dou}, {Garcia}, {Garcia Lopez}, {Gao}, {Gendron},
  {Genzel}, {Gillessen}, {Gordo}, {Habibi}, {Haubois}, {Haug}, {Hau{\ss}mann},
  {Henning}, {Hippler}, {Horrobin}, {Hubert}, {Hubin}, {Jimenez Rosales},
  {Jochum}, {Jocou}, {Kaufer}, {Kellner}, {Kendrew}, {Kervella}, {Kok},
  {Kulas}, {Lacour}, {Lapeyr{\`e}re}, {Lazareff}, {Le Bouquin}, {L{\'e}na},
  {Lippa}, {Lenzen}, {M{\'e}rand}, {M{\"u}ler}, {Neumann}, {Ott}, {Palanca},
  {Paumard}, {Pasquini}, {Perraut}, {Perrin}, {Pfuhl}, {Plewa}, {Rabien},
  {Ram{\'\i}rez}, {Ramos}, {Rau}, {Rodr{\'\i}guez-Coira}, {Rohloff}, {Rousset},
  {Sanchez-Bermudez}, {Scheithauer}, {Sch{\"o}ller}, {Schuler}, {Spyromilio},
  {Straub}, {Straubmeier}, {Sturm}, {Tacconi}, {Tristram}, {Vincent}, {von
  Fellenberg}, {Wank}, {Waisberg}, {Widmann}, {Wieprecht}, {Wiest},
  {Wiezorrek}, {Woillez}, {Yazici}, {Ziegler}, \& {Zins}}]{GravityCol2018}
{Gravity Collaboration}, {Abuter}, R., {Amorim}, A., {et~al.} 2018, \aap, 615,
  L15, \dodoi{10.1051/0004-6361/201833718}

\bibitem[{{Green}(2018)}]{Green2018}
{Green}, G. 2018, The Journal of Open Source Software, 3, 695,
  \dodoi{10.21105/joss.00695}

\bibitem[{{Gunn} {et~al.}(2006){Gunn}, {Siegmund}, {Mannery}, {Owen}, {Hull},
  {Leger}, {Carey}, {Knapp}, {York}, {Boroski}, {Kent}, {Lupton}, {Rockosi},
  {Evans}, {Waddell}, {Anderson}, {Annis}, {Barentine}, {Bartoszek}, {Bastian},
  {Bracker}, {Brewington}, {Briegel}, {Brinkmann}, {Brown}, {Carr},
  {Czarapata}, {Drennan}, {Dombeck}, {Federwitz}, {Gillespie}, {Gonzales},
  {Hansen}, {Harvanek}, {Hayes}, {Jordan}, {Kinney}, {Klaene}, {Kleinman},
  {Kron}, {Kresinski}, {Lee}, {Limmongkol}, {Lindenmeyer}, {Long}, {Loomis},
  {McGehee}, {Mantsch}, {Neilsen}, {Neswold}, {Newman}, {Nitta}, {Peoples},
  {Pier}, {Prieto}, {Prosapio}, {Rivetta}, {Schneider}, {Snedden}, \&
  {Wang}}]{Gunn2006}
{Gunn}, J.~E., {Siegmund}, W.~A., {Mannery}, E.~J., {et~al.} 2006, \aj, 131,
  2332, \dodoi{10.1086/500975}

\bibitem[{{Hasselquist} {et~al.}(2016){Hasselquist}, {Shetrone}, {Cunha},
  {Smith}, {Holtzman}, {Lawler}, {Allende Prieto}, {Beers}, {Chojnowski},
  {Fern{\'a}ndez-Trincado}, {Garc{\'\i}a-Hern{\'a}ndez}, {Hearty}, {Majewski},
  {Pereira}, {Placco}, {Villanova}, \& {Zamora}}]{Hasselquist2016}
{Hasselquist}, S., {Shetrone}, M., {Cunha}, K., {et~al.} 2016, \apj, 833, 81,
  \dodoi{10.3847/1538-4357/833/1/81}

\bibitem[{{Hasselquist} {et~al.}(2021){Hasselquist}, {Hayes}, {Lian},
  {Weinberg}, {Zasowski}, {Horta}, {Beaton}, {Feuillet}, {Garro}, {Gallart},
  {Smith}, {Holtzman}, {Minniti}, {Lacerna}, {Shetrone}, {J{\"o}nsson},
  {Cioni}, {Fillingham}, {Cunha}, {O'Connell}, {Fern{\'a}ndez-Trincado},
  {Mu{\~n}oz}, {Schiavon}, {Almeida}, {Anguiano}, {Beers}, {Bizyaev},
  {Brownstein}, {Cohen}, {Frinchaboy}, {Garc{\'\i}a-Hern{\'a}ndez}, {Geisler},
  {Lane}, {Majewski}, {Nidever}, {Nitschelm}, {Povick}, {Price-Whelan},
  {Roman-Lopes}, {Rosado}, {Sobeck}, {Stringfellow}, {Valenzuela}, {Villanova},
  \& {Vincenzo}}]{Hasselquist2021}
{Hasselquist}, S., {Hayes}, C.~R., {Lian}, J., {et~al.} 2021, \apj, 923, 172,
  \dodoi{10.3847/1538-4357/ac25f9}

\bibitem[{{Hattori} {et~al.}(2018{\natexlab{a}}){Hattori}, {Valluri}, {Bell},
  \& {Roederer}}]{Hattori2018A}
{Hattori}, K., {Valluri}, M., {Bell}, E.~F., \& {Roederer}, I.~U.
  2018{\natexlab{a}}, \apj, 866, 121, \dodoi{10.3847/1538-4357/aadee5}

\bibitem[{{Hattori} {et~al.}(2018{\natexlab{b}}){Hattori}, {Valluri}, \&
  {Castro}}]{Hattori2018}
{Hattori}, K., {Valluri}, M., \& {Castro}, N. 2018{\natexlab{b}}, \apj, 869,
  33, \dodoi{10.3847/1538-4357/aaed22}

\bibitem[{{Hawkins} {et~al.}(2015){Hawkins}, {Jofr{\'e}}, {Masseron}, \&
  {Gilmore}}]{Hawkins2015}
{Hawkins}, K., {Jofr{\'e}}, P., {Masseron}, T., \& {Gilmore}, G. 2015, \mnras,
  453, 758, \dodoi{10.1093/mnras/stv1586}

\bibitem[{{Hawkins} \& {Wyse}(2018)}]{Hawkins2018}
{Hawkins}, K., \& {Wyse}, R. F.~G. 2018, \mnras, 481, 1028,
  \dodoi{10.1093/mnras/sty2282}

\bibitem[{{Hayes} {et~al.}(2018){Hayes}, {Majewski}, {Shetrone},
  {Fern{\'a}ndez-Alvar}, {Allende Prieto}, {Schuster}, {Carigi}, {Cunha},
  {Smith}, {Sobeck}, {Almeida}, {Beers}, {Carrera}, {Fern{\'a}ndez-Trincado},
  {Garc{\'\i}a-Hern{\'a}ndez}, {Geisler}, {Lane}, {Lucatello}, {Matthews},
  {Minniti}, {Nitschelm}, {Tang}, {Tissera}, \& {Zamora}}]{Hayes2018}
{Hayes}, C.~R., {Majewski}, S.~R., {Shetrone}, M., {et~al.} 2018, \apj, 852,
  49, \dodoi{10.3847/1538-4357/aa9cec}

\bibitem[{{Heber} {et~al.}(2008){Heber}, {Edelmann}, {Napiwotzki}, {Altmann},
  \& {Scholz}}]{Heber2008}
{Heber}, U., {Edelmann}, H., {Napiwotzki}, R., {Altmann}, M., \& {Scholz},
  R.~D. 2008, \aap, 483, L21, \dodoi{10.1051/0004-6361:200809767}

\bibitem[{{Helmi} {et~al.}(2018){Helmi}, {Babusiaux}, {Koppelman}, {Massari},
  {Veljanoski}, \& {Brown}}]{Helmi2018}
{Helmi}, A., {Babusiaux}, C., {Koppelman}, H.~H., {et~al.} 2018, \nat, 563, 85,
  \dodoi{10.1038/s41586-018-0625-x}

\bibitem[{{Herzog-Arbeitman} {et~al.}(2018){Herzog-Arbeitman}, {Lisanti}, \&
  {Necib}}]{HerzogArbeitman2018}
{Herzog-Arbeitman}, J., {Lisanti}, M., \& {Necib}, L. 2018, \jcap, 2018, 052,
  \dodoi{10.1088/1475-7516/2018/04/052}

\bibitem[{{Hills}(1988)}]{Hills1988}
{Hills}, J.~G. 1988, \nat, 331, 687, \dodoi{10.1038/331687a0}

\bibitem[{{Holtzman} {et~al.}(2015){Holtzman}, {Shetrone}, {Johnson}, {Allende
  Prieto}, {Anders}, {Andrews}, {Beers}, {Bizyaev}, {Blanton}, {Bovy},
  {Carrera}, {Chojnowski}, {Cunha}, {Eisenstein}, {Feuillet}, {Frinchaboy},
  {Galbraith-Frew}, {Garc{\'{\i}}a P{\'e}rez}, {Garc{\'{\i}}a-Hern{\'a}ndez},
  {Hasselquist}, {Hayden}, {Hearty}, {Ivans}, {Majewski}, {Martell},
  {Meszaros}, {Muna}, {Nidever}, {Nguyen}, {O'Connell}, {Pan}, {Pinsonneault},
  {Robin}, {Schiavon}, {Shane}, {Sobeck}, {Smith}, {Troup}, {Weinberg},
  {Wilson}, {Wood-Vasey}, {Zamora}, \& {Zasowski}}]{Holtzman2015}
{Holtzman}, J.~A., {Shetrone}, M., {Johnson}, J.~A., {et~al.} 2015, \aj, 150,
  148, \dodoi{10.1088/0004-6256/150/5/148}

\bibitem[{{Irrgang} {et~al.}(2018){Irrgang}, {Kreuzer}, \&
  {Heber}}]{Irrgang2018}
{Irrgang}, A., {Kreuzer}, S., \& {Heber}, U. 2018, \aap, 620, A48,
  \dodoi{10.1051/0004-6361/201833874}

\bibitem[{{Irrgang} {et~al.}(2013){Irrgang}, {Wilcox}, {Tucker}, \&
  {Schiefelbein}}]{Irrgang2013}
{Irrgang}, A., {Wilcox}, B., {Tucker}, E., \& {Schiefelbein}, L. 2013, \aap,
  549, A137, \dodoi{10.1051/0004-6361/201220540}

\bibitem[{{J{\"o}nsson} {et~al.}(2018){J{\"o}nsson}, {Allende Prieto},
  {Holtzman}, {Feuillet}, {Hawkins}, {Cunha}, {M{\'e}sz{\'a}ros},
  {Hasselquist}, {Fern{\'a}ndez-Trincado}, {Garc{\'\i}a-Hern{\'a}ndez},
  {Bizyaev}, {Carrera}, {Majewski}, {Pinsonneault}, {Shetrone}, {Smith},
  {Sobeck}, {Souto}, {Stringfellow}, {Teske}, \& {Zamora}}]{Jonsson2018}
{J{\"o}nsson}, H., {Allende Prieto}, C., {Holtzman}, J.~A., {et~al.} 2018, \aj,
  156, 126, \dodoi{10.3847/1538-3881/aad4f5}

\bibitem[{{J{\"o}nsson} {et~al.}(2020){J{\"o}nsson}, {Holtzman}, {Allende
  Prieto}, {Cunha}, {Garc{\'\i}a-Hern{\'a}ndez}, {Hasselquist}, {Masseron},
  {Osorio}, {Shetrone}, {Smith}, {Stringfellow}, {Bizyaev}, {Edvardsson},
  {Majewski}, {M{\'e}sz{\'a}ros}, {Souto}, {Zamora}, {Beaton}, {Bovy}, {Donor},
  {Pinsonneault}, {Poovelil}, \& {Sobeck}}]{Jonsson2020}
{J{\"o}nsson}, H., {Holtzman}, J.~A., {Allende Prieto}, C., {et~al.} 2020, \aj,
  160, 120, \dodoi{10.3847/1538-3881/aba592}

\bibitem[{{Kobayashi} {et~al.}(2020{\natexlab{a}}){Kobayashi}, {Karakas}, \&
  {Lugaro}}]{Kobayashi2020A}
{Kobayashi}, C., {Karakas}, A.~I., \& {Lugaro}, M. 2020{\natexlab{a}}, \apj,
  900, 179, \dodoi{10.3847/1538-4357/abae65}

\bibitem[{{Kobayashi} {et~al.}(2020{\natexlab{b}}){Kobayashi}, {Leung}, \&
  {Nomoto}}]{Kobayashi2020}
{Kobayashi}, C., {Leung}, S.-C., \& {Nomoto}, K. 2020{\natexlab{b}}, \apj, 895,
  138, \dodoi{10.3847/1538-4357/ab8e44}

\bibitem[{{Koppelman} {et~al.}(2019){Koppelman}, {Helmi}, {Massari},
  {Price-Whelan}, \& {Starkenburg}}]{Koppelman2019}
{Koppelman}, H.~H., {Helmi}, A., {Massari}, D., {Price-Whelan}, A.~M., \&
  {Starkenburg}, T.~K. 2019, \aap, 631, L9, \dodoi{10.1051/0004-6361/201936738}

\bibitem[{{Kreuzer} {et~al.}(2020){Kreuzer}, {Irrgang}, \&
  {Heber}}]{Kreuzer2020}
{Kreuzer}, S., {Irrgang}, A., \& {Heber}, U. 2020, \aap, 637, A53,
  \dodoi{10.1051/0004-6361/202037747}

\bibitem[{{Lane} {et~al.}(2022){Lane}, {Bovy}, \& {Mackereth}}]{Lane2022}
{Lane}, J. M.~M., {Bovy}, J., \& {Mackereth}, J.~T. 2022, \mnras, 510, 5119,
  \dodoi{10.1093/mnras/stab3755}

\bibitem[{{Leung} \& {Bovy}(2019)}]{Leung2019}
{Leung}, H.~W., \& {Bovy}, J. 2019, \mnras, 489, 2079,
  \dodoi{10.1093/mnras/stz2245}

\bibitem[{{Li} {et~al.}(2021){Li}, {Luo}, {Lu}, {Zhang}, {Li}, {Wang}, {Zuo},
  {Xiang}, {Ting}, {Marchetti}, {Li}, {Wang}, {Zhang}, {Hattori}, {Zhao},
  {Zhang}, \& {Zhao}}]{Li2021}
{Li}, Y.-B., {Luo}, A.~L., {Lu}, Y.-J., {et~al.} 2021, \apjs, 252, 3,
  \dodoi{10.3847/1538-4365/abc16e}

\bibitem[{{Mackereth} {et~al.}(2019){Mackereth}, {Schiavon}, {Pfeffer},
  {Hayes}, {Bovy}, {Anguiano}, {Allende Prieto}, {Hasselquist}, {Holtzman},
  {Johnson}, {Majewski}, {O'Connell}, {Shetrone}, {Tissera}, \&
  {Fern{\'a}ndez-Trincado}}]{Mackereth2019}
{Mackereth}, J.~T., {Schiavon}, R.~P., {Pfeffer}, J., {et~al.} 2019, \mnras,
  482, 3426, \dodoi{10.1093/mnras/sty2955}

\bibitem[{{Majewski} {et~al.}(2017){Majewski}, {Schiavon}, {Frinchaboy},
  {Allende Prieto}, {Barkhouser}, {Bizyaev}, {Blank}, {Brunner}, {Burton},
  {Carrera}, {Chojnowski}, {Cunha}, {Epstein}, {Fitzgerald}, {Garc{\'\i}a
  P{\'e}rez}, {Hearty}, {Henderson}, {Holtzman}, {Johnson}, {Lam}, {Lawler},
  {Maseman}, {M{\'e}sz{\'a}ros}, {Nelson}, {Nguyen}, {Nidever}, {Pinsonneault},
  {Shetrone}, {Smee}, {Smith}, {Stolberg}, {Skrutskie}, {Walker}, {Wilson},
  {Zasowski}, {Anders}, {Basu}, {Beland}, {Blanton}, {Bovy}, {Brownstein},
  {Carlberg}, {Chaplin}, {Chiappini}, { ein}, {Elsworth}, {Feuillet},
  {Fleming}, {Galbraith-Frew}, {Garc{\'\i}a}, {Garc{\'\i}a-Hern{\'a}ndez},
  {Gillespie}, {Girardi}, {Gunn}, {Hasselquist}, {Hayden}, {Hekker}, {Ivans},
  {Kinemuchi}, {Klaene}, {Mahadevan}, {Mathur}, {Mosser}, {Muna}, {Munn},
  {Nichol}, {O'Connell}, {Parejko}, {Robin}, {Rocha-Pinto}, {Schultheis},
  {Serenelli}, {Shane}, {Silva Aguirre}, {Sobeck}, {Thompson}, {Troup},
  {Weinberg}, \& {Zamora}}]{Majewski2017}
{Majewski}, S.~R., {Schiavon}, R.~P., {Frinchaboy}, P.~M., {et~al.} 2017, \aj,
  154, 94, \dodoi{10.3847/1538-3881/aa784d}

\bibitem[{{Marchetti}(2021)}]{Marchetti2021}
{Marchetti}, T. 2021, \mnras, 503, 1374, \dodoi{10.1093/mnras/stab599}

\bibitem[{{Marchetti} {et~al.}(2019){Marchetti}, {Rossi}, \&
  {Brown}}]{Marchetti2019}
{Marchetti}, T., {Rossi}, E.~M., \& {Brown}, A.~G.~A. 2019, \mnras, 490, 157,
  \dodoi{10.1093/mnras/sty2592}

\bibitem[{{Martell} {et~al.}(2017){Martell}, {Sharma}, {Buder}, {Duong},
  {Schlesinger}, {Simpson}, {Lind}, {Ness}, {Marshall}, {Asplund},
  {Bland-Hawthorn}, {Casey}, {De Silva}, {Freeman}, {Kos}, {Lin}, {Zucker},
  {Zwitter}, {Anguiano}, {Bacigalupo}, {Carollo}, {Casagrande}, {Da Costa},
  {Horner}, {Huber}, {Hyde}, {Kafle}, {Lewis}, {Nataf}, {Navin}, {Stello},
  {Tinney}, {Watson}, \& {Wittenmyer}}]{Martell2017}
{Martell}, S.~L., {Sharma}, S., {Buder}, S., {et~al.} 2017, \mnras, 465, 3203,
  \dodoi{10.1093/mnras/stw2835}

\bibitem[{{McMillan}(2017)}]{McMillan2017}
{McMillan}, P.~J. 2017, \mnras, 465, 76, \dodoi{10.1093/mnras/stw2759}

\bibitem[{{M{\'e}sz{\'a}ros} {et~al.}(2012){M{\'e}sz{\'a}ros}, {Allende
  Prieto}, {Edvardsson}, {Castelli}, {Garc{\'\i}a P{\'e}rez}, {Gustafsson},
  {Majewski}, {Plez}, {Schiavon}, {Shetrone}, \& {de Vicente}}]{Meszaros2012}
{M{\'e}sz{\'a}ros}, S., {Allende Prieto}, C., {Edvardsson}, B., {et~al.} 2012,
  \aj, 144, 120, \dodoi{10.1088/0004-6256/144/4/120}

\bibitem[{{Miyamoto} \& {Nagai}(1975)}]{Miyamoto1975}
{Miyamoto}, M., \& {Nagai}, R. 1975, \pasj, 27, 533

\bibitem[{{Monty} {et~al.}(2020){Monty}, {Venn}, {Lane}, {Lokhorst}, \&
  {Yong}}]{Monty2020}
{Monty}, S., {Venn}, K.~A., {Lane}, J. M.~M., {Lokhorst}, D., \& {Yong}, D.
  2020, \mnras, 497, 1236, \dodoi{10.1093/mnras/staa1995}

\bibitem[{{Morton}(2015)}]{Morton2015}
{Morton}, T.~D. 2015, {isochrones: Stellar model grid package}.
\newblock \doeprint{1503.010}

\bibitem[{{Myeong} {et~al.}(2019){Myeong}, {Vasiliev}, {Iorio}, {Evans}, \&
  {Belokurov}}]{Myeong2019}
{Myeong}, G.~C., {Vasiliev}, E., {Iorio}, G., {Evans}, N.~W., \& {Belokurov},
  V. 2019, \mnras, 488, 1235, \dodoi{10.1093/mnras/stz1770}

\bibitem[{{Nidever} {et~al.}(2015){Nidever}, {Holtzman}, {Allende Prieto},
  {Beland}, {Bender}, {Bizyaev}, {Burton}, {Desphande}, {Fleming},
  {Garc{\'{\i}}a P{\'e}rez}, {Hearty}, {Majewski}, {M{\'e}sz{\'a}ros}, {Muna},
  {Nguyen}, {Schiavon}, {Shetrone}, {Skrutskie}, {Sobeck}, \&
  {Wilson}}]{Nidever2015}
{Nidever}, D.~L., {Holtzman}, J.~A., {Allende Prieto}, C., {et~al.} 2015, \aj,
  150, 173, \dodoi{10.1088/0004-6256/150/6/173}

\bibitem[{{Nissen} \& {Schuster}(2010)}]{Nissen2010}
{Nissen}, P.~E., \& {Schuster}, W.~J. 2010, \aap, 511, L10,
  \dodoi{10.1051/0004-6361/200913877}

\bibitem[{{Nomoto} {et~al.}(2013){Nomoto}, {Kobayashi}, \&
  {Tominaga}}]{Nomoto2013}
{Nomoto}, K., {Kobayashi}, C., \& {Tominaga}, N. 2013, \araa, 51, 457,
  \dodoi{10.1146/annurev-astro-082812-140956}

\bibitem[{{Paxton} {et~al.}(2011){Paxton}, {Bildsten}, {Dotter}, {Herwig},
  {Lesaffre}, \& {Timmes}}]{Paxton2011}
{Paxton}, B., {Bildsten}, L., {Dotter}, A., {et~al.} 2011, \apjs, 192, 3,
  \dodoi{10.1088/0067-0049/192/1/3}

\bibitem[{{Paxton} {et~al.}(2013){Paxton}, {Cantiello}, {Arras}, {Bildsten},
  {Brown}, {Dotter}, {Mankovich}, {Montgomery}, {Stello}, {Timmes}, \&
  {Townsend}}]{Paxton2013}
{Paxton}, B., {Cantiello}, M., {Arras}, P., {et~al.} 2013, \apjs, 208, 4,
  \dodoi{10.1088/0067-0049/208/1/4}

\bibitem[{{Paxton} {et~al.}(2015){Paxton}, {Marchant}, {Schwab}, {Bauer},
  {Bildsten}, {Cantiello}, {Dessart}, {Farmer}, {Hu}, {Langer}, {Townsend},
  {Townsley}, \& {Timmes}}]{Paxton2015}
{Paxton}, B., {Marchant}, P., {Schwab}, J., {et~al.} 2015, \apjs, 220, 15,
  \dodoi{10.1088/0067-0049/220/1/15}

\bibitem[{{Pereira} {et~al.}(2012){Pereira}, {Jilinski}, {Drake}, {de Castro},
  {Ortega}, {Chavero}, \& {Roig}}]{Pereira2012}
{Pereira}, C.~B., {Jilinski}, E., {Drake}, N.~A., {et~al.} 2012, \aap, 543,
  A58, \dodoi{10.1051/0004-6361/201219122}

\bibitem[{{Piffl} {et~al.}(2014){Piffl}, {Scannapieco}, {Binney}, {Steinmetz},
  {Scholz}, {Williams}, {de Jong}, {Kordopatis}, {Matijevi{\v{c}}},
  {Bienaym{\'e}}, {Bland-Hawthorn}, {Boeche}, {Freeman}, {Gibson}, {Gilmore},
  {Grebel}, {Helmi}, {Munari}, {Navarro}, {Parker}, {Reid}, {Seabroke},
  {Watson}, {Wyse}, \& {Zwitter}}]{Piffl2014}
{Piffl}, T., {Scannapieco}, C., {Binney}, J., {et~al.} 2014, \aap, 562, A91,
  \dodoi{10.1051/0004-6361/201322531}

\bibitem[{{Poveda} {et~al.}(1967){Poveda}, {Ruiz}, \& {Allen}}]{Poveda1967}
{Poveda}, A., {Ruiz}, J., \& {Allen}, C. 1967, Boletin de los Observatorios
  Tonantzintla y Tacubaya, 4, 86

\bibitem[{Price-Whelan(2018)}]{adrian-price-whelan_2018}
Price-Whelan, A. 2018, adrn/pyia v0.2, a Python package for working with data
  from the Gaia mission, \dodoi{10.5281/zenodo.1228136}

\bibitem[{{Przybilla} {et~al.}(2008){Przybilla}, {Fernanda Nieva}, {Heber}, \&
  {Butler}}]{Przybilla2008}
{Przybilla}, N., {Fernanda Nieva}, M., {Heber}, U., \& {Butler}, K. 2008,
  \apjl, 684, L103, \dodoi{10.1086/592245}

\bibitem[{{Queiroz} {et~al.}(2018){Queiroz}, {Anders}, {Santiago}, {Chiappini},
  {Steinmetz}, {Dal Ponte}, {Stassun}, {da Costa}, {Maia}, {Crestani}, {Beers},
  {Fern{\'a}ndez-Trincado}, {Garc{\'\i}a-Hern{\'a}ndez}, {Roman-Lopes}, \&
  {Zamora}}]{Queiroz2018}
{Queiroz}, A.~B.~A., {Anders}, F., {Santiago}, B.~X., {et~al.} 2018, \mnras,
  476, 2556, \dodoi{10.1093/mnras/sty330}

\bibitem[{{Queiroz} {et~al.}(2020){Queiroz}, {Anders}, {Chiappini},
  {Khalatyan}, {Santiago}, {Steinmetz}, {Valentini}, {Miglio}, {Bossini},
  {Barbuy}, {Minchev}, {Minniti}, {Garc{\'\i}a Hern{\'a}ndez}, {Schultheis},
  {Beaton}, {Beers}, {Bizyaev}, {Brownstein}, {Cunha},
  {Fern{\'a}ndez-Trincado}, {Frinchaboy}, {Lane}, {Majewski}, {Nataf},
  {Nitschelm}, {Pan}, {Roman-Lopes}, {Sobeck}, {Stringfellow}, \&
  {Zamora}}]{Queiroz2020}
{Queiroz}, A.~B.~A., {Anders}, F., {Chiappini}, C., {et~al.} 2020, \aap, 638,
  A76, \dodoi{10.1051/0004-6361/201937364}

\bibitem[{{Reggiani} {et~al.}(2022){Reggiani}, {Ji}, {Schlaufman}, {Frebel},
  {Necib}, {Nelson}, {Hawkins}, \& {Galarza}}]{Reggiani2022}
{Reggiani}, H., {Ji}, A.~P., {Schlaufman}, K.~C., {et~al.} 2022, arXiv
  e-prints, arXiv:2203.16364.
\newblock \doarXiv{2203.16364}

\bibitem[{{Riello} {et~al.}(2021){Riello}, {De Angeli}, {Evans}, {Montegriffo},
  {Carrasco}, {Busso}, {Palaversa}, {Burgess}, {Diener}, {Davidson}, {Rowell},
  {Fabricius}, {Jordi}, {Bellazzini}, {Pancino}, {Harrison}, {Cacciari}, {van
  Leeuwen}, {Hambly}, {Hodgkin}, {Osborne}, {Altavilla}, {Barstow}, {Brown},
  {Castellani}, {Cowell}, {De Luise}, {Gilmore}, {Giuffrida}, {Hidalgo},
  {Holland}, {Marinoni}, {Pagani}, {Piersimoni}, {Pulone}, {Ragaini}, {Rainer},
  {Richards}, {Sanna}, {Walton}, {Weiler}, \& {Yoldas}}]{Riello2021}
{Riello}, M., {De Angeli}, F., {Evans}, D.~W., {et~al.} 2021, \aap, 649, A3,
  \dodoi{10.1051/0004-6361/202039587}

\bibitem[{{Rybizki} {et~al.}(2020){Rybizki}, {Demleitner}, {Bailer-Jones},
  {Tio}, {Cantat-Gaudin}, {Fouesneau}, {Chen}, {Andrae}, {Girardi}, \&
  {Sharma}}]{Rybizki2020}
{Rybizki}, J., {Demleitner}, M., {Bailer-Jones}, C., {et~al.} 2020, \pasp, 132,
  074501, \dodoi{10.1088/1538-3873/ab8cb0}

\bibitem[{{Rybizki} {et~al.}(2021){Rybizki}, {Green}, {Rix}, {El-Badry},
  {Demleitner}, {Zari}, {Udalski}, {Smart}, \& {Gould}}]{Rybizki2021}
{Rybizki}, J., {Green}, G., {Rix}, H.-W., {et~al.} 2021, arXiv e-prints,
  arXiv:2101.11641.
\newblock \doarXiv{2101.11641}

\bibitem[{{Santana} {et~al.}(2021){Santana}, {Beaton}, {Covey}, {O'Connell},
  {Longa-Pe{\~n}a}, {Cohen}, {Fern{\'a}ndez-Trincado}, {Hayes}, {Zasowski},
  {Sobeck}, {Majewski}, {Chojnowski}, {De Lee}, {Oelkers}, {Stringfellow},
  {Almeida}, {Anguiano}, {Donor}, {Frinchaboy}, {Hasselquist}, {Johnson},
  {Kollmeier}, {Nidever}, {Price-Whelan}, {Rojas-Arriagada}, {Schultheis},
  {Shetrone}, {Simon}, {Aerts}, {Borissova}, {Drout}, {Geisler}, {Law},
  {Medina}, {Minniti}, {Monachesi}, {Mu{\~n}oz}, {Poleski}, {Roman-Lopes},
  {Schlaufman}, {Stutz}, {Teske}, {Tkachenko}, {Van Saders}, {Weinberger}, \&
  {Zoccali}}]{Santana2021}
{Santana}, F.~A., {Beaton}, R.~L., {Covey}, K.~R., {et~al.} 2021, \aj, 162,
  303, \dodoi{10.3847/1538-3881/ac2cbc}

\bibitem[{{Schlafly} \& {Finkbeiner}(2011)}]{Schlafly2011}
{Schlafly}, E.~F., \& {Finkbeiner}, D.~P. 2011, \apj, 737, 103,
  \dodoi{10.1088/0004-637X/737/2/103}

\bibitem[{{Shen} {et~al.}(2018){Shen}, {Boubert}, {G{\"a}nsicke}, {Jha},
  {Andrews}, {Chomiuk}, {Foley}, {Fraser}, {Gromadzki}, {Guillochon}, {Kotze},
  {Maguire}, {Siebert}, {Smith}, {Strader}, {Badenes}, {Kerzendorf}, {Koester},
  {Kromer}, {Miles}, {Pakmor}, {Schwab}, {Toloza}, {Toonen}, {Townsley}, \&
  {Williams}}]{Shen2018}
{Shen}, K.~J., {Boubert}, D., {G{\"a}nsicke}, B.~T., {et~al.} 2018, \apj, 865,
  15, \dodoi{10.3847/1538-4357/aad55b}

\bibitem[{{Shetrone} {et~al.}(2015){Shetrone}, {Bizyaev}, {Lawler}, {Allende
  Prieto}, {Johnson}, {Smith}, {Cunha}, {Holtzman}, {Garc{\'{\i}}a P{\'e}rez},
  {M{\'e}sz{\'a}ros}, {Sobeck}, {Zamora}, {Garc{\'{\i}}a-Hern{\'a}ndez},
  {Souto}, {Chojnowski}, {Koesterke}, {Majewski}, \& {Zasowski}}]{Shetrone2015}
{Shetrone}, M., {Bizyaev}, D., {Lawler}, J.~E., {et~al.} 2015, \apjs, 221, 24,
  \dodoi{10.1088/0067-0049/221/2/24}

\bibitem[{{Shipp} {et~al.}(2021){Shipp}, {Erkal}, {Drlica-Wagner}, {Li},
  {Pace}, {Koposov}, {Cullinane}, {Da Costa}, {Ji}, {Kuehn}, {Lewis}, {Mackey},
  {Simpson}, {Wan}, {Zucker}, {Bland-Hawthorn}, {Ferguson}, {Lilleengen}, \&
  {Lilleengen}}]{Shipp2021}
{Shipp}, N., {Erkal}, D., {Drlica-Wagner}, A., {et~al.} 2021, \apj, 923, 149,
  \dodoi{10.3847/1538-4357/ac2e93}

\bibitem[{{Smith} {et~al.}(2021){Smith}, {Bizyaev}, {Cunha}, {Shetrone},
  {Souto}, {Allende Prieto}, {Masseron}, {M{\'e}sz{\'a}ros}, {J{\"o}nsson},
  {Hasselquist}, {Osorio}, {Garc{\'\i}a-Hern{\'a}ndez}, {Plez}, {Beaton},
  {Holtzman}, {Majewski}, {Stringfellow}, \& {Sobeck}}]{Smith2021}
{Smith}, V.~V., {Bizyaev}, D., {Cunha}, K., {et~al.} 2021, \aj, 161, 254,
  \dodoi{10.3847/1538-3881/abefdc}

\bibitem[{{Soubiran} {et~al.}(2018){Soubiran}, {Jasniewicz}, {Chemin},
  {Zurbach}, {Brouillet}, {Panuzzo}, {Sartoretti}, {Katz}, {Le Campion},
  {Marchal}, {Hestroffer}, {Th{\'e}venin}, {Crifo}, {Udry}, {Cropper},
  {Seabroke}, {Viala}, {Benson}, {Blomme}, {Jean-Antoine}, {Huckle}, {Smith},
  {Baker}, {Damerdji}, {Dolding}, {Fr{\'e}mat}, {Gosset}, {Guerrier}, {Guy},
  {Haigron}, {Jan{\ss}en}, {Plum}, {Fabre}, {Lasne}, {Pailler}, {Panem},
  {Riclet}, {Royer}, {Tauran}, {Zwitter}, {Gueguen}, \& {Turon}}]{Soubiran2018}
{Soubiran}, C., {Jasniewicz}, G., {Chemin}, L., {et~al.} 2018, \aap, 616, A7,
  \dodoi{10.1051/0004-6361/201832795}

\bibitem[{{Unwin} {et~al.}(2008){Unwin}, {Shao}, {Tanner}, {Allen}, {Beichman},
  {Boboltz}, {Catanzarite}, {Chaboyer}, {Ciardi}, {Edberg}, {Fey}, {Fischer},
  {Gelino}, {Gould}, {Grillmair}, {Henry}, {Johnston}, {Johnston}, {Jones},
  {Kulkarni}, {Law}, {Majewski}, {Makarov}, {Marcy}, {Meier}, {Olling}, {Pan},
  {Patterson}, {Pitesky}, {Quirrenbach}, {Shaklan}, {Shaya}, {Strigari},
  {Tomsick}, {Wehrle}, \& {Worthey}}]{Unwin2008}
{Unwin}, S.~C., {Shao}, M., {Tanner}, A.~M., {et~al.} 2008, \pasp, 120, 38,
  \dodoi{10.1086/525059}

\bibitem[{{van der Marel} \& {Kallivayalil}(2014)}]{vanderMarel2014}
{van der Marel}, R.~P., \& {Kallivayalil}, N. 2014, \apj, 781, 121,
  \dodoi{10.1088/0004-637X/781/2/121}

\bibitem[{{Wilson} {et~al.}(2019){Wilson}, {Hearty}, {Skrutskie}, {Majewski},
  {Holtzman}, {Eisenstein}, {Gunn}, {Blank}, {Henderson}, {Smee}, {Nelson},
  {Nidever}, {Arns}, {Barkhouser}, {Barr}, {Beland}, {Bershady}, {Blanton},
  {Brunner}, {Burton}, {Carey}, {Carr}, {Colque}, {Crane}, {Damke}, {Davidson},
  {Dean}, {Di Mille}, {Don}, {Ebelke}, {Evans}, {Fitzgerald}, {Gillespie},
  {Hall}, {Harding}, {Harding}, {Hammond}, {Hancock}, {Harrison}, {Hope},
  {Horne}, {Karakla}, {Lam}, {Leger}, {MacDonald}, {Maseman}, {Matsunari},
  {Melton}, {Mitcheltree}, {O'Brien}, {O'Connell}, {Patten}, {Richardson},
  {Rieke}, {Rieke}, {Roman-Lopes}, {Schiavon}, {Sobeck}, {Stolberg}, {Stoll},
  {Tembe}, {Trujillo}, {Uomoto}, {Vernieri}, {Walker}, {Weinberg}, {Young},
  {Anthony-Brumfield}, {Bizyaev}, {Breslauer}, {De Lee}, {Downey}, {Halverson},
  {Huehnerhoff}, {Klaene}, {Leon}, {Long}, {Mahadevan}, {Malanushenko},
  {Nguyen}, {Owen}, {S{\'a}nchez-Gallego}, {Sayres}, {Shane}, {Shectman},
  {Shetrone}, {Skinner}, {Stauffer}, \& {Zhao}}]{Wilson2019}
{Wilson}, J.~C., {Hearty}, F.~R., {Skrutskie}, M.~F., {et~al.} 2019, \pasp,
  131, 055001, \dodoi{10.1088/1538-3873/ab0075}

\bibitem[{{Yoon} {et~al.}(2016){Yoon}, {Beers}, {Placco}, {Rasmussen},
  {Carollo}, {He}, {Hansen}, {Roederer}, \& {Zeanah}}]{Yoon2016}
{Yoon}, J., {Beers}, T.~C., {Placco}, V.~M., {et~al.} 2016, \apj, 833, 20,
  \dodoi{10.3847/0004-637X/833/1/20}

\bibitem[{{Yu} \& {Madau}(2007)}]{Yu2007}
{Yu}, Q., \& {Madau}, P. 2007, \mnras, 379, 1293,
  \dodoi{10.1111/j.1365-2966.2007.12034.x}

\bibitem[{{Zasowski} {et~al.}(2013){Zasowski}, {Johnson}, {Frinchaboy},
  {Majewski}, {Nidever}, {Rocha Pinto}, {Girardi}, {Andrews}, {Chojnowski},
  {Cudworth}, {Jackson}, {Munn}, {Skrutskie}, {Beaton}, {Blake}, {Covey},
  {Deshpande}, {Epstein}, {Fabbian}, {Fleming}, {Garcia Hernandez}, {Herrero},
  {Mahadevan}, {M{\'e}sz{\'a}ros}, {Schultheis}, {Sellgren}, {Terrien}, {van
  Saders}, {Allende Prieto}, {Bizyaev}, {Burton}, {Cunha}, {da Costa},
  {Hasselquist}, {Hearty}, {Holtzman}, {Garc{\'\i}a P{\'e}rez}, {Maia},
  {O'Connell}, {O'Donnell}, {Pinsonneault}, {Santiago}, {Schiavon}, {Shetrone},
  {Smith}, \& {Wilson}}]{Zasowski2013}
{Zasowski}, G., {Johnson}, J.~A., {Frinchaboy}, P.~M., {et~al.} 2013, \aj, 146,
  81, \dodoi{10.1088/0004-6256/146/4/81}

\bibitem[{{Zasowski} {et~al.}(2017){Zasowski}, {Cohen}, {Chojnowski},
  {Santana}, {Oelkers}, {Andrews}, {Beaton}, {Bender}, {Bird}, {Bovy},
  {Carlberg}, {Covey}, {Cunha}, {Dell'Agli}, {Fleming}, {Frinchaboy},
  {Garc{\'{\i}}a-Hern{\'a}ndez}, {Harding}, {Holtzman}, {Johnson}, {Kollmeier},
  {Majewski}, {M{\'e}sz{\'a}ros}, {Munn}, {Mu{\~n}oz}, {Ness}, {Nidever},
  {Poleski}, {Rom{\'a}n-Z{\'u}{\~n}iga}, {Shetrone}, {Simon}, {Smith},
  {Sobeck}, {Stringfellow}, {Szigeti{\'a}ros}, {Tayar}, \&
  {Troup}}]{Zasowski2017}
{Zasowski}, G., {Cohen}, R.~E., {Chojnowski}, S.~D., {et~al.} 2017, \aj, 154,
  198, \dodoi{10.3847/1538-3881/aa8df9}

\end{thebibliography}
\bibliographystyle{aasjournal}



\end{document}